\documentclass[english,11pt]{article}
\usepackage[utf8]{inputenc}
\usepackage{graphicx}
\usepackage{color}
\usepackage{float}
\usepackage{slashed}
\usepackage{subfigure}
\usepackage{amsthm,amsmath,amssymb,mathrsfs}
\usepackage{setspace}
\allowdisplaybreaks
\usepackage{jheppub}
\usepackage{xcolor} 
\usepackage{babel}
\usepackage{booktabs}
\usepackage{mathtools}
\usepackage{siunitx}
\usepackage{cancel}
\def\thefootnote{\fnsymbol{footnote}}

\usepackage[compat=1.1.0]{tikz-feynman}

\usepackage[normalem]{ulem}

\newcommand{\vecsigma}{\boldsymbol{\sigma}}
\newcommand{\vecr}{\bold{r}}
\newcommand{\vecp}{\bold{p}}
\newcommand{\vecP}{\bold{P}}
\newcommand{\vecq}{\bold{q}}
\newcommand{\veck}{\bold{k}}
\newcommand{\vecx}{\bold{x}}
\newcommand{\vecv}{\bold{v}}
\newcommand{\absk}{\left|\bold{k}\right|}
\newcommand{\absq}{\left|\bold{q}\right|}
\newcommand{\vbkg}{V_{\rm bkg}}
\newcommand{\db}{\lambda_{\rm dB}}
\newcommand{\VCPV}{V_{\cancel{\rm CP}}}
\newcommand{\gCPV}{g_{\cancel{\rm CP}}}

\begin{document}

\title{
    Axion forces in axion backgrounds
}

\author[a]{Yuval Grossman,}
\emailAdd{yg73@cornell.edu}
\author[a]{Bingrong Yu,}
\emailAdd{bingrong.yu@cornell.edu}
\author[a,b]{Siyu Zhou}
\emailAdd{sz682@cornell.edu}
\affiliation[a]{Department of Physics, LEPP, Cornell University, Ithaca, NY 14853, USA}
\affiliation[b]{University of Science and Technology of China, Hefei, Anhui 230026, China}

\abstract{Axions can naturally be very light due to the protection of an (approximate) shift symmetry. Because of their pseudoscalar nature, the long-range force mediated by the axion at tree level is spin dependent, which cannot lead to observable effects between two unpolarized macroscopic objects. At the one-loop level, however, the exchange of two axions does mediate a spin-independent force. This force is coherently enhanced in the presence of an axion background.
In this work, we study the two-axion exchange force in a generic axion background. 
We find that the breaking of the axion shift symmetry plays a crucial role in determining this force. The background-induced axion force $V_{\rm bkg}$ vanishes in the shift-symmetry restoration limit. The shift symmetry can be broken either explicitly by non-perturbative effects or effectively by the axion background. When the shift symmetry is broken,
$V_{\rm bkg}$ scales as $1/r$ and could be further enhanced by a large occupation number of the background axions. 
We investigate possible experimental probes of this effect in two distinct scenarios: an axion dark matter background and a solar axion flux, using fifth-force searches and atomic spectroscopy experiments. In the axion dark matter case, we find that the background-induced axion force can place strong constraints on axion couplings and masses, comparable to existing astrophysical bounds.}

\maketitle
\def\thefootnote{\arabic{footnote}}
\section{Introduction}
\label{sec:intro}
The axion appears as a pseudo-Nambu-Goldstone boson (pNGB) of a spontaneously broken $U(1)$ global symmetry~\cite{Wilczek:1977pj,Weinberg:1977ma}. It is well motivated both in particle physics to solve the strong CP problem~\cite{Peccei:1977ur,Peccei:1977hh} and in cosmology to serve as the candidate of cold dark matter (DM)~\cite{Dine:1982ah,Preskill:1982cy,Arvanitaki:2009fg,Arias:2012az}, see \cite{Marsh:2015xka,DiLuzio:2020wdo} for recent reviews of axion physics and its cosmology.
As a generic feature of pNGB, the axion mass is protected by the shift symmetry. 
Nonzero axion mass originates from non-perturbative instanton effects that lead to
\begin{align}
\label{eq:mass}
m_a \sim \frac{\mu_0^2}{f_a}\;,    
\end{align}
where $\mu_0$ is the scale of the non-perturbative effects that explicitly break the shift symmetry (for the QCD axion, $\mu_0\sim \Lambda_{\rm QCD}\approx 200~{\rm MeV}$), and $f_a$ (known as the axion decay constant) is the scale of the spontaneous $U(1)$ symmetry breaking, which characterizes the (inverse) axion coupling strength to the ordinary matter. We learn from Eq.~(\ref{eq:mass}) that $m_a$ can naturally be small if $\mu_0 \ll f_a$, as is usually assumed in axion models. Moreover, the radiative corrections to the axion mass are all suppressed by high powers of $f_a$, thus making the axion mass protected from quantum corrections. 

Given that the axion can naturally be very light from the theoretical point of view, one may expect a long-range force between macroscopic objects mediated by axions. This ``axion force'' already exists at the tree level, i.e., by the exchange of one axion~\cite{Moody:1984ba,Daido:2017hsl}. 
In this work, we consider only the CP-conserving axion interactions. 
Due to the pseudoscalar nature of the axion couplings,  the one-axion force between two fermions is spin-dependent and only results in the dipole-dipole interaction:
  \begin{align}
      V_a (\vecr) \sim \frac{1}{f_a^2}\left(\vecsigma_1\cdot\nabla\right)\left(\vecsigma_2\cdot \nabla\right) \frac{e^{-m_a r}}{4\pi r}\;,\label{eq:Va}
  \end{align}
where $\vecsigma_1$ and $\vecsigma_2$ are the spin vectors of the two external fermions. 
Note that since we do not include CP-violating axion couplings such as $a\bar{\psi}\psi$, there are no monopole-monopole or monopole-dipole interactions at the tree level. (For the searches of the axion force when there exist CP-violating interactions, see \cite{Georgi:1986kr,Stadnik:2013raa,Arvanitaki:2014dfa,ARIADNE:2017tdd,OHare:2020wah,Okawa:2021fto,Arvanitaki:2024dev,Fiorillo:2025zzx}.)

In the massless axion limit, Eq.~(\ref{eq:Va}) scales as $V_a\sim 1/(f_a^2 r^3)$. However, this force vanishes for unpolarized sources as the spin prefactor in Eq.~(\ref{eq:Va}) averages to zero (see App.~\ref{app:vacuum}). As a result, it cannot lead to observable effects in the traditional fifth-force experiments with unpolarized sources. 
This is in contrast with the CP-even scalar, where in the latter case the scalar-mediated force at the tree level is spin-independent and the fifth-force searches can probe very small scalar couplings in the small mass region~\cite{Hees:2018fpg,Gan:2023wnp}. The spin-dependent axion force in Eq.~(\ref{eq:Va}) can only be tested when the source has some net polarization; see \cite{Vasilakis:2008yn,Hunter:2013hza,Heckel:2013ina,Kotler:2015ura,Terrano:2015sna} for the detection of spin-dependent interactions with polarized sources (a recent review on this topic can be found in \cite{Cong:2024qly}). However, one should note that the dipole-dipole axion force scales as $1/r^3$, which is much smaller than the $1/r$ Yukawa force at long distances. As a result, the bounds from fifth-force searches on the axion interaction via Eq.~(\ref{eq:Va}) are much weaker than those
of a CP-even scalar.

The spin-independent axion force starts to appear at the one-loop level by the exchange of two axions, which is a quantum effect. This two-axion force was first calculated in \cite{Grifols:1994zz,Ferrer:1998ue} (see also \cite{Fischbach:1999iz}), in which they obtained $V_{2a}\sim 1/(f_a^4 r^5)$ for massless axions. 
Although this effect is coherent over macroscopic objects, it is heavily suppressed due to the $r^{-5}$ factor. 

Recently, it was pointed out in \cite{Bauer:2023czj,Rostagni:2024trx} that breaking the axion shift symmetry can enhance the two-axion force. More specifically, for the QCD axion, non-perturbative effect below the confinement scale induces a shift-symmetry-breaking coupling of the form $\mu a^2 \bar{N}N/f_a^2$ between axions and nucleons, where $\mu\sim {\cal O}(10)~{\rm MeV}$. Such a coupling leads to an axion force scaling as $V_{2a}\sim \mu^2/(f_a^4 r^3)$, which dominates over the one in the shift-invariant limit. Combined with the fifth-force searches at the laboratory length scale, it gives a constraint on the axion coupling $f_a \gtrsim 10~{\rm GeV}$~\cite{Bauer:2023czj}.

An important feature of the two-axion force is that it can be influenced by an axion background. 
In vacuum, the interaction between test particles is mediated by two virtual axions, both of which are generated from quantum fluctuations. However, when there exists an axion background, the interaction can also be mediated by the scattering between test particles and background axions. Intuitively, it is equivalent to putting one of the axion propagators on-shell.
Such an effect is coherent over background axions if their de Broglie wavelength is larger than the distance between the test particles. The background-induced axion force may change the $r$-dependence of its vacuum counterpart and lead to dominant observable effects.  

The background effect is a generic feature of quantum forces, which is essentially a result of coherent scattering. In recent years, there has been growing interest in studying this phenomenon, including the detection of the neutrino-mediated force in various neutrino backgrounds in the Standard Model framework~\cite{Ghosh:2022nzo,Ghosh:2024qai,Blas:2022ovz} and the applications of background effects in the search for light DM~\cite{Fukuda:2021drn,Kim:2022ype,Banerjee:2022sqg,Bouley:2022eer,Evans:2023uxh,Beadle:2023flm,Kim:2023pvt,Arza:2023wou,Day:2023mkb,VanTilburg:2024xib,Barbosa:2024zfz,Bauer:2024yow,Bauer:2024hfv,Evans:2024dty,Li:2024bbe,Zhou:2025wax,Banerjee:2025dlo,delCastillo:2025rbr,Gan:2025nlu,Gan:2025icr}; for earlier studies on this topic, see \cite{Horowitz:1993kw,Ferrer:1998ju,Ferrer:1999ad,Ferrer:1998rw,Ferrer:2000hm}. In particular, recent work \cite{VanTilburg:2024xib,Barbosa:2024zfz} calculated the DM background-induced forces where the DM particle couples quadratically to fermions and investigated their implications in fifth-force search experiments. 

The background-induced force from isotropic and thermal backgrounds was studied using the formalism of thermal field theory in~\cite{Horowitz:1993kw,Ferrer:1998ju,Ferrer:1999ad,Ferrer:1998rw,Ferrer:2000hm}. The case of anisotropic and non-thermal backgrounds was first studied in \cite{Ghosh:2022nzo} in the context of neutrino-mediated forces. Notably, it was found that there is a large enhancement along the direction parallel to the preferred direction of the anisotropic background, where the force scales as $1/r$~\cite{Ghosh:2022nzo}. However, it was also found that the finite spread of momentum and configuration space leads to a smearing at large distances where the background particles are no longer coherent, reducing the experimental senstivities to detect the force~\cite{Ghosh:2022nzo}. Later, this decoherence effect was studied in great detail in \cite{VanTilburg:2024xib} for bosonic DM mediators, where intuitive understandings were provided. Motivated by \cite{Ghosh:2022nzo,VanTilburg:2024xib}, in this work, we study the background effects on the axion-mediated forces, allowing for both isotropic and anisotropic axion backgrounds, and pay particular attention to the decoherence effect caused by the phase-space spread of background axions as well as the finite size of the experimental objects.

Various axion background profiles can remain today based on well-motivated particle physics and cosmological models. A huge occupation number of the background axions is expected to exist as relics if the axions make up the cold DM~\cite{Arias:2012az}. Axions can also be copiously produced in the early universe via different mechanisms and comprise the cosmic axion background at present day~\cite{Dror:2021nyr}. Moreover, an energetic axion flux with energies of order keV can be produced from the Sun through axion coupling to photons~\cite{CAST:2007jps}. 

Since the axion background may significantly change the axion force and affect the detection of axions, in this work, we aim to perform a systematic study on the axion forces in various axion backgrounds. Our main findings are:
\begin{itemize}
\item  
The magnitude of the background-induced axion force $\vbkg$ is determined by the degree of shift symmetry breaking in the theory. In particular, in the shift-symmetry restoration limit, $\vbkg$ vanishes at the leading order of the non-relativistic approximation of the external particles.
\item 
The existence of an axion background can effectively violate the shift symmetry, which enhances $V_{\rm bkg}$ if the energy scale of the background is larger than the axion mass. 
\item $\vbkg$ has a different scaling behavior from its vacuum counterpart $V_{2a}$. In particular, unlike $V_{2a}$, $\vbkg$ is generically not exponentially suppressed at long distances $r\gg 1/m_a$.
\item In the coherent region ($r \ll \db$), $\vbkg$ has a generic scaling behavior $\vbkg \sim n_a/r$ that is independent of the details of backgrounds, where $\db$ and $n_a$ are the typical de Broglie wavelength and the number density of background axions, respectively. 
\end{itemize}
Given the above features, we find that $\vbkg$ can dominate over $V_{2a}$ in a large region of the parameter space of axion mass and couplings.

Some of the results we obtain are not new and have appeared previously in the literature. Others, while not entirely new, are derived here using a different method that provides new insights. Finally, some results are entirely new. As we go through the paper, we elaborate on each case and clarify how our findings relate to existing work.

This paper is organized as follows. In Sec.~\ref{sec:formalism}, we derive the general formalism of the axion force in the existence of an arbitrary axion background, with both shift-invariant and breaking interactions taken into account. Then we apply it to studying the axion forces in three typical axion backgrounds in Sec.~\ref{sec:examples}. We discuss the specific experimental probes of the axion forces in Sec.~\ref{sec:detection}. A comparison between our results and the existing literature is provided in Sec.~\ref{sec:compare} and our main conclusion is summarized in Sec.~\ref{sec:conclusions}. Necessary technical details are provided in several appendices.

\section{The formalism of background-induced axion forces}
\label{sec:formalism}

In this section, we derive the general formulae of the two-axion-exchange force in an arbitrary axion background. We first focus on the scenario where the axion interactions are subject to shift-invariant symmetry but still allow for a finite axion mass. We show that $V_{\rm bkg}$ vanishes in the shift-invariant limit, regardless of the details of the background. 
Then we introduce sources of shift-symmetry breaking and calculate their effects on the axion force.

As we show below, the background-induced axion force between two fermions $\psi_1$ and $\psi_2$ has a generic form:
\begin{align}
 \boxed{
V_{\rm bkg}(r) \sim \frac{n_a}{4\pi r} \times \frac{m_1 m_2}{f_a^4}\times \frac{\text{order parameter}}{\text{background scale}}\times \text{decoherence factor}
\;.}
\label{eq:parametric}
\end{align}
In this expression,
\begin{itemize}
\item $n_a$ is the number density of background axions;
\item 
$m_i$ is the mass of $\psi_i$;
\item the ``order parameter'' is a dimensionless quantity (normalized by the fermion mass) that characterizes the magnitude of shift-symmetry breaking in a theory, which tends to zero in the shift-invariant limit;
    \item the ``background scale'' is the typical energy scale of the axion background; it is reduced to the axion mass for non-relativistic axion backgrounds (such as the axion cold DM);
\item 
the dimensionless ``decoherence factor''  describes the deviation of the axion force from a $1/r$ potential and can never exceed 1.
For $r\gtrsim \lambda_{\rm dB}$, $V_{\rm bkg}$ is more suppressed than $1/r$ due to the decoherence effect, while in the coherent limit $r\ll \lambda_{\rm dB}$, the decoherence factor is reduced to 1 regardless of the details of the background. 
\end{itemize}
In the following parts of this section, we discuss each term in Eq.~(\ref{eq:parametric}) with more details.

\begin{figure}[t]
    \centering
    \subfigure[Box 1]{
    \includegraphics[scale=0.9]
    {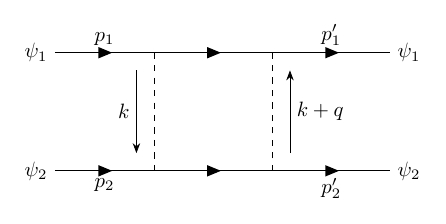}
    \label{subfig:box1}
    }
    \hspace{1cm}
    \subfigure[Box 2]{
    \includegraphics[scale=0.9]
    {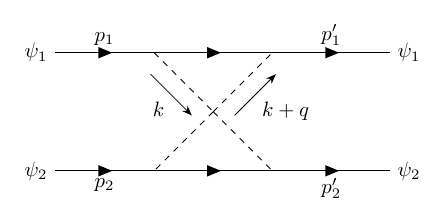}
    \label{subfig:box2}
    }\\
    \subfigure[Triangle 1]{
    \includegraphics[scale=0.9]
    {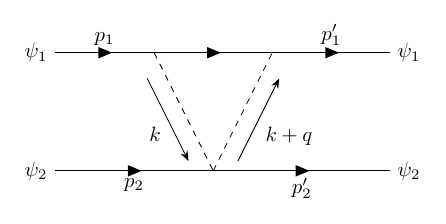}
    \label{subfig:tri1}
    }
    \hspace{1cm}
    \subfigure[Triangle 2]{
    \includegraphics[scale=0.9]
    {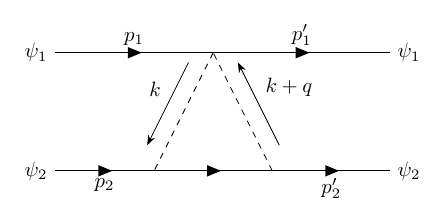}
    \label{subfig:tri2}
    }
    \subfigure[Bubble]{
    \includegraphics[scale=0.9]
    {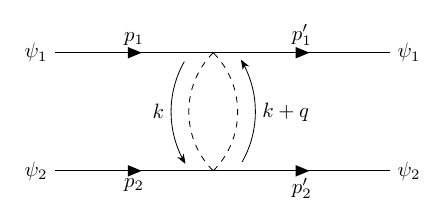}
    \label{subfig:bub}
    }
    \caption{Feynman diagrams of two-axion exchange between two fermions $\psi_1$ and $\psi_2$. Each of them can be affected by an axion background.}
    \label{fig:Feyn}
\end{figure}

\begin{figure}[t]
    \centering
    \subfigure[$\mathcal{M}_t^1\mathcal{M}_t^2$]{
    \includegraphics[scale=0.75]
    {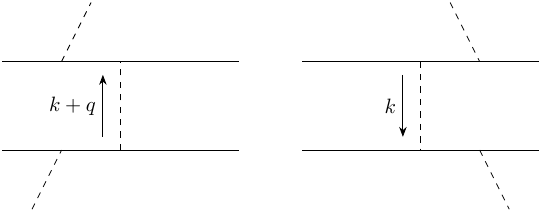}
    \label{subfig:bkg-tt}
    }
    \hspace{0.8cm}
    \subfigure[$\mathcal{M}_u^1\mathcal{M}_u^2$]{
    \includegraphics[scale=0.75]
    {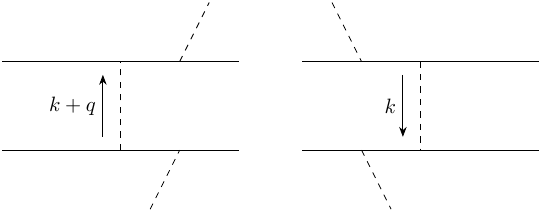}
    \label{subfig:bkg-uu}
    }\\
    \subfigure[$\mathcal{M}_t^1\mathcal{M}_u^2$]{
    \includegraphics[scale=0.75]
    {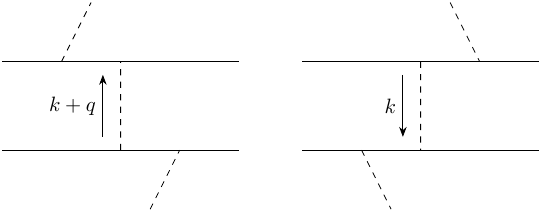}
    \label{subfig:bkg-tu}
    }
    \hspace{0.8cm}
    \subfigure[$\mathcal{M}_u^1\mathcal{M}_t^2$]{
    \includegraphics[scale=0.75]
    {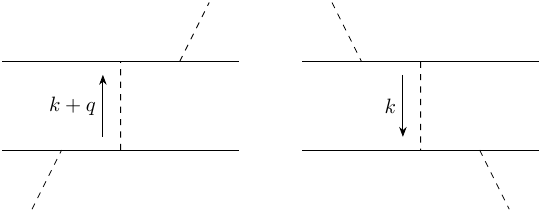}
    \label{subfig:bkg-ut}
    }
    \caption{Effective box diagrams in the medium of axions that contribute to the background-induced axion force through coherent scattering. The four diagrams in the first line correspond to  the ``Box 1'' diagram in Fig.~\ref{fig:Feyn}, while those in the second line correspond to the ``Box 2'' diagram in Fig.~\ref{fig:Feyn}. See App.~\ref{app:coherent-scattering} for detailed calculations.}
    \label{fig:background boxes}
\end{figure}
\subsection{Shift-invariant axion interaction}

We start from the effective interaction between the axion, $a$, and a fermion, $\psi$:
\begin{align}
\mathcal{L}_{\rm eff} = \frac{1}{2}\left(\partial_{\mu}a\right)\left(\partial^{\mu}a\right)-\frac{1}{2}m_{a}^{2}a^{2}+\bar{\psi}\left(i\slashed{\partial}-m_\psi\right)\psi+{\cal L}_{\rm int}\;.\label{eq:Leff} 
\end{align}
The axion-fermion interaction term ${\cal L}_{\rm int}$ can be expressed in two equivalent forms (see App.~\ref{app:mass-vs-derivative} for details).
In the \emph{derivative basis}, the interaction term is written as
\begin{align}
{\cal L}_{\rm int}^{\rm d}=\frac{c_{\psi}}{2}\frac{\partial_{\mu}a}{f_{a}}\bar{\psi}\gamma^{\mu}\gamma_{5}\psi
\;.\label{eq:Lintd}   
\end{align}
which has an explicit shift invariance: $a \to a + {\rm constant}$. Equivalently, one can work it to the \emph{pseudoscalar basis}: 
\begin{align}
{\cal L}_{\rm int}^{\rm p} = - i c_\psi m_\psi \frac{a}{f_a}  \bar{\psi} \gamma_5 \psi  + \frac{1}{2} c_\psi^2 m_\psi \frac{a^2}{f_a^2} \bar{\psi}\psi + {\cal O}\left(\frac{a^3}{f_a^3}\right). \label{eq:Leffp}
\end{align}
In Eq.~(\ref{eq:Leffp}), the shift invariance is kept  up to ${\cal O}(1/f_a^2)$. Since we are interested in the two-axion effect in this work, it is sufficient to neglect the ${\cal O}(1/f_a^3)$ terms in the pseudoscalar basis. 

The axion mass in Eq.~(\ref{eq:Leff}) is the only term that breaks shift symmetry; it may come from unspecific nonperturbative effects at high-energy scales. In the following discussion we treat $m_a$ as a free parameter for the purpose of generality.

At the tree-level, the derivative coupling in Eq.~(\ref{eq:Lintd}) induces an axion force between two fermions that is spin-dependent (see App.~\ref{app:vacuum}). We are interested in the spin-independent axion force, which starts to appear at the one-loop level (see Fig.~\ref{fig:Feyn}). Note that in the derivative basis there is only linear axion coupling, so only the two box diagrams in Fig.~\ref{fig:Feyn}  contribute, while in the pseudoscalar basis all five diagrams in Fig.~\ref{fig:Feyn} are relevant because there are both linear and quadratic couplings. However, the physical observable should not be affected by the basis we choose.

\subsubsection{Calculation of the amplitude}

To extract the axion force, one needs to calculate the amplitude of the elastic scattering $\psi_1 (p_1) + \psi_2 (p_2) \to \psi_1 (p_1') + \psi_2 (p_2')$ mediated by two axions with a momentum transfer $q=p_1'-p_1=p_2-p_2'$. The effective potential mediated by axions is given by the Fourier transform of the scattering amplitude in the non-relativistic (NR) limit,
\begin{align}
    V(\vecr) = - \frac{1}{4m_1 m_2}\int \frac{{\rm d}^3 \vecq}{\left(2\pi\right)^3}e^{i\vecq\cdot\vecr} {\cal M}_{\rm NR} (\vecq)\;.\label{eq:fourier}
\end{align}

In the NR limit, $q^\mu\approx (0,\vecq)$, $4m_1m_2$ is the normalization factor, and ${\cal M}_{\rm NR}$ denotes the scattering amplitude in the NR limit. The scattering amplitude includes contributions from the relevant diagrams in Fig.~\ref{fig:Feyn}, which can be calculated using quantum field theories. For each of the diagrams, as we show in a while, the amplitude can be decomposed into the production of two tree-level Compton scattering amplitudes.

When there exists an axion background, the interaction can also be mediated by coherent scattering between $\psi_i$ and on-shell background axions (see Fig.~\ref{fig:background boxes}). To include the background effect, one can replace the vacuum axion propagator with the modified propagator:
\begin{align}
D_{\rm mod}(k) = \frac{i}{k^2-m_a^2} + 2 \pi \delta\left(k^2-m_a^2\right)\Theta(k^0)f(\veck)\equiv D_{\rm vac}(k) + D_{\rm bkg}(k)\;,
\label{eq:mod-propagator}
\end{align}
where $\Theta$ is the step function and $f(\veck)$ is the phase-space distribution of background axions, normalized as
\begin{align}
    n_a = \int \frac{{\rm d}^3\veck}{\left(2\pi\right)^3} f(\veck)\;.
\end{align}

The modified propagator in Eq.~(\ref{eq:mod-propagator}) contains two pieces: the first part is the axion propagator in vacuum; the second part comes from the background correction, where the existence of $\delta$-function implies that the background axions are on-shell. In addition, $D_{\rm bkg}$ is proportional to the phase-space number density, indicating that the background correction is a coherent effect, i.e., it is proportional to the axion number density at the amplitude level. The formalism in Eq.~(\ref{eq:mod-propagator}) has been widely used in the literature to calculate the background corrections in various cases~\cite{Notzold:1987ik,Horowitz:1993kw,Ferrer:1998ju,Ferrer:1999ad,Ferrer:2000hm,Ghosh:2022nzo,Ghosh:2024qai,Blas:2022ovz,Barbosa:2024zfz}. It provides a compact and model-independent way to parametrize the background effect. More specifically, for each of the diagram in Fig.~\ref{fig:Feyn},
\begin{itemize}
    \item when both axion propagators take $D_{\rm vac}$, it gives the two-axion force in vacuum $V_{2a}$;
    \item when one of the propagators takes $D_{\rm vac}$ and the other takes $D_{\rm bkg}$, it gives the background-induced axion force $\vbkg$;
    \item when both propagators take $D_{\rm bkg}$,  it does not contribute to the interaction between $\psi_1$ and $\psi_2$. 
\end{itemize}

We emphasize that Eq.~(\ref{eq:mod-propagator}) is essentially a result of coherent scattering between $\psi_i$ and the background axions, as shown in Fig.~\ref{fig:background boxes}. As a result, its validity does \emph{not} require the background axions to be in thermal equilibrium. In App.~\ref{app:coherent-scattering}, we show how to derive the same result of the axion force using coherent scattering without assuming Eq.~(\ref{eq:mod-propagator}). This is inspired by the method proposed in \cite{VanTilburg:2024xib}, which derived the background-induced force for quadratically coupled meditors (corresponding to the bubble diagram) in the language of the coherent scattering.

With the modified propagator in hand, we proceed to calculate the scattering amplitude in Eq.~(\ref{eq:fourier}). For the relevant diagrams in Fig.~\ref{fig:Feyn}, the total amplitude is given by
\begin{align}
i\mathcal{M}_{\rm tot}=\int\frac{\text{d}^{4}k}{\left(2\pi\right)^{4}}\int\frac{\text{d}^{4}k^{\prime}}{\left(2\pi\right)^{4}} &\left(2\pi\right)^{4}\delta^{(4)}\left(q+k-k^{\prime}\right)D_{\rm mod}(k)D_{\rm mod}(k')\nonumber\\
&\times i\mathcal{M}_{C}\left(p_{1},p_{1}^{\prime};-k,-k^{\prime}\right)i\mathcal{M}_{C}\left(p_{2},p_{2}^{\prime};k,k^{\prime}\right).
\label{eq:amplitude-tot}
\end{align}
Here ${\cal M}_C\left(p_{\rm in},p_{\rm out};k_{\rm in},k_{\rm out}\right)$ denotes the amplitude of the tree-level Compton scattering $\psi(p_{\rm in})+a(k_{\rm in}) \to \psi(p_{\rm out}) + a(k_{\rm out})$, which is calculated in App.~\ref{app:compton}:
\begin{align}
  {\cal M}_C\left(p_{\rm in},p_{\rm out};k_{\rm in},k_{\rm out}\right)=\frac{c_{\psi}^{2}}{f_a^{2}}m_{\psi}^{2}\bar{u}\left(p_{\rm out}\right)\left(\frac{\slashed{k}_{\rm in}}{k_{\rm in}^2+2k_{\rm in} \cdot p_{\rm in}}-\frac{\slashed{k}_{\rm out}}{k_{\rm out}^2-2k_{\rm out}\cdot p_{\rm in}}-\frac{1}{m_{\psi}}\right)u\left(p_{\rm in}\right),\label{eq:compton}
\end{align}
where $u(p)$ is the wavefunction of the external fermion with momentum $p$. As expected, we confirm in App.~\ref{app:compton} that the calculations of the Compton scattering amplitude in both derivative basis and pseudoscalar basis provide the same result (\ref{eq:compton}) up to ${\cal O}\left(1/f_a^2\right)$. Consequently, the amplitude in Eq.~(\ref{eq:amplitude-tot}) agrees in both bases as well.

For the purpose of illustration, it is instructive to split ${\cal M}_C$ into three parts:
\begin{align}
{\cal M}_C \equiv {\cal M}_t^{\rm p} + {\cal M}_u^{\rm p} + {\cal M}_c^{\rm p}\;,\label{eq:split} 
\end{align}
according to the three terms in the bracket of Eq.~(\ref{eq:compton}). Here ${\cal M}_t^{\rm p}$, ${\cal M}_u^{\rm p}$, and ${\cal M}_c^{\rm p}$ correspond to the amplitude of the $t$-channel, $u$-channel, and contact scattering in the pseudoscalar basis, respectively, as shown in Fig.~\ref{fig:compton} and Eqs.~(\ref{eq:Mtp})-(\ref{eq:Mcp}). Then the multiplication of two ${\cal M}_C$ in Eq.~(\ref{eq:amplitude-tot}) can be schematically written as:
\begin{align}
{\cal M}_C {\cal M}_C &\sim \underbrace{\left({\cal M}_t^{\rm p}{\cal M}_t^{\rm p} + {
\cal M
}_u^{\rm p}{
\cal M
}_u^{\rm p}\right)}_\text{Box 1} + \underbrace{\left({\cal M}_t^{\rm p} {\cal M}_u^{\rm p}+{\cal M}_u^{\rm p} {\cal M}_t^{\rm p}\right)}_\text{Box 2} + \underbrace{{\cal M}_c^{\rm p}{\cal M}_c^{\rm p}}_\text{Bubble}\nonumber\\
&+\underbrace{\left({\cal M}_t^{\rm p} + {\cal M}_u^{\rm p}\right) {\cal M}_c^{\rm p}}_\text{Triangle 1}+\underbrace{{\cal M}_c^{\rm p}\left({\cal M}_t^{\rm p} + {\cal M}_u^{\rm p}\right)}_\text{Triangle 2}\;, \label{eq:schematical}     
\end{align}
where the text indicates the corresponding loop amplitude in Fig.~\ref{fig:Feyn} in the pseudoscalar basis. 
Therefore, the multiplication of the two ${\cal M}_C$ in Eq.~(\ref{eq:amplitude-tot}) indeed captures all the relevant loop diagrams in Fig.~\ref{fig:Feyn}, with ${\cal M}_C$ given by Eq.~(\ref{eq:compton}).

Next we turn to the axion propagators in Eq.~(\ref{eq:amplitude-tot}). When both of the propagators take the vacuum part, it gives the amplitude of the two-axion force in vacuum~\cite{Ferrer:1998ue,Bauer:2023czj}. The effect from background corrections is extracted by taking the ``crossing term'', namely, only one of the propagators is on-shell:
\begin{align}
i\mathcal{M}_{\rm bkg}=\int\frac{\text{d}^{4}k}{\left(2\pi\right)^{4}}\int\frac{\text{d}^{4}k^{\prime}}{\left(2\pi\right)^{4}} &\left(2\pi\right)^{4}\delta^{(4)}\left(q+k-k^{\prime}\right)\left[D_{\rm bkg}(k)D_{\rm vac}(k')+D_{\rm vac}(k)D_{\rm bkg}(k')\right]\nonumber\\
&\times i\mathcal{M}_{C}\left(p_{1},p_{1}^{\prime};-k,-k^{\prime}\right)i\mathcal{M}_{C}\left(p_{2},p_{2}^{\prime};k,k^{\prime}\right).
\label{eq:amplitude-bkg}
\end{align}
Substituting Eq.~(\ref{eq:mod-propagator}) into Eq.~(\ref{eq:amplitude-bkg}) one obtains:
	\begin{align}
		\mathcal{M}_{\rm bkg}=-\int&\frac{\text{d}^{3}\veck}{\left(2\pi\right)^{3}}f(\veck)\int {\rm d}k^0 \delta\left(k^2-m_a^2\right)\Theta\left(k^0\right) \left[\frac{\mathcal{M}_{C}\left(p_{1},p_{1}^{\prime};-k,-k-q\right)\mathcal{M}_{C}\left(p_{2},p_{2}^{\prime};k,k+q\right)}{(k+q)^2-m_a^2}\right.\nonumber\\
		&\left.+\frac{\mathcal{M}_{C}\left(p_{1},p_{1}^{\prime};-k+q,-k\right)\mathcal{M}_{C}\left(p_{2},p_{2}^{\prime};k-q,k\right)}{(k-q)^2-m_a^2}\right],
	\end{align}
where we have performed the shift $k\rightarrow k-q$ in the second term in the bracket.

The existence of an axion background effectively violates the Lorentz symmetry through the time-independent phase-space distribution $f(\veck)$. Therefore, it is most convenient to first integrate out the time component $k^0$ using the $\delta$-function, leading to
\begin{align}
		\mathcal{M}_{\rm bkg}=-\int\frac{\text{d}^{3}k}{\left(2\pi\right)^{3}}&\frac{f(\veck)}{2E_{\veck}}\left[\frac{\mathcal{M}_{C}\left(p_{1},p_{1}^{\prime};-k,-k-q\right)\mathcal{M}_{C}\left(p_{2},p_{2}^{\prime};k,k+q\right)}{(k+q)^2-m_a^2}\right.\nonumber\\
		&\left.+\frac{\mathcal{M}_{C}\left(p_{1},p_{1}^{\prime};-k+q,-k\right)\mathcal{M}_{C}\left(p_{2},p_{2}^{\prime};k-q,k\right)}{(k-q)^2-m_a^2}\right]\Bigg|_{k^0 = E_{\veck}},
	\label{eq:amplitude-bkg-simplified}
\end{align}
where $E_{\veck} \equiv (\veck^2+m_a^2)^{1/2}$. According to Eq.~(\ref{eq:schematical}), the contraction of two ${\cal M}_C$ involves multiple terms, which can be calculated separately. So we can write
\begin{align}
{\cal M}_{\rm bkg} ={\cal M}_\text{box,$1$}^{\rm p} + {\cal M}_\text{box,$2$}^{\rm p}  +
{\cal M}_\text{tri,$1$}^{\rm p} + {\cal M}_\text{tri,$2$}^{\rm p}  + {\cal M}_\text{bub}^{\rm p} \;, \label{eq:Mbkg-split}
\end{align}
where each term denotes the  contribution from the corresponding diagram in Fig.~\ref{fig:Feyn} in the pseudoscalar basis. Taking the NR limit  and expanding the wavefunctions up to the leading order of velocity and performing the Fourier transform in Eq.~(\ref{eq:fourier}), we obtain the (spin-independent) background-induced axion force (see App.~\ref{app:integral} for details):
\begin{align}
    V_{\rm bkg}^{\rm inv}(\vecr) = -\frac{c_1^2 c_2^2}{4\pi r}\frac{m_1 m_2}{f_a^4}\int\frac{\text{d}^3\veck}{(2\pi)^3}\frac{f(\veck)}{2E_{\veck}}{\cal Q}_{\rm inv}\left[\cos\left(\left|\veck\right| r-\veck\cdot\vecr\right)+\cos\left(\left|\veck\right|r+\veck\cdot\vecr\right)\right], \label{eq:Vbkg} 
\end{align}
where  $r\equiv \left|\vecr\right|$, $c_i\equiv c_{\psi_i}$ is the coupling in Eq.~(\ref{eq:Leff}) between axion and $\psi_i$, and
\begin{align}
 Q_{\rm inv} = \sum_j{\cal Q}_j\;, \qquad j= \text{box,1,\, box,2,\, tri,1,\, tri,2,\, bub}\;,\label{eq:Qsum}    
\end{align}
where
\begin{align}
{\cal Q}_\text{box,1} &=  \frac{2E_{\veck}^2 m_1 m_2\left(4 E_{\veck}^2 m_1 m_2 - m_a^4\right)}{\left(4m_1^2 E_{\veck}^2-m_a^4\right)\left(4m_2^2 E_{\veck}^2-m_a^4\right)}\;,\label{eq:Qbox1}\\
{\cal Q}_\text{box,2} &=  \frac{2E_{\veck}^2 m_1 m_2\left(4 E_{\veck}^2 m_1 m_2 + m_a^4\right)}{\left(4m_1^2 E_{\veck}^2-m_a^4\right)\left(4m_2^2 E_{\veck}^2-m_a^4\right)}\;,\label{eq:Qbox2}\\
{\cal Q}_\text{tri,1} & = \frac{-4m_1^2 E_{\veck}^2}{4m_1^2 E_{\veck}^2-m_a^4}\;,\label{eq:Qtri1}\\
{\cal Q}_\text{tri,2} & = \frac{-4m_2^2 E_{\veck}^2}{4m_2^2 E_{\veck}^2-m_a^4}\;,\label{eq:Qtri2}\\
{\cal Q}_\text{bub} & = 1\;.\label{eq:Qbub}
\end{align}
\renewcommand\arraystretch{1.8}
\begin{table}[t]
	\centering
\begin{tabular}{c|c|c|c|c|c|c}
\hline\hline
Order Parameters & ${\cal Q}_\text{box,1}$ & ${\cal Q}_\text{box,2}$ & ${\cal Q}_\text{tri,1}$ & ${\cal Q}_\text{tri,2}$ & ${\cal Q}_\text{bub}$ & ${\cal Q}_\text{inv}$\\
\hline
Full Expressions & (\ref{eq:Qbox1}) & (\ref{eq:Qbox2}) & (\ref{eq:Qtri1}) & (\ref{eq:Qtri2}) & (\ref{eq:Qbub}) & (\ref{eq:Qinv})\\
\hline
$m_i \gg m_a$ & $1/2$ & $1/2$ & $-1$ & $-1$ & 1 & $\frac{m_a^8}{16m_1^2 m_2^2 E_{\veck}^4}$ \\
\hline
$m_i \ll m_a$ & $-\frac{2E_{\veck}^2 m_1 m_2}{m_a^4}$ & $\frac{2E_{\veck}^2 m_1 m_2}{m_a^4}$ & $\frac{4m_1^2 E_{\veck}^2}{m_a^4}$ & $\frac{4m_2^2 E_{\veck}^2}{m_a^4}$  & 1 & 1 \\
\hline\hline
\end{tabular}
\caption{\label{tab:order-parameter}The order parameters ${\cal Q}_j$ (in the pseudoscalar basis) that break the shift symmetry of the diagram in Fig.~\ref{fig:Feyn}. The background-induced axion force from each diagram is proportional to the corresponding order parameter. The order parameter with shift-invariant axion interaction (\ref{eq:Leff}) is given by the sum of five terms: ${\cal Q}_{\rm inv}=\sum_j {\cal Q}_j$. The axion massless limit and  heavy limit have been taken in the last two lines, respectively.}
\end{table}
\renewcommand\arraystretch{1.0}

The result of the axion force in Eq.~(\ref{eq:Vbkg}) and the ${\cal Q}$-factors in Eqs.~(\ref{eq:Qbox1})-(\ref{eq:Qbub}) are valid up to leading order in the NR approximation, and are applicable for any axion background. Their derivations are provided in App.~\ref{app:integral}. The values of ${\cal Q}$-factors in the hierarchical limit ($m_a \ll m_i$ or $m_a \gg m_i$) are collected in Tab.~\ref{tab:order-parameter}. 

\subsubsection{Order parameters}
\label{subsubsec:order-parameter}
The ${\cal Q}$-factors play the role of the order parameters, which characterize the degree of shift-symmetry breaking. To see this, we add Eqs.~(\ref{eq:Qbox1})-(\ref{eq:Qbub}) together and find that the terms that are not proportional to the power of the axion mass get exactly cancelled:
\begin{align}
 {\cal Q}_{\rm inv} = \sum_j {\cal Q}_j = \frac{m_a^8}{\left(4m_1^2 E_{\veck}^2-m_a^4\right)\left(4m_2^2 E_{\veck}^2-m_a^4\right)}\;. \label{eq:Qinv}  
\end{align}
When there exists only shift-invariant interaction between axion and fermions [Eq.~(\ref{eq:Leff})], the axion mass is the only source that breaks the shift symmetry. As a result, the order parameter ${\cal Q}_{\rm inv}$ vanishes as $m_a \to 0$.

The background-induced force in Eq.~(\ref{eq:Vbkg}) scales as $V_{\rm bkg}^{\rm inv}\sim {\cal Q}_{\rm inv}/r$ at the leading order of the NR approximation, and thus vanishes as $m_a \to 0$.
One may wonder why the $1/r$ long-range potential from the background correction disappears in the $m_a \to 0$ limit that corresponds to the restoration limit of the shift symmetry. This can be understood in the following way: First of all, as shown in App.~\ref{app:coherent-scattering}, the background effect only exists when there are at least two axions involved in the scattering process. The effective Hamiltonian responsible for the interaction between the fermion $\psi$ and two axions can be written as 
\begin{align}
{\cal H}_{\rm eff} = \frac{1}{\Lambda}\bar{\psi}\Gamma\psi a^2\;, \label{eq:Heff}  
\end{align}
where $\Lambda$ is the cutoff scale depending on the fermion mass and the axion decay constant, $\Gamma$ denotes the possible Lorentz structure that is determined by calculating the elastic scattering $\psi + a \to \psi + a$ in Fig.~\ref{fig:compton}. Note that Eq.~(\ref{eq:Heff}) could result from the quadratic vertex in Fig.~\ref{subfig:contact}, as well from the combination of two linear vertices in Fig.~\ref{subfig:t-channel} or \ref{subfig:u-channel}.
The $1/r$ potential is a typical tree-level effect, which (intuitively) is obtained  by replacing one of the axion fields with its classical field value $a_0$ in Eq.~(\ref{eq:Heff}): 
\begin{align}
{\cal H}_{\rm eff}\to {\cal H}_\text{eff,bkg} = \frac{a_0}{\Lambda}\bar{\psi}\Gamma\psi a\;.    
\end{align}
Then one can treat $a_0/\Lambda$ as the constant coupling and the remaining part as an analogy of the Yukawa potential, getting $V_{\rm bkg}\sim (a_0/\Lambda)^2/r$. 
However, in the absence of the shift-symmetry breaking, the classical field value $a_0$ can always be shifted to zero using the degree of freedom of shift invariance, which then forces $V_{\rm bkg}$ to vanish.

In general, the breaking of the shift symmetry is manifested in three effects  that induce a nonvanishing long-range potential $V_{\rm bkg}$: ($a$) the axion has a nonzero mass; ($b$) the axion-fermion coupling explicitly breaks the shift symmetry; ($c$) the existence of the axion background sets a typical energy scale that effectively breaks the shift symmetry --- this effect is more significant than the axion mass effect when background axions are relativistic. In scenario ($a$), $V_{\rm bkg}$ is suppressed by the power of the axion mass, making it difficult to detect. In Sec.~\ref{subsec:shift-break}, we will focus on scenarios ($b$) and ($c$).

\subsubsection{Decoherence factor}
\label{subsubsec:decoherence}
There is a decoherence factor in Eq.~(\ref{eq:Vbkg}):
\begin{align}
{\cal D} \equiv \frac{1}{2} \left[\cos\left(\absk r-\veck\cdot\vecr\right)+\cos\left(\absk r+\veck\cdot\vecr\right)\right].\label{eq:Dfactor}
\end{align}
It describes the departure of $V_{\rm bkg}$ from a $1/r$ potential at long distances. 

Suppose that the phase-space distribution $f(\veck)$ has a density around a peaked value $\veck=\veck_0$, then the typical de Broglie wavelength of the background axions is given by $\lambda_{\rm dB}\sim 1/|\veck_0|$. The background axions inside the volume $\lambda_{\rm dB}^3$ can coherently scatter with the test particles. This can be seen from Eq.~(\ref{eq:Dfactor}) by noticing that ${\cal D}$ approaches 1 for $r\ll \lambda_{\rm dB}$, and we have $V_{\rm bkg}\sim 1/r$ in this case. 

On the other hand, the background axion field at two places departing from $r\gtrsim \lambda_{\rm dB}$ generically has different phases. As a result, their contributions to $V_{\rm bkg}$ are destructive, making it more suppressed than $1/r$. This is described by the cosine terms in Eq.~(\ref{eq:Dfactor}), which have oscillating behaviors when the argument is larger than ${\cal O}(1)$ and kills the leading $1/r$ term after integrating over phase space. In most cases, due to the decoherence factor in Eq.~(\ref{eq:Dfactor}), $V_{\rm bkg}$ is additionally suppressed by a factor of $(\lambda_{\rm dB}/r)^n$ compared to $1/r$ at long distances $r \gg \lambda_{\rm dB}$, where the power index $n$ depends on the specific form of $f(\veck)$. We will show some concrete examples in Sec.~\ref{sec:examples}.

Similar decoherence effects of background-induced forces at long distances have also been noticed recently in \cite{Ghosh:2022nzo,Ghosh:2024qai,Blas:2022ovz} for neutrinos and in \cite{VanTilburg:2024xib,Barbosa:2024zfz} for quadratically coupled scalar DM. We note that the form of the decoherence factor in Eq.~(\ref{eq:Dfactor}) is slightly different from that in \cite{VanTilburg:2024xib,Barbosa:2024zfz}: Ref.~\cite{VanTilburg:2024xib} obtained ${\cal D}=\cos\left(\absk r-\veck\cdot \vecr\right)$ while in Ref.~\cite{Barbosa:2024zfz} it is given by ${\cal D}=\cos\left(2\veck\cdot\vecr\right)$.
Technically, the difference with respect to Ref.~\cite{VanTilburg:2024xib} arises since in Ref.~\cite{VanTilburg:2024xib} the decoherence factor was obtained by using retarded propagators for the off-shell mediator, while we used Feynman propagators (see App.~\ref{app:integral} for more details). With regard to Ref.~\cite{Barbosa:2024zfz}, they also used Feynman propagators, but the disagreement emerges from a different order of integration. While that does not affect the result in case of isotropic backgrounds, it does lead to a deviation for anisotropic backgrounds.\footnote{We thank Sergio Barbosa and Sylvain Fichet for the correspondence about \cite{Barbosa:2024zfz}, in which they agreed on that point.}
Note that the three expressions for ${\cal D}$ reduce to 1 in the coherent region $r\ll \lambda_{\rm dB}$; they also agree with each other after integrating over the phase space for isotropic backgrounds. There is, however, an ${\cal O}(1)$ difference between them for anisotropic backgrounds at $r\gtrsim \lambda_{\rm dB}$.

\subsubsection{Further discussions}
We close this subsection with a brief remark about other scenarios in which the formulae derived in this subsection can be used. Although Eqs.~(\ref{eq:Vbkg})-(\ref{eq:Qbub}) are applicable for the axion force with only shift-invariant interactions, one can also use them to conveniently extract the background-induced force in theories with explicit shift-symmetry breaking:
\begin{itemize}
    \item By replacing ${\cal Q}_{\rm inv}$ with ${\cal Q}_{\rm bub}$ in Eq.~(\ref{eq:Vbkg}), one gets:
    \begin{align}
      V_{\rm bkg}^{\rm bub} (\vecr) = -\frac{c_1^2 c_2^2}{4\pi r}\frac{m_1 m_2}{f_a^4}\int\frac{\text{d}^3\veck}{(2\pi)^3}\frac{f(\veck)}{E_{\veck}}\,{\cal D}\;.    
    \end{align}
This is the background-induced force in a theory where the mediator is only quadratically coupled to the SM fermions: ${\cal L}_{\rm quad}= c_i^2 m_i a^2\bar{\psi}_i\psi_i/(2f_a^2)$, as studied in \cite{VanTilburg:2024xib,Barbosa:2024zfz}.

    \item In addition, replace ${\cal Q}_{\rm inv}$ with the two box terms ${\cal Q}_\text{box,1}+{\cal Q}_\text{box,2}$ in Eq.~(\ref{eq:Vbkg}):
\begin{align}
V_{\rm bkg}^{\rm box} (\vecr) &= -\frac{c_1^2 c_2^2}{4\pi r}\frac{m_1 m_2}{f_a^4}\int\frac{\text{d}^3\veck}{(2\pi)^3}\frac{f(\veck)}{E_{\veck}}\frac{16 E_\veck^4 m_1^2 m_2^2}{\left(4m_1^2 E_\veck^4 - m_a^4\right)\left(4m_2^2 E_\veck^2 - m_a^4\right)}\,{\cal D}.
\end{align}
This corresponds to a theory where there is a pure pseudoscalar coupling: ${\cal L}_{\rm pseudo}= -ic_i m_i a \bar{\psi}_i\gamma_5\psi_i/f_a$. Note that in this case, the shift symmetry is explicitly broken by the fermion mass from the coupling. In the massless axion limit $m_a \ll m_i$, we have ${\cal Q}_\text{box,1}+{\cal Q}_\text{box,2} \approx 1$ and $V_{\rm bkg}^{\rm box} (r) \approx V_{\rm bkg}^{\rm bub} (r)$.  
\end{itemize}

\subsection{Breaking of the shift symmetry}
\label{subsec:shift-break}
We have seen from Sec.~\ref{subsubsec:order-parameter} that shift symmetry breaking is crucial to having a long-range $1/r$ axion force. In this subsection, we consider two scenarios of the axion shift symmetry breaking beyond the axion mass term and calculate their effects on $V_{\rm bkg}$: ($i$) the non-perturbative effect induces a quadratic coupling between axions and fermions; 
($ii$) the axion background modifies the dispersion relation of the axion propagator by enforcing it to be on-shell, which effectively violates the shift symmetry. 

\subsubsection{Quadratic coupling from non-perturbative effects}
The axion quadratic terms explicitly break the shift symmetry. Note that although there is a quadratic coupling $a^2 \bar{\psi}\psi$ in Eq.~(\ref{eq:Leffp}), its contribution to the axion force is cancelled by the linear coupling, which coexists in the pseudoscalar basis. As a result, cancellation occurs among the order parameters in Eqs.~(\ref{eq:Qbox1})-(\ref{eq:Qbub}), making the remaining term in Eq.~(\ref{eq:Qinv}) suppressed by the axion mass in the leading order of the NR approximation. This cancellation is expected because the shift symmetry is explicitly conserved in the derivative basis Eq.~(\ref{eq:Lintd}), which is equivalent to Eq.~(\ref{eq:Leffp}) up to ${\cal O}(1/f_a^2)$. In the following, we consider additional sources of quadratic coupling that are not related to the linear coupling.

We consider couplings that emerge from non-perturbative effects from the axion potential. They become relevant when the energy scale is below $\mu_0\sim (m_a f_a)^{1/2}$ [see Eq.~(\ref{eq:mass})], which may induce a quadratic coupling between the axion and the SM fermions:
\begin{align}
\label{eq:muterm}
{\cal L}_\mu = \frac{\mu}{2f_a^2}a^2 \bar{\psi}\psi\;.   
\end{align}
Here $\mu$ is a parameter with mass dimension one.
For the QCD axion, the above effective coupling appears below the QCD confinement scale, and $\mu$ can be calculated using chiral perturbation theory (with $\psi$ the nucleon): $\mu\equiv\mu_{\rm QCD}\approx 15~{\rm MeV}$~\cite{Fukuda:2021drn,Bauer:2023czj,VanTilburg:2024xib}. In the more general cases of axion-like particles, $\mu$ is a free parameter that is determined by unknown ultraviolet (UV) physics. Since the coupling in Eq.~(\ref{eq:muterm}) is induced below the non-perturbative scale, one may expect an upper bound of $\mu$:
\begin{align}
\mu \lesssim \sqrt{m_a f_a}\;. \label{eq:mubound}   
\end{align}
The point to note is that $\mu$ can be much larger than the axion mass, which leads to a significant shift-symmetry-breaking effect compared to the mass term.

In \cite{Fukuda:2021drn}, it was pointed out that the quadratic coupling (\ref{eq:muterm}) is important for the direct detection of axion DM through coherent scattering between axions and nucleons. Eq.~(\ref{eq:muterm}) provides additional contributions to the diagrams in Fig.~\ref{fig:Feyn} and thus affects the axion force. The two-axion force in vacuum induced from Eq.~(\ref{eq:muterm}) was calculated in \cite{Bauer:2023czj}:
\begin{align}
V_{2a}(r) = -\frac{\mu^2}{32 \pi^3 f_a^4}\frac{m_a}{r^2}K_1\left(2m_a r\right) \qquad (\text{vacuum two-axion force})\;, \label{eq:V2a}   
\end{align}
where $K_1$ is the modified Bessel function. The short- and long-range limits turn out to be:
\begin{align}
\text{for $r\ll m_a^{-1}$}:\quad  V_{2a}(r) &=  - \frac{\mu^2}{64 \pi^3 f_a^4 r^3}\;,\label{eq:V2a-small-r}\\
\text{for $r\gg m_a^{-1}$}:\quad V_{2a}(r) & = - \frac{\mu^2}{64 \pi^3 f_a^4 r^3} \sqrt{\pi m_a r}\,e^{-2m_a r}\;.\label{eq:V2a-large-r}
\end{align}
At short distances, $V_{2a}\sim 1/r^3$ and is not suppressed by the axion mass. When the distance is larger than the inverse of the axion mass, $V_{2a}$ starts to decrease exponentially --- this is a typical feature of the quantum force where the propagators are off-shell.

Next we calculated the background-induced axion force caused by Eq.~(\ref{eq:muterm}) and expressed it using the generic form of Eq.~(\ref{eq:parametric}):
\begin{align}
V_{\text{bkg}}^{\slashed{\rm shift}}\left(\vecr;\mu\right)=-\frac{1}{4\pi r}\frac{m_1 m_2}{f_a^4}\int\frac{\text{d}^{3}\veck}{\left(2\pi\right)^{3}}\frac{f(\veck)}{E_{\veck}}\left[{\cal Q}_\text{tri}^\slashed{\rm shift}(\mu)+{\cal Q}_\text{bub}^\slashed{\rm shift}(\mu)\right]{\cal D}\;, \label{eq:Vbkgmu}  
\end{align}
where ${\cal Q}_{\rm tri}^{\slashed{\rm shift}}$ and ${\cal Q}_{\rm bub}^{\slashed{\rm shift}}$ are the order parameters induced by Eq.~(\ref{eq:muterm}) that break the shift invariance. The first term denotes the contribution from two triangle diagrams, Fig.~\ref{subfig:tri1} and Fig.~\ref{subfig:tri2}, which is a mixing effect between the quadratic coupling in Eq.~(\ref{eq:muterm}) and the linear coupling in Eq.~(\ref{eq:Leff}): 
\begin{align}
{\cal Q}_{\rm tri}^{\slashed{\rm shift}}(\mu)&=-\frac{2\mu m_a^4}{m_1 m_2}\frac{4E_{\veck}^2 m_1 m_2 \left(c_1^2 m_2 + c_2^2 m_1\right)-m_a^4\left(c_1^2 m_1 + c_2^2 m_2\right)}{\left(4 E_{\veck}^2 m_1^2-m_a^4\right)\left(4E_{\veck}^2 m_2^2-m_a^4\right)}\nonumber\\
&\approx-\frac{\mu m_a^4}{2m_1 m_2 E_{\veck}^2}\left(\frac{c_1^2}{m_1}+\frac{c_2^2}{m_2}\right),
\end{align}
where in the second line we used $m_a\ll m_i$. The second term comes from the bubble diagram, Fig.~\ref{subfig:bub}, that is purely induced by the quardratic coupling in Eq.~(\ref{eq:muterm}):
\begin{align}
{\cal Q}_{\rm bub}^{\slashed{\rm shift}}(\mu)&= \frac{\mu^2}{m_1 m_2}\;.\label{eq:Qmu}    
\end{align}
Since ${\cal Q}_{\rm tri}^{\slashed{\rm shift}}$ is suppressed by the fourth power of the axion mass compared to ${\cal Q}_{\rm bub}^{\slashed{\rm shift}}$, it can be neglected for the practical detection of light axions. Then Eq.~(\ref{eq:Vbkgmu}) is simplified to
\begin{align}
V_{\rm bkg}^{\slashed{\rm shift}}\left(\vecr;\mu\right) = -\frac{\mu^2}{4\pi r f_a^4}\int\frac{\text{d}^{3}\veck}{\left(2\pi\right)^{3}}\frac{f(\veck)}{2E_{\veck}}\left[\cos\left(\absk r-\veck\cdot\vecr\right)+\cos\left(\absk r+\veck\cdot\vecr\right)\right].\label{eq:Vbkgmu2}    
\end{align}
Therefore, at the leading order, the magnitude of $V_{\rm bkg}$ is controlled by the non-perturbative scale $\mu$. 
\vspace{0.2cm}

In the following, we discuss two special cases of interest.

$(i)$ For the isotropic axion background $f(\veck) = f(\kappa)$ with $\kappa\equiv \absk$, one can first integrate out the angular part of the phase space, leading to
    \begin{align}
    V_{\rm bkg}^{\rm iso}\left(r;\mu\right) = -\frac{\mu^2}{16\pi^3 r^2 f_a^4}\int_0^\infty {\rm d}\kappa  \frac{\kappa f(\kappa)}{\sqrt{\kappa^2+m_a^2}}\sin\left(2\kappa r\right).\label{eq:Vbkgmuiso}
    \end{align}
This agrees with the result in \cite{VanTilburg:2024xib,Barbosa:2024zfz} with effective quadratic couplings of scalars for isotropic backgrounds.

$(ii)$ In the coherent region $r\ll \lambda_{\rm dB}$, we have ${\cal D} \to 1$, and Eq.~(\ref{eq:Vbkgmu2}) is reduced to 
\begin{align}
V_{\rm bkg}^{\rm coh}\left(r;\mu\right) &= -\frac{\mu^2}{4\pi r f_a^4}\int\frac{\text{d}^{3}\veck}{\left(2\pi\right)^{3}}\frac{f(\veck)}{E_{\veck}}=-\frac{\mu^2 \tilde{n}_a}{4\pi r f_a^4 m_a}\;,
\end{align}
where $\tilde{n}_a$ is the effective number density of background axions, defined as:
\begin{align}
    \tilde{n}_a \equiv \int\frac{\text{d}^{3}\veck}{\left(2\pi\right)^{3}}\frac{f(\veck)}{E_{\veck}/m_a}\;.
\end{align}
In particular, for NR axion backgrounds $E_{\veck}\approx m_a$, then $\tilde{n}_a$ is reduced to the usual number density $n_a$, and we have
\begin{align}
V_\text{bkg,NR}^\text{coh}\left(r;\mu\right) =  -\frac{\mu^2 \rho_a}{4\pi r f_a^4 m_a^2}\;,\label{eq:Vcoh}
\end{align}
where $\rho_{a}=m_a n_a$ is the energy density of the NR axion background. In this region, $V_{\rm bkg}$ is insensible to the specific distribution function $f(\veck)$ and is enhanced by $1/m_a^2$ for a fixed $\rho_a$. The result of Eq.~(\ref{eq:Vcoh}) agrees with that derived in \cite{Hees:2018fpg,Banerjee:2022sqg} using classical field theories.

From the theoretical point of view, one cannot keep increasing the magnitude of $V_\text{bkg,NR}^\text{coh}$ by infinitely decreasing $m_a$ in Eq.~(\ref{eq:Vcoh}). To see this, we recall that the NR energy density $\rho_a$ can be connected to the axion classical field value $a_0$ via $\rho_a \sim m_a^2 a_0^2$. However, the validity of the axion effective interaction requires $a_0 \lesssim f_a$ (otherwise, the axion potential is anharmonic, and one needs to include the nonlinear corrections from axion self-interactions).
As a result, we have $\rho_a \lesssim m_a^2 f_a^2$ and Eq.~(\ref{eq:Vcoh}) is bounded from
\begin{align}
\left|V_\text{bkg,NR}^\text{coh}\left(r;\mu\right)\right| \lesssim \frac{\mu^2}{4\pi r f_a^2}\;.\label{eq:EFT bound}
\end{align}

Note that under the assumption of $a_0 \lesssim f_a$, the contribution to $\vbkg$ from $n$-axion exchange (with $n>2$) is more suppressed compared to that from two-axion exchange, and hence can be neglected.

A comparison between $\vbkg$ and $V_{2a}$ is shown in Fig.~\ref{fig:compare}, see Sec.~\ref{sec:detection} for more discussions.

\subsubsection{Effective shift-symmetry breaking from axion backgrounds}
\label{subsubsec:effectivebreaking}
Suppose there is an axion background with the phase-space distribution function $f\left(\veck;\veck_0\right)$, in which we assume the momentum distribution is peaked around $\veck=\veck_0$. It sets a typical energy scale $\left(\veck_0^2 + m_a^2\right)^{1/2}$. If the background axions are relativistic, i.e.,  $|\veck_0|\gg m_a$, then $|\veck_0|$ plays the role of the order parameter to effectively break the shift symmetry.

This can be seen by including the relativistic correction to Eq.~(\ref{eq:Vbkg}). Note that to get Eq.~(\ref{eq:Vbkg}) we have taken the NR limit, which means that we neglected all terms suppressed by higher orders of $\vecq^2$. Terms with higher orders of $\vecq^2$ result in  derivative operators after the Fourier transform. In vacuum, the derivative operators increase the power index $n$ of a $1/r^n$ potential, making it more suppressed at long distances. However, when there is a background, the derivative can act on the decoherence factor $\cos\left(|\veck|\,r\pm \veck\cdot\vecr\right)$ that is not a polynomial function of $r$; in this case, the final effect is to replace $\vecq^2$ with $\veck_0^2$. Therefore, terms with higher orders of $\vecq^2$ can still lead to a $1/r$ potential in the axion background, which is proportional to the power of the typical energy scale of the background. 

The relativistic correction of $V_{\rm bkg}$ is calculated in App.~\ref{appsub:relativistic-correction}, with the final expression given by Eq.~(\ref{eq:relativistic}). Note that we have taken $m_a = 0$ because we assume that the axion background is relativistic and we are interested in the effect of shift symmetry breaking that comes purely from the background. The result in Eq.~(\ref{eq:relativistic}) contains multiple terms of $1/r^5$, $1/r^4$, $1/r^3$, $1/r^2$, and $1/r$. For $r\gg 1/|\veck_0|$, which is satisfied for most fifth-force experiments with a relativistic axion background, the $1/r$ term becomes dominant, then we obtain:
\begin{align}
V_\text{bkg}^{\text{rel}}\left(\vecr\right) =  - \frac{c_1^2 c_2^2}{64\pi m_1m_2f_a^4} \times \frac{1}{r} \times \int\frac{\text{d}^3\veck}{(2\pi)^3}\frac{|\veck|^4}{E_{\veck}}f\left(\veck;\veck_0\right)\left(1-3c^2+3c^4\right){\cal D}\;, \label{eq:Vbkgk0}
\end{align}
where the index ``rel'' implies the relativistic correction from the axion background, ${\cal D}$ is given by Eq.~(\ref{eq:Dfactor}) and $c\equiv\cos(\hat{\veck}\cdot\hat{\vecr})$ denotes the cosine of the angle between $\veck$ and $\vecr$. After the phase-space integral, the momentum $\veck$ in Eq.~(\ref{eq:Vbkgk0}) is typically replaced by its central value $\veck_0$.
Comparing with the parametric form in Eq.~(\ref{eq:parametric}), we find the order parameter in a relativistic axion background is given by
\begin{align}
{\cal Q}^{}_{\rm rel}\left(\veck_0\right)= \frac{\left|\veck_0\right|^4}{16 m_1^2 m_2^2}\;. \label{eq:Qrel}
\end{align}
Note that the NR approximation of the external fermions $\psi_i$ holds when $|\veck_0| \ll m_i$, which causes a suppression to ${\cal Q}^{}_{\rm rel}$. Nevertheless, the order parameter in Eq.~(\ref{eq:Qrel}) can still be much larger than that in Eq.~(\ref{eq:Qinv}) given an energetic axion background satisfying $|\veck_0|\gg m_a$. A typical example is the solar axion flux with $|\veck_0| \sim {\rm keV}$, as will be discussed in Sec.~\ref{subsec:solar}.

\subsection{Summary}
We have derived the general formulae for the background-induced axion force $\vbkg$, with a generic form described by Eq.~(\ref{eq:parametric}). In the absence of shift-symmetry-breaking interactions, cancellations occur among different diagrams, causing a suppression to the final result. 
In the NR limit, $\vbkg$ is given by Eq.~(\ref{eq:Vbkg}), where the axion mass plays the role of order parameter that breaks the shift symmetry. There are two possibilities to enhance $\vbkg$. First, the non-perturbative effect from the axion potential induces a quadratic coupling that explicitly breaks the shift symmetry, leading to a much larger $\vbkg$ in Eq.~(\ref{eq:Vbkgmu2}). Second, the axion background effectively breaks the shift symmetry; this effect appears as relativistic corrections of the NR scattering amplitude. In an energetic axion background with a typical momentum $|\veck_0|\gg m_a$, $\vbkg$ is enhanced by a factor of $|\veck_0|^4/m_a^4$, as shown in Eq.~(\ref{eq:Vbkgk0}).

\section{Typical axion backgrounds}
\label{sec:examples}
In this section, we apply the general formulae derived in Sec.~\ref{sec:formalism} to several well-motivated axion backgrounds: Maxwell-Boltzmann (MB) distribution, Bose-Einstein (BE) distribution, and relativistic axion flux. They correspond to the ideal situations of three typical axion sources, respectively: cold axion DM, thermal axion relics, and solar axion flux. For each example, we calculate the specific $\vbkg$ using Eqs.~(\ref{eq:Vbkg}), (\ref{eq:Vbkgmu2}) and (\ref{eq:Vbkgk0}). We leave the discussion of experimental probes to Sec.~\ref{sec:detection}.

\subsection{Maxwell-Boltzmann distribution}
\label{subsec:MB}
As a first example, we consider the following isotropic MB distribution:
\begin{align}
f_{\rm MB}(\kappa) = n_a \left(2\pi\right)^{3/2}\kappa_0^{-3}e^{-\kappa^2/(2\kappa_0^2)}\;,\label{eq:fMB}    
\end{align}
where $n_a = \int {\rm d}^3\veck\, f_{\rm MB}(\kappa)/(2\pi)^3$ is the number density and $\kappa_0$ characterizes the momentum dispersion. Eq.~(\ref{eq:fMB}) coincides with the NR limit of the thermal MB distribution with an effective temperature $T_{a}$: $e^{-E_\veck/T_{a}} \to e^{-\kappa^2/(2 m_a T_a)}$. Note that the dynamics of cold DM in the galaxy is dominated by gravitational interactions, and the DM particles are not in thermal equilibrium. However, $N$-body simulation indicates that the DM phase-space distribution in the galaxy can still be approximately described by the Gaussian form in Eq.~(\ref{eq:fMB})~\cite{Freese:1987wu,Hoeft:2003ea,Faltenbacher:2006rb,Vogelsberger:2008qb}. (The more realistic DM distribution has anisotropy due to the motion of the Earth, which will be discussed in Sec.~\ref{subsec:axion-DM}). The momentum dispersion is estimated by $\kappa_0 \approx m_a v_a/\sqrt{2}$ with $v_a \sim {\cal O}(10^{-3})$ the local velocity of the DM halo.

We first calculate $\vbkg$ with the shift-invariant interaction. For NR axion background, assuming $m_a\ll m_i$, the order parameter (\ref{eq:Qinv}) is reduced to $Q_{\rm inv}\approx m_a^4/(16 m_1^2 m_2^2)$. Substituting it into Eq.~(\ref{eq:Vbkg}) and performing the integral one obtains
\begin{align}
V_\text{MB}^\text{inv}(r) &= - \frac{c_1^2 c_2^2 m_a^3}{256\pi^3 r^2 f_a^4 m_1 m_2}\int_0^\infty {\rm d}\kappa \kappa f_{\rm MB}(\kappa)\sin\left(2\kappa r\right)\nonumber\\
&=-\frac{c_1^2 c_2^2 \rho_a m_a^2}{64\pi r f_a^4 m_1 m_2} e^{-2\kappa_0^2 r^2}\;,\label{eq:VbkginvMB}
\end{align}
where in the second line we have used the relation $\rho_a = m_a n_a$ for NR axions. Note that the potential in Eq.~(\ref{eq:VbkginvMB}) is suppressed by $m_a^2/(m_1 m_2)$ for light axions. Due to the decoherence factor in Eq.~(\ref{eq:Dfactor}), the isotropic MB distribution leads to an exponential suppression to $\vbkg$ at large distances $r\gg \lambda_{\rm dB}\sim1/\kappa_0$; while for $r\ll 1/\kappa_0$, $\vbkg$ exhibits the generic $1/r$ behavior. Similar results were also derived in \cite{VanTilburg:2024xib,Barbosa:2024zfz} with effective quadratic couplings.

 We proceed to calculate $\vbkg$ with explicit shift-symmetry breaking using Eq.~(\ref{eq:Vbkgmuiso}). It turns out to be
\begin{align}
V_\text{MB}^{\slashed{\rm shift}}\left(r;\mu\right) =  -\frac{\mu^2 \rho_a}{4\pi r f_a^4 m_a^2} e^{-2\kappa_0^2 r^2}\;.\label{eq:VbkgmuMB}    
\end{align}
Note that Eq.~(\ref{eq:VbkgmuMB}) can also be obtained from Eq.~(\ref{eq:VbkginvMB}) by simply replacing the order parameter ${\cal Q}_{\rm inv}=m_a^4/(16 m_1^2 m_2^2)$ with that in Eq.~(\ref{eq:Qmu}). 

Since the axion background considered in this case is NR, the relativistic correction from Eq.~(\ref{eq:Vbkgk0}) is more suppressed than that in Eq.~(\ref{eq:VbkginvMB}) and hence can be neglected.

\subsection{Bose-Einstein distribution}
\label{subsec:BE}
If axions were thermally produced in the early universe and then relativistically decoupled from the SM bath at some temperature $T_{\rm d}$, they remain as the thermal relics to the present day that obey the BE distribution~\cite{Dror:2021nyr}:
\begin{align}
f_{\rm BE}(\kappa) = \frac{1}{e^{\kappa/T_a}-1}\;,   \label{eq:fBE}
\end{align}
where $T_a$ is the axion effective temperature at the present day, given by
\begin{align}
T_a = \left[\frac{g_{*S}\left(T_0\right)}{g_{*S}\left(T_{\rm d}\right)}\right]^{\frac{1}{3}} T_0\equiv \xi_a T_0\;,\label{eq:Ta} 
\end{align}
where $T_0\approx 2.7~{\rm K}\approx 0.23~{\rm meV}$ is the temperature of today's cosmic microwave background (CMB), $g_{*S}(T)$ is the relativistic degree of freedom relevant to the entropy at temperature $T$. The discrepancy between $T_a$ and $T_{\rm d}$ appears because the SM thermal bath can be reheated after axion decoupling (with entropy conserved), as characterized by the factor $\xi_a \leq 1$. The precise value of $\xi_a$ depends on $T_{\rm d}$, which is determined by the couplings between axions and SM particles. If axions were relativistic when decoupling, their contribution to the effective neutrino number $N_{\rm eff}$ at the CMB epoch can be written as
\begin{align}
\Delta N_{\rm eff} = \frac{4}{7}\left(\frac{11}{4}\right)^{\frac{4}{3}} \left(\frac{T_a}{T_0}\right)^4.    
\end{align}
The recent CMB constraint $\Delta N_{\rm eff} \lesssim 0.285$~\cite{Planck:2018vyg} requires $\xi_a \lesssim 0.6$ or $T_a \lesssim 0.14~{\rm meV}$.
As an estimate, we take $T_a = 10^{-4}~{\rm eV}$ in the following discussion.

The relativistic correction (\ref{eq:Vbkgk0}) is only relevant for $m_a \ll T_a$. However, even in this case, it is still highly suppressed by $T_a^4/(m_1^2 m_2^2)$ and thus can be neglected. On the other hand, using the BE distribution, Eqs.~(\ref{eq:Vbkg}) and  (\ref{eq:Vbkgmu2}) are reduced to
\begin{align}
 V_{\rm BE}^{\rm inv}\left(r\right) &= - \frac{c_1^2 c_2^2 m_a^8}{256\pi^3 r^2 f_a^4 m_1 m_2}\int_0^\infty {\rm d}\kappa \frac{\kappa \sin\left(2\kappa r\right)}{\left(\kappa^2+m_a^2\right)^{5/2}\left(e^{\kappa/T_a}-1\right)}\;,\\
 V_\text{BE}^{\slashed{\rm shift}}\left(r;\mu\right) &= - \frac{\mu^2}{16\pi^3 r^2 f_a^4}\int_0^\infty {\rm d}\kappa \frac{\kappa \sin\left(2\kappa r\right)}{\left(\kappa^2+m_a^2\right)^{1/2}\left(e^{\kappa/T_a}-1\right)}\;.
\end{align}
We discuss two hierarchical limits:

$(i)$ Relativistic relic axions with $m_a \ll 10^{-4}~{\rm eV}$. In this case, $V_{\rm BE}^{\rm inv}$ is negligible because of the suppression of the axion mass, while $V_\text{BE}^{\slashed{\rm shift}}$ is given by
\begin{align}
V_\text{BE}^{\slashed{\rm shift}}\left(r;\mu\right) &= -\frac{\mu^2}{64\pi^3 r^3 f_a^4}\left[2\pi r T_a \coth\left(2\pi r T_a\right)-1\right]\nonumber\\
&=-\frac{\mu^2}{64\pi^3 f_a^4} 
        \begin{cases}
            4\pi^2 T_a^2/(3r)& \text{for $r\ll T_a^{-1}$}\\
            2\pi T_a/r^2& \text{for $r\gg T_a^{-1}$}
        \end{cases}\;.\label{eq:VBErel}
    \end{align} 
The typical de Broglie wavelength of axion relics is characterized by the axion temperature $\db\sim1/T_a$, while the number density scales as $n_a \sim T_a^3$. Comparing Eq.~(\ref{eq:VBErel}) with (\ref{eq:parametric}), we find that the decoherence factor causes an additional suppression $1/(T_ a r)$ compared to a $1/r$ potential at long distances $r\gg \db$. 

$(ii)$ NR relic axions with $m_a \gg 10^{-4}~{\rm eV}$. In this case, we arrive at
  \begin{align}
     V_\text{BE}^{\slashed{\rm shift}}\left(r;\mu\right) &= \frac{16 \mu^2  m_1 m_2}{m_a^4}V_{\rm BE}^{\rm inv}(r)\nonumber\\
     &=-\frac{\mu^{2}T_a^2}{32\pi^{3}f_{a}^{4} m_a r^{2}} \left[i\psi ^{(1)}\left(1+2 i r T_a\right)-i\psi ^{(1)}\left(1-2 i r T_a\right)\right]\nonumber\\
     &=-\frac{\mu^{2}}{32\pi^{3}f_{a}^{4} m_a}
     \begin{cases}
         8\zeta(3) T_a^3/r&\text{for $r\ll T_a^{-1}$}\\
          T_a/r^3& \text{for $r\gg T_a^{-1}$}
     \end{cases}\;,\label{eq:VBENR}
  \end{align}
where $\zeta(3)\approx 1.202$, $\psi^{(n)}$ is the $n$-th ordered polygamma function, defined as 
\begin{align*}
\psi^{(n)}(x)\equiv \frac{{\rm d}^{n+1}}{{\rm d}x^{n+1}}\log \Gamma(x)    
\end{align*}
with $\Gamma(x)$ being the gamma function. From Eq.~(\ref{eq:VBENR}), it can be seen that for NR relic axions, the suppression caused by the decoherence effect is $1/(T_a r)^2$ when $r\gg \db$.

If the relic axions are NR today, their energy density is given by 
\begin{align}
\rho_a = m_a n_a = m_a \int \frac{{\rm d}^3 \veck}{\left(2\pi\right)^3}\frac{1}{e^{\kappa/T_a}-1}=\frac{\zeta(3)}{\pi^2}m_a T_a^3\;.    
\end{align}
By asking $\rho_a$ to not exceed the total relic abundance of DM, one obtains a conservative upper bound on the axion mass if axions comprise the thermal relics at present:
\begin{align}
    m_a \lesssim 80~{\rm eV}\;,
\end{align}
which may be relaxed if $T_a$ is below $10^{-4}~{\rm eV}$.

\subsection{Relativistic axion flux}
\label{subsec:flux}
We consider an energetic axion flux with good direction and typical momentum $\kappa_0\gg m_a$. The phase-space distribution function can be written as
\begin{align}
f_{\rm flux}\left(\veck\right)=\left(2\pi\right)^3\frac{\text{d}\Phi}{\text{d}\kappa}\delta\left(\hat{\veck}-\hat{\veck}_0
\right),\label{eq:flux}
\end{align}
where the unit vector $\hat{\veck}_0$ denotes the direction of the flux, ${\rm d}\Phi/{\rm d}\kappa$ is the differential flux (i.e., the flux per energy) with the peaked value at $\kappa=\kappa_0$ and normalized as
\begin{align}
\int \frac{{\rm d}^3\veck}{\left(2\pi\right)^3}f_{\rm flux}\left(\veck\right) = \Phi\;, 
\end{align}
with $\Phi$ the total flux. 

Since the flux is relativistic, one can neglect the contribution from Eq.~(\ref{eq:Vbkg}) that is suppressed by the axion mass. Substituting Eq.~(\ref{eq:flux}) into Eqs.~(\ref{eq:Vbkgmu2}) and (\ref{eq:Vbkgk0}), we obtain
\begin{align}
V_\text{flux}^{\slashed{\rm shift}}\left(r,\alpha;\mu\right) & = -\frac{\mu^{2}}{4\pi r f_{a}^{4}}\int_0^\infty\frac{\text{d}\kappa}{\kappa}\frac{\text{d}\Phi}{\text{d}\kappa}\cos\left(\kappa r \right)\cos\left(\kappa r \cos\alpha\right),\label{eq:Vfluxmu}\\
V_\text{flux}^\text{rel}\left(r,\alpha\right) &=  - \frac{c_1^2 c_2^2}{64\pi r m_1m_2f_a^4} \int_0^\infty {\rm d}\kappa \kappa^3 \frac{\text{d}\Phi}{\text{d}\kappa}\left(1-3\cos^2\alpha+3\cos^4\alpha\right)\cos\left(\kappa r \right)\cos\left(\kappa r \cos\alpha\right),\label{eq:Vfluxk0}
\end{align}
where $\alpha$ is the angle between $\hat{\veck}_0$ and $\vecr$. Note that the momentum spread and the finite size of the object will lead to oscillations when the argument of the cosine function is much larger than 1, which makes the potential more suppressed than $1/r$. The condition to avoid the decoherence suppression is given by
\begin{align}
\alpha^2 \lesssim \pi/\Delta\left(\kappa r\right),  
\end{align}
where $\Delta\left(\kappa r\right)$ denotes the spread of $\kappa r$. For example, for $\Delta \left(\kappa r\right) = {\rm keV}\cdot {\rm cm}$, this condition implies $\alpha\lesssim 10^{-4}$.
In the ideal case of the small-angle limit, Eqs.~(\ref{eq:Vfluxmu})-(\ref{eq:Vfluxk0}) are reduced to
\begin{align}
V_{\rm flux}^{\slashed{\rm shift}}\left(r;\mu\right) &=  -\frac{\mu^{2}}{8\pi r f_{a}^{4}}\int_0^\infty\frac{\text{d}\kappa}{\kappa}\frac{\text{d}\Phi}{\text{d}\kappa}\;,\label{eq:Vfluxmu2}\\
V_\text{flux}^\text{rel}\left(r\right) &=  - \frac{c_1^2 c_2^2}{128\pi r m_1m_2f_a^4} \int_0^\infty {\rm d}\kappa \kappa^3 \frac{\text{d}\Phi}{\text{d}\kappa}\;.\label{eq:Vfluxk02}
\end{align}

The relativistic axion beam with a fixed direction corresponds to a realistic scenario --- the axion flux emitted from the Sun, which will be discussed in Sec.~\ref{subsec:solar}.

\section{Experimental probes of the axion forces}
\label{sec:detection}

\renewcommand\arraystretch{1.1}
\begin{table}[t]
    \centering
    \begin{tabular}{c c  c c}
    \hline
    Experiments & $\left.\delta V\right/V_{\text{grav}}$ & $\langle r \rangle$ & Refs\\
    \hline
    MICROSCOPE & $3.8\times10^{-15}$ & $\sim7100$km & \citep{MICROSCOPE:2022doy}\\
    Washington2007 & $1.8\times10^{-13}$ & $\sim6400$km & \citep{Schlamminger:2007ht}\\
    Washington1999 & $3.0\times10^{-9}$ & $\sim0.3$m & \citep{Smith:1999cr}\\
    Irvine1985 & $7\times10^{-5}$ & $2-5$cm & \citep{Hoskins:1985tn}\\
    HUST2012 & $10^{-3}$ & $\sim2$mm & \citep{Yang:2012zzb}\\
    E\"{o}t-Wash2006 & $2\times10^{-3}$ & $\sim0.5$mm & \citep{Kapner:2006si}\\
    HUST2020 & $3\times10^{-3}$ & $\sim0.3$mm & \citep{Tan:2020vpf}\\
    Washington2020 & $\sim1$ & $52\mu$m & \citep{Lee:2020zjt}\\
     Future levitated optomechanics & $\sim 10^4$ & $1\mu$m & \citep{Moore:2020awi}\\
     IUPUI2014 & $2.7\times 10^7$ & $560$nm & \citep{Chen:2014oda}\\
    \hline
    \end{tabular}
    \caption{The sensitivities of typical fifth-force search experiments. Here $V_{\rm grav}$ is the Newtonian potential, while $\delta V$ is the contribution from long-range forces other than gravity. The third column lists the typical length scale of the corresponding experiment.}
    \label{table:deltav/v}
\end{table}
\renewcommand\arraystretch{1}

\begin{figure}[t]
    \centering
    \includegraphics[scale=0.7]
    {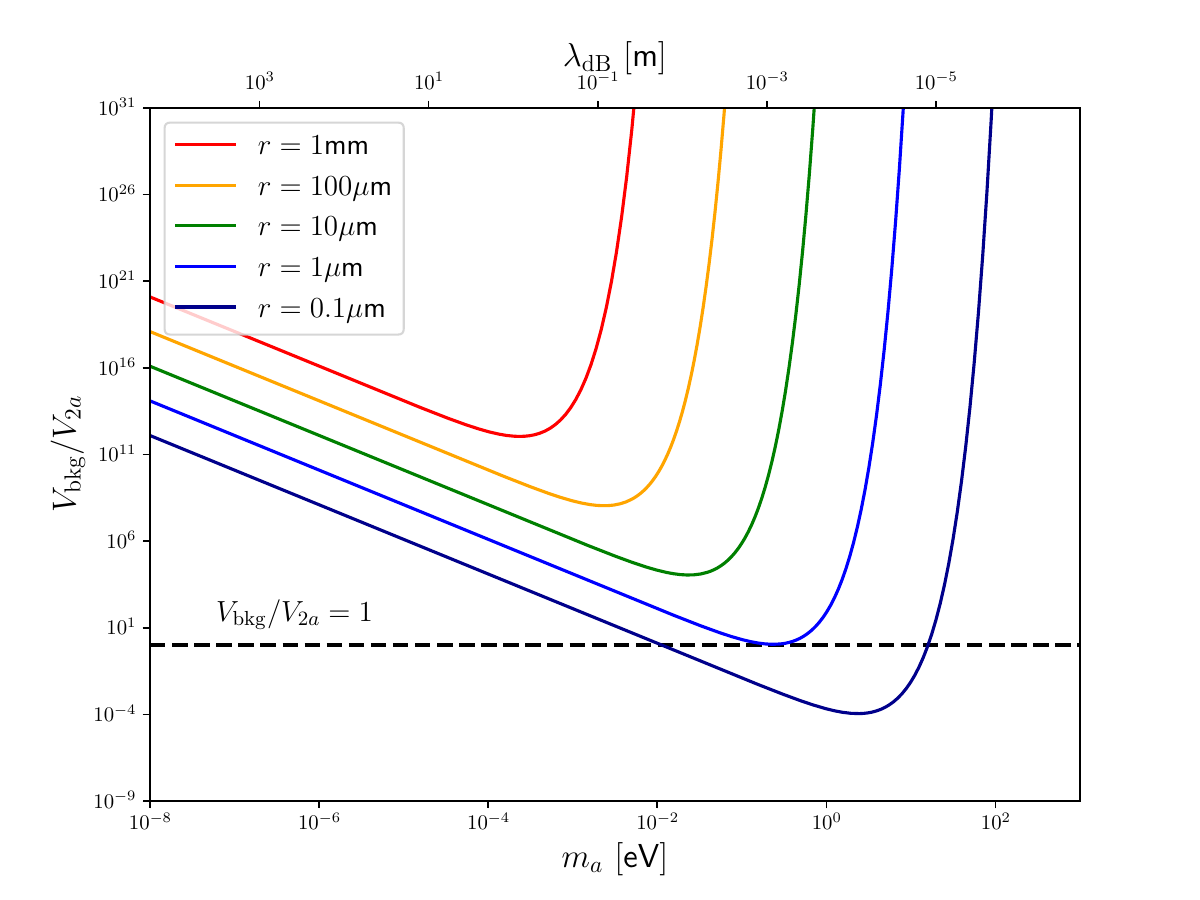}
    \caption{\label{fig:compare}The comparison between the background-induced axion force $V_{\rm bkg}$ in the coherent region [Eq.~(\ref{eq:Vcoh})] and its vacuum counterpart $V_{2a}$ [Eq.~(\ref{eq:V2a})]. Note that the dependence on both $\mu$ and $f_a$ is cancelled in the ratio $V_{\rm bkg}/V_{2a}$, indicating that these two terms are of the same order from the view of effective field theories. We have fixed $\rho_a = \rho_{\rm DM} \approx 0.4~{\rm GeV}/{\rm cm}^3$ as the local density of DM. We also take the typical de Broglie wavelength as $\lambda_{\rm dB} = 1/(m_a v_a)$, with $v_a \approx 10^{-3}$ the average velocity of the axion DM. Different curves in the plot correspond to the comparison at different length scales.}
\end{figure}

Here we move to discuss ways to probe the axion force. One way to probe the force is through fifth-force detection experiments.
The long-range axion force between two objects contributes as a correction to gravity, which may leave imprints in the precision tests of gravitational inverse-square law and equivalence principle (see e.g.~\cite{Adelberger:2009zz,Wagner:2012ui,franklin2016rise} for a review). 

Assuming the existence of axions and axion backgrounds, the (spin-independent) long-range potential between two objects can be written as
\begin{align}
V(r) &= V_{\rm grav}(r) + V_{2a}(r) + \vbkg(r) \equiv V_{\rm grav}(r) + \delta V(r)\;,
\end{align}
where $V_{\rm grav}$ denotes the Newtonian potential, $V_{2a}$ is the vacuum two-axion potential given in Eq.~(\ref{eq:V2a}), and $V_{\rm bkg}$ is the background-induced axion force. In Table \ref{table:deltav/v}, we collect the sensitivities and length scales of typical experiments that aim to detect the fifth force.

The vacuum axion force $V_{2a}$ suffers from exponential suppression at long distances $r\gg 1/m_a$. On the other hand, $\vbkg$ remains $1/r$ as long as $r \ll \db$. Note that for cold DM with local velocity $v_a \sim 10^{-3}$, we have $\db \sim 1/(m_a v_a)\sim 10^3/m_a$, which is three orders of magnitude larger than the range of $V_{2a}$.
In Fig.~\ref{fig:compare}, we compare the magnitude of $V_{\rm bkg}$ with $V_{2a}$ in the coherent region assuming a cold DM background. 
As one can see, for small axion masses, $V_{\rm bkg}$ is enhanced by $1/m_a^2$, while for large axion masses $m_a \gg 1/r$, $V_{2a}$ is exponentially suppressed. As a result, $V_{\rm bkg}$ dominates over $V_{2a}$ at the length scales of most fifth-force search experiments. 
At $r\gtrsim \db$, $\vbkg$ is additionally suppressed due to the decoherence effect, but in most cases (except for the isotropic MB discussed in Sec.~\ref{subsec:MB}) the suppression is less than the exponential suppression. Based on the above argument, we will neglect the contribution from $V_{2a}$ and assume $\delta V = \vbkg$. 

This section is organized as follows. In Sec.~\ref{subsec:axion-DM}, we calculate the effect of the background-induced axion force $\vbkg$ on fifth-force detection experiments, assuming that axions constitute the DM background. In Sec.~\ref{subsec:atom}, we discuss an alternative detection strategy based on atomic spectroscopy, again under the assumption that axions form the DM. Finally, in Sec.~\ref{subsec:solar}, we consider a different source of the axion background --- the solar axion flux --- and estimate the corresponding experimental sensitivities.

\subsection{Axion dark matter}
\label{subsec:axion-DM}
Since the axion DM background is non-relativistic, the relevant $\vbkg$ comes from Eq.~(\ref{eq:Vbkgmu2}), induced by the quadratic coupling and dominant for light axions, and from Eq.~(\ref{eq:Vbkg}), induced by the axion mass and only relevant for heavy axions. We will refer to them as ``$\mu$-term'' and ``mass term'' in the following discussion, respectively.

\subsubsection{Coherent limit}
Before diving into the specific phase-space distribution, let us first make the most optimistic estimate, i.e., take the coherent limit $r\ll \db$. In this case, $\vbkg$ induced from $\mu$-term is given by Eq.~(\ref{eq:Vcoh}), a typical $1/r$ long-range potential. We further assume that the size of the object is smaller than $\db$, such that all particles in the object can contribute coherently to $\vbkg$. Comparing the axion force with the gravity, we obtain
\begin{align}
\frac{\delta V_\text{$\mu$-term}}{V_{\rm grav}} &= \frac{\mu^2 \rho_{a} N_1 N_2}{4\pi r f_a^4 m_a^2} \Bigg/ \frac{G_N M_1 M_2}{r}=\frac{2\mu^2 \rho_{a} M_{\rm Pl}^2}{f_a^4 m_a^2 m_N^2}\;,\label{eq:deltaVoverVgrav}
\end{align}
where $M_i$ and $N_i$ are the mass and number of particles of object $i=1,2$, $M_1/N_1 \approx M_2/N_2 \approx m_N \approx 1~{\rm GeV}$ is the average nucleon mass, and $M_{\rm Pl}=1/\sqrt{8\pi G_N}\approx 2.4 \times 10^{18}~{\rm GeV}$ is the reduced Planck mass with $G_N$ the gravitational constant. We assume that axions constitute the entire cold DM and take $\rho_a = \rho_{\rm DM}\approx 0.4~{\rm GeV/cm^3}$, the DM energy density in the local galaxy.
For QCD axion, we have $\mu=\mu_{\rm QCD}\approx 15~{\rm MeV}$ and $m_a f_a \approx 5.7 \times 10^{-3}~{\rm GeV^2}$~\cite{GrillidiCortona:2015jxo}, leading to
\begin{align}
    \left(\frac{\delta V_\text{$\mu$-term}}{V_{\rm grav}}\right)_\text{QCD axion} \approx  10^{-15} \left(\frac{5\times 10^{5}~{\rm GeV}}{f_a}\right)^{2},
\end{align}
where $\delta V/V_{\rm grav}\sim 10^{-15}$ is the sensitivity of MICROSCOPE experiment~\cite{MICROSCOPE:2022doy}.
For generic axion-like particles, $\mu$ is determined by UV models responsible for the shift symmetry breaking. Here we take the upper bound of $\mu\lesssim\sqrt{m_a f_a}$ [see Eq.~(\ref{eq:mubound})] as an estimate:
\begin{align}
\left(\frac{\delta V_\text{$\mu$-term}}{V_{\rm grav}}\right)_\text{ALP} \approx 10^{-15}\left(\frac{10^{-21}~{\rm eV}}{m_a}\right)\left(\frac{3\times10^{13}~{\rm GeV}}{f_a}\right)^3.
\end{align}

Although axions are usually considered to be light particles (sub-eV), there are still possibilities that axions can have a much heavier mass, see \cite{Holdom:1982ex,Rubakov:1997vp,Fukuda:2015ana,Dimopoulos:2016lvn,Gherghetta:2016fhp,Agrawal:2017ksf,Hook:2019qoh,Gherghetta:2020keg,Kelly:2020dda,Dunsky:2023ucb} for the theoretical efforts. In this case, the $\vbkg$ induced by the mass term is not negligible. Yet, we still assume that the axion is lighter than the nucleon, $m_a\lesssim m_N$. 
Taking the coherent limit in Eq.~(\ref{eq:Vbkg}) to maximize the effect, we obtain
\begin{align}
\frac{\delta V_\text{mass-term}}{V_{\rm grav}} &= \frac{c_N^4 m_a^2 \rho_a M_{\rm Pl}^2}{8 f_a^4 m_N^4}\approx 10^{-15} \left(\frac{m_a}{{\rm GeV}}\right)^2\left(\frac{2\times 10^2~{\rm GeV}}{f_a/c_N}\right)^4,
\end{align}
where $c_N$ is the dimensionless axion coupling in Eq.~(\ref{eq:Lintd}) with $\psi$ to be the nucleon. 

\subsubsection{Decoherence effect}
The realistic distribution of axion DM may lead to additional suppression compared to a $1/r$ potential due to the decoherence factor in Eq.~(\ref{eq:Dfactor}). As mentioned in Sec.~\ref{subsec:MB}, the actual distribution of DM in the local galaxy is not isotropic; it has a preferred direction because of the Earth's motion relative to the DM halo. This can be described by the boosted Maxwell-Boltzmann (BMB) distribution: 
\begin{align}
f_{\rm BMB}\left(\veck\right) &= f_{\rm MB}\left(\veck - m_a\vecv_a\right)\nonumber\\
&=n_a \left(2\pi\right)^{3/2}\kappa_0^{-3}\exp\left(-\frac{\left|\veck-m_a \vecv_a\right|^2}{2\kappa_0^2}\right),\label{eq:fBMB}
\end{align}
where $f_{\rm MB}$ is given in Eq.~(\ref{eq:fMB}), $\vecv_a$ is the velocity of the Earth relative to the galaxy center with $v_a\equiv |\vecv_a|\approx 220~{\rm km/s}$, and $\kappa_0 = m_a v_a/\sqrt{2}$ is the momentum dispersion according to the  Standard Halo Model~\cite{ParticleDataGroup:2024cfk}.

\subsubsection*{Point-like object}
We first assume that the size of the test objects is negligible.
Combining Eq.~(\ref{eq:fBMB}) with Eqs.~(\ref{eq:Vbkg}) and (\ref{eq:Vbkgmu2}), the relevant background-induced axion forces between two point-like objects are given by
\begin{align}
&\text{$\mu$-term}:\quad
V_\text{bkg,point}^{\slashed{\rm shift}}\left(\vecr;\mu\right) = -\frac{\mu^2 \rho_a}{4\pi  f_a^4 m_a^2}\frac{{\cal F}_{\rm PS}\left(\vecr\right)}{r}\;,\\
&\text{mass-term}:\;\;
V_\text{bkg,point}^{\rm inv}\left(\vecr\right) = -\frac{c_N^4 \rho_a m_a^2}{64\pi f_a^4 m_N^2}\frac{{\cal F}_{\rm PS}\left(\vecr\right)}{r}\;,
\end{align}
where the dimensionless phase-space form factor is defined as
\begin{align}
{\cal F}_{\rm PS}\left(\vecr\right)\equiv \int\frac{\text{d}^3\veck}{(2\pi)^3}\frac{f_{\rm BMB}(\veck)}{n_a}\,{\cal D}\;, \label{eq:FPSdef} 
\end{align}
with ${\cal D}$ given by Eq.~(\ref{eq:Dfactor}). 
${\cal F}_{\rm PS}$ characterizes the decoherence effect from the phase-space spread. Note that the form factor defined in Eq.~(\ref{eq:FPSdef}) is slightly different from that defined in \cite{VanTilburg:2024xib}: our decoherence factor ${\cal D}$ is invariant under the parity reflection $\vecr \to -\vecr$, while \cite{VanTilburg:2024xib} used an asymmetric decoherence factor ${\cal D}_{\rm asy}=\cos\left(|\veck| r -\veck\cdot\vecr\right)$ that does not have a parity symmetry (see Sec.~\ref{sec:compare} for more discussions on the physical origin of two decoherence factors).

By explicit calculation, we obtain (see App.~\ref{app:decoherence} for details)
\begin{align}
{\cal F}_{\rm PS}\left(r,\alpha\right) = \frac{e^{-1}}{\sqrt{2\pi}\,\kappa_0^3}&\int_0^\infty {\rm d}\kappa \kappa^2\cos\left(\kappa r\right)e^{-\frac{\kappa^2}{2\kappa_0^2}}\nonumber\\
\times&\int_{-1}^1 {\rm d}z\, e^{\frac{\sqrt{2 }\kappa z}{\kappa_0}}\cos\left(\kappa r \cos\alpha z\right)J_0\left(\kappa r \sin\alpha \sqrt{1-z^2}\right),\label{eq:FPS}     
\end{align}
where $\alpha$ is the angle between $\vecr$ and $\vecv_a$, and $J_0$ is the Bessel function of the first kind. In the parallel limit $\alpha \to 0$, we find an analytical expression for ${\cal F}_{\rm PS}$:
\begin{align}
{\cal F}_{\rm PS}\left(r,\alpha \to 0\right) = \frac{1}{2+\kappa_0^2 r^2}\left\{1+e^{-2\kappa_0^2 r^2}\left[\left(1+\kappa_0^2 r^2\right)\cos\left(2\sqrt{2}\kappa_0 r\right)
-\frac{\kappa_0 r}{\sqrt{2}}\sin\left(2\sqrt{2}\kappa_0 r\right)\right]
\right\}.\label{eq:FPSparallel}    
\end{align}
The de Broglie wavelength of the axion DM is given by $\db \sim 1/\kappa_0$. For $r \ll 1/\kappa_0$, there is ${\cal F}_{\rm PS} \to 1$, and we recover the result in the coherent limit as expected. On the other hand, for $r\gg 1/\kappa_0$, Eq.~(\ref{eq:FPSparallel}) is reduced to
\begin{align}
{\cal F}_{\rm PS}\left(r\gg \kappa_0^{-1}, \alpha \to 0\right) = 1/\left(\kappa_0^2 r^2\right).  
\end{align}
So, the phase-space decoherence effect causes a quadratic suppression of the axion force at long distances. Note that a similar $1/r^2$ suppression was also found in \cite{VanTilburg:2024xib}, which used the asymmetric decoherence factor. Although there is an ${\cal O}(1)$ numerical difference between our result and that in \cite{VanTilburg:2024xib}, the qualitative result does not change.

\begin{figure}[t]
    \centering
    \includegraphics[width=0.52\linewidth]{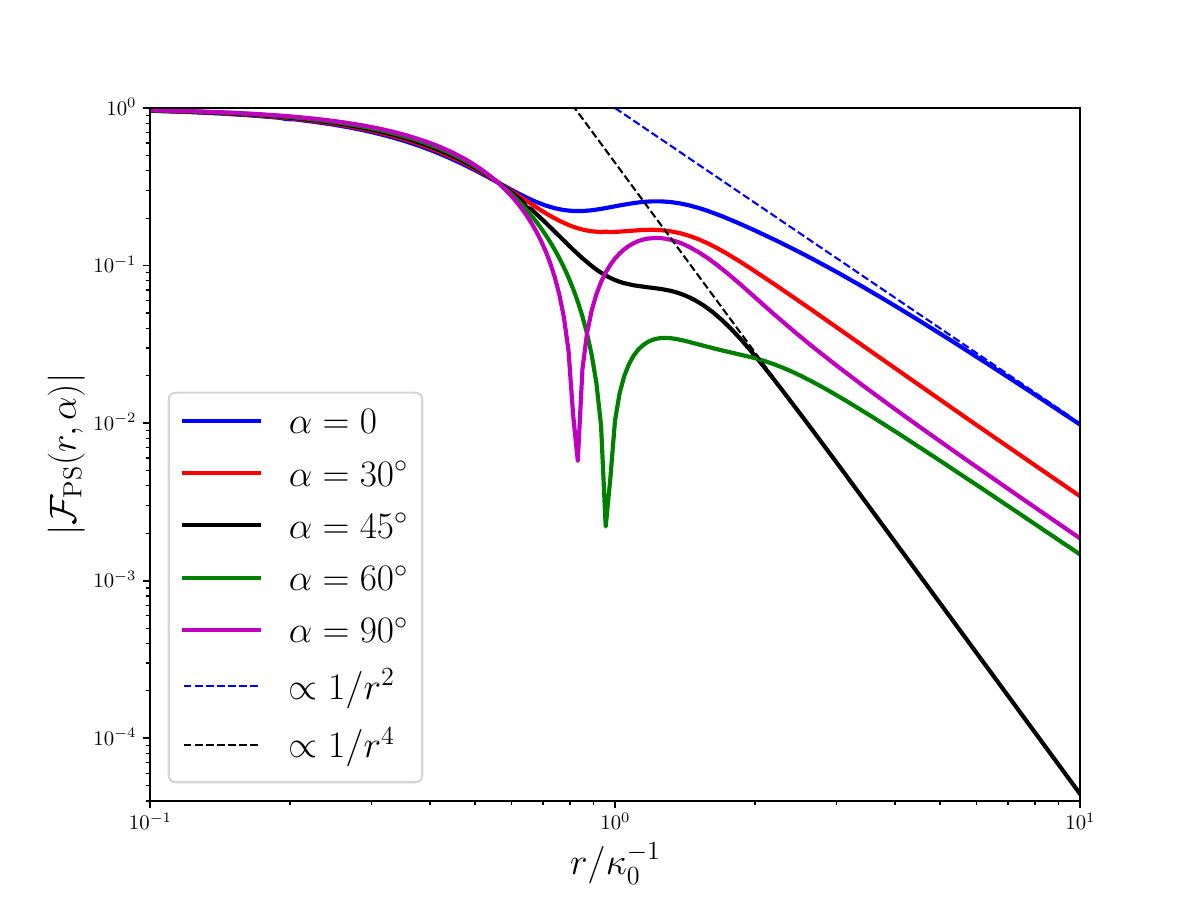}
    \includegraphics[width=0.45\linewidth]{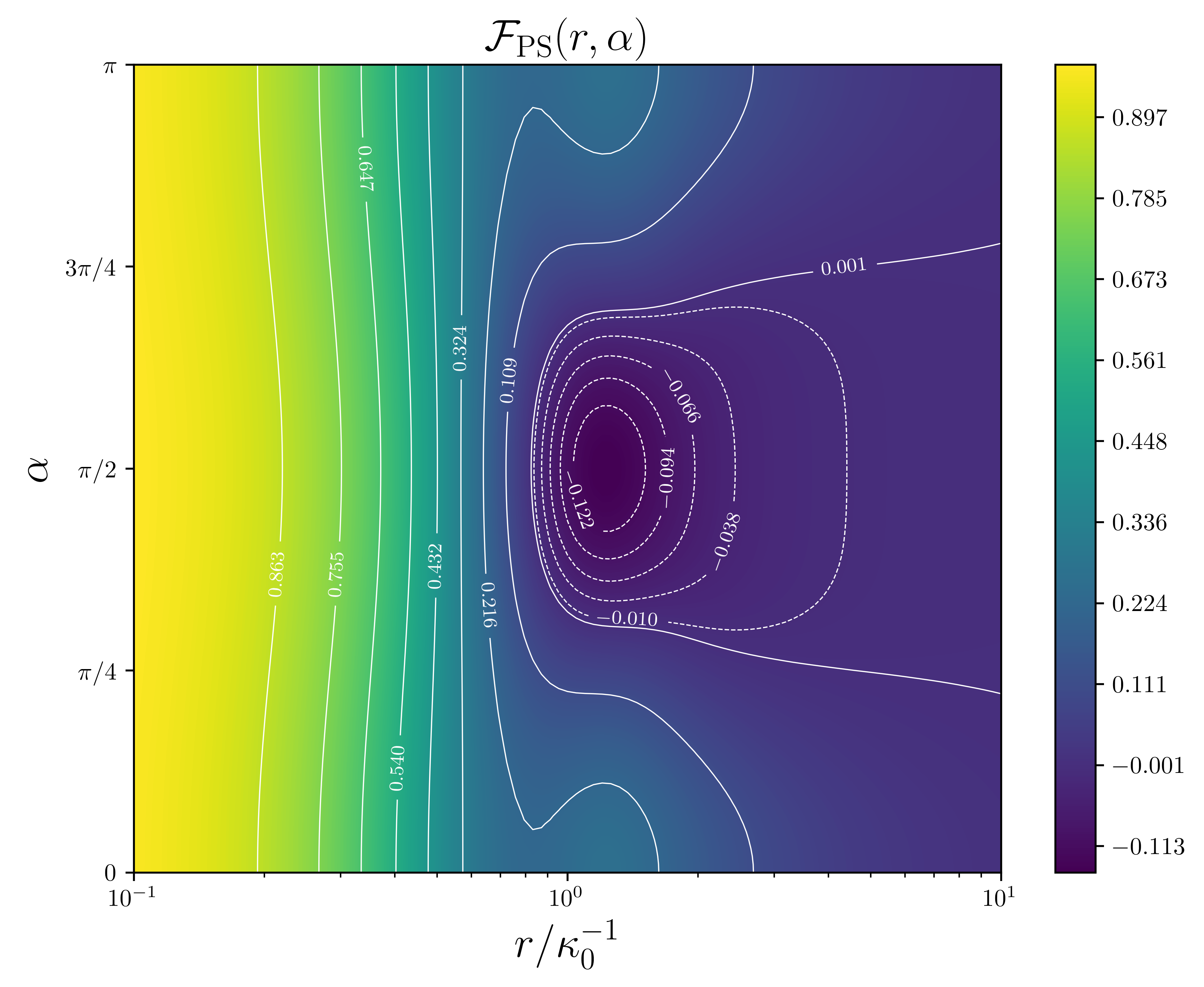}\quad
    
    \caption{\label{fig:FPS}The phase-space form factor ${\cal F}_{\rm PS}$. \emph{Left panel}: the scaling behavior of the magnitude of ${\cal F}_{\rm PS}$ with $r$ for different values of $\alpha$. Note the cusps correspond to the place where ${\cal F}_{\rm PS}$ changes sign. \emph{Right panel}: the contour plot of ${\cal F}_{\rm PS}$ as a function of $r$ and $\alpha$.}
\end{figure}

For a generic value of $\alpha$, the integral in Eq.~(\ref{eq:FPS}) can only be calculated numerically. In the left panel of Fig.~\ref{fig:FPS}, we show ${\cal F}_{\rm PS}$ as a function of $r$ for $\alpha=0$, $30^{\circ}$, $45^{\circ}$, $60^{\circ}$ and $90^{\circ}$. In the coherent limit $r\ll 1/\kappa_0$, ${\cal F}_{\rm PS} \to 1$ for any value of $\alpha$. The distinguishable effects from different angles start to appear in the non-coherent region. 
For $\alpha \in [0,45^{\circ})$, ${\cal F}_{\rm PS}$ is positive across all distances and scales as $1/r^2$ at $r\gg 1/\kappa_0$, and ${\cal F}_{\rm PS}$ decreases monotonically as one increases $\alpha$; when $\alpha$ is around $45^{\circ}$, there is partial cancellation happening between the two terms in Eq.~(\ref{eq:Dfactor}) after the phase-space integral, making ${\cal F}_{\rm PS}$ more suppressed and scaling as $1/r^4$; for $\alpha \in (45^{\circ},90^{\circ}]$, ${\cal F}_{\rm PS}$ changes sign around $r\sim 1/\kappa_0$ and again scales as $1/r^2$ at $r\gg 1/\kappa_0$, and $|{\cal F}_{\rm PS}|$ increases monotonically when  $\alpha$ varies from $45^{\circ}$ to $90^{\circ}$. 

In the right panel of Fig.~\ref{fig:FPS}, we make the contour plot of ${\cal F}_{\rm PS}$ with different values of $r$ and $\alpha$. 
The contour lines also indicate that $|{\cal F}_{\rm PS}|$ attains its lowest around $\alpha=45^{\circ}$ and $\alpha=135^{\circ}$. 
Clearly, ${\cal F}_{\rm PS}$ is symmetric under the transformation $\alpha \to \pi-\alpha$, as also implied by Eq.~(\ref{eq:FPS}). This can be understood in the following way: although the axion DM wind picks a preferred direction by $\vecv_a$, there is no net momentum transfer between the two-object system and the axion background --- $\psi_1$ borrows some momentum from the background but $\psi_2$ returns the same momentum back to the background, and vice versa (see Fig.~\ref{fig:background boxes}). As a result, the magnitude of the background-induced force from object 1 to object 2 should be equal to that from object 2 to object 1 according to momentum conservation. Therefore, the parity symmetry is kept at the potential level: $\vbkg(\vecr)=\vbkg(-\vecr)$. Since $\vbkg \propto {\cal F}_{\rm PS}$, the form factor is also invariant under reflection.

In conclusion, the above decoherence effect from phase-space spread is crucial to consider in the detection of axion DM, which may decrease the sensitivity compared to the coherent limit. In particular, the different scaling behavior between $\alpha \approx 45^{\circ}$ and the other values of $\alpha$ provides an interesting possibility to probe the periodic and time-varying signals due to the Earth's motion relative to the axion DM wind.

\subsubsection*{Finite-size object}
\begin{figure}[t]
    \centering
    \includegraphics[scale=1]
    {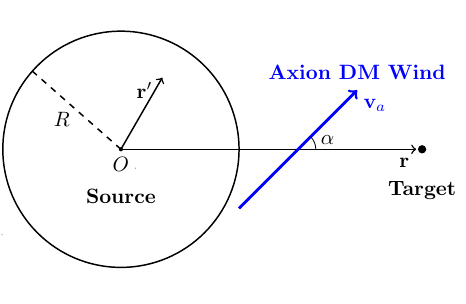}
    \vspace{-0.5cm}
    \caption{\label{fig:finite-size}The schematic plot of the axion force in the background of the axion DM wind between a finite-size source and a point-like target.}
\end{figure}

In actual fifth-force detection experiments, the object has a finite size. We simplify the geometric setup by treating the target as a point-like object while the source is a ball with radius $R$ and constant density (this corresponds to the MICROSCOPE-like experiments). We choose the center of the ball as the reference point and denote the vector from the center of the ball to the point-like object as $\vecr$ with $r\equiv |\vecr|>R$, as shown in Fig.~\ref{fig:finite-size}. Then the background-induced axion forces between them turn out to be 
\begin{align}
V_\text{bkg,ball}^{\slashed{\rm shift}}\left(\vecr;\mu\right) &= -\frac{\mu^2 \rho_a}{4\pi  f_a^4 m_a^2}\frac{N_{\rm p} N_{\rm b}}{V_{\rm b}}\int_{\rm ball} {\rm d}^3\vecr'\, \frac{{\cal F}_{\rm PS}\left(\vecr-\vecr'\right)}{\left|\vecr -\vecr'\right|}\;,\\
V_\text{bkg,ball}^{{\rm inv}}\left(\vecr\right) &= -\frac{c_N^4 \rho_a m_a^2}{64\pi  f_a^4 m_N^2}\frac{N_{\rm p} N_{\rm b}}{V_{\rm b}}\int_{\rm ball} {\rm d}^3\vecr'\, \frac{{\cal F}_{\rm PS}\left(\vecr-\vecr'\right)}{\left|\vecr -\vecr'\right|}\;, 
\end{align}
where $N_{\rm p}$ and $N_{\rm b}$ are the number of nucleons in the point-like object and the ball, respectively, and $V_{\rm b}=4\pi R^3/3$ is the volume of the ball. The physical observable in the fifth-force experiments is proportional to the correction to gravity: $\delta V/V_{\rm grav}$. Comparing the axion forces with gravity, one obtains
\begin{align}
\frac{\delta V_\text{$\mu$-term}}{V_{\rm grav}} &= \frac{2\mu^2 \rho_{a} M_{\rm Pl}^2}{f_a^4 m_a^2 m_N^2}\,{\cal F}_{\rm tot}\;,\label{eq:deltaVmu}\\
\frac{\delta V_\text{mass-term}}{V_{\rm grav}} &= \frac{c_N^4\rho_{a} m_a^2 M_{\rm Pl}^2}{8 f_a^4  m_N^2}\,{\cal F}_{\rm tot}\;,\label{eq:deltaVmass}
\end{align}
where ${\cal F}_{\rm tot}$ is the total form factor that includes the decoherence effect from both phase space and the finite size:
\begin{align}
{\cal F}_{\rm tot}\left(\vecr\right) \equiv \frac{r}{V_{\rm b}}\int_{\rm ball} {\rm d}^3\vecr'\,\frac{{\cal F}_{\rm PS}\left(\vecr-\vecr'\right)}{\left|\vecr -\vecr'\right|}\;. \label{eq:Ftotdef}   
\end{align}
${\cal F}_{\rm tot}$ can be calculated numerically using Eqs.~(\ref{eq:Ftotcompute})-(\ref{eq:alphaeff}) and (\ref{eq:FPS}).

\begin{figure}[t]
    \centering
    \includegraphics[scale=0.37]
    {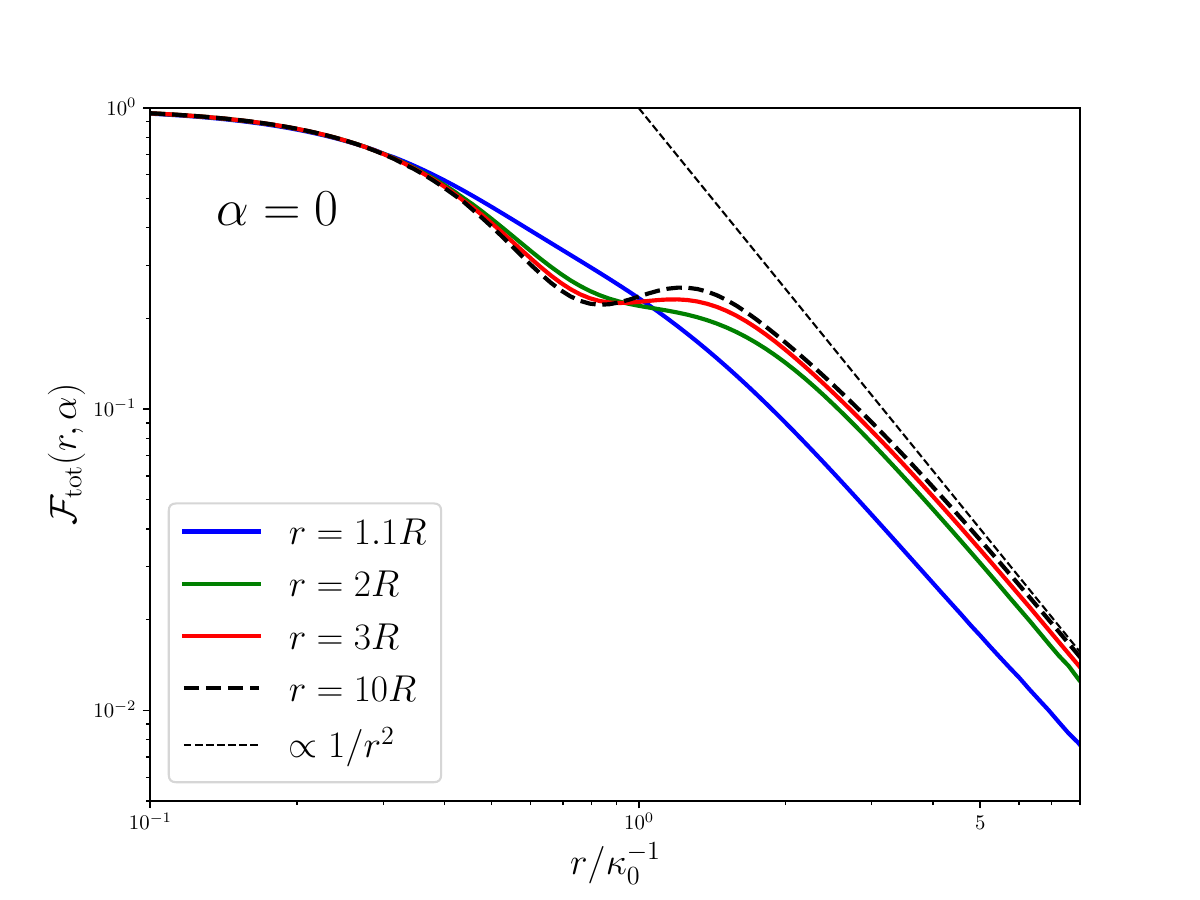}
    \includegraphics[scale=0.37]
    {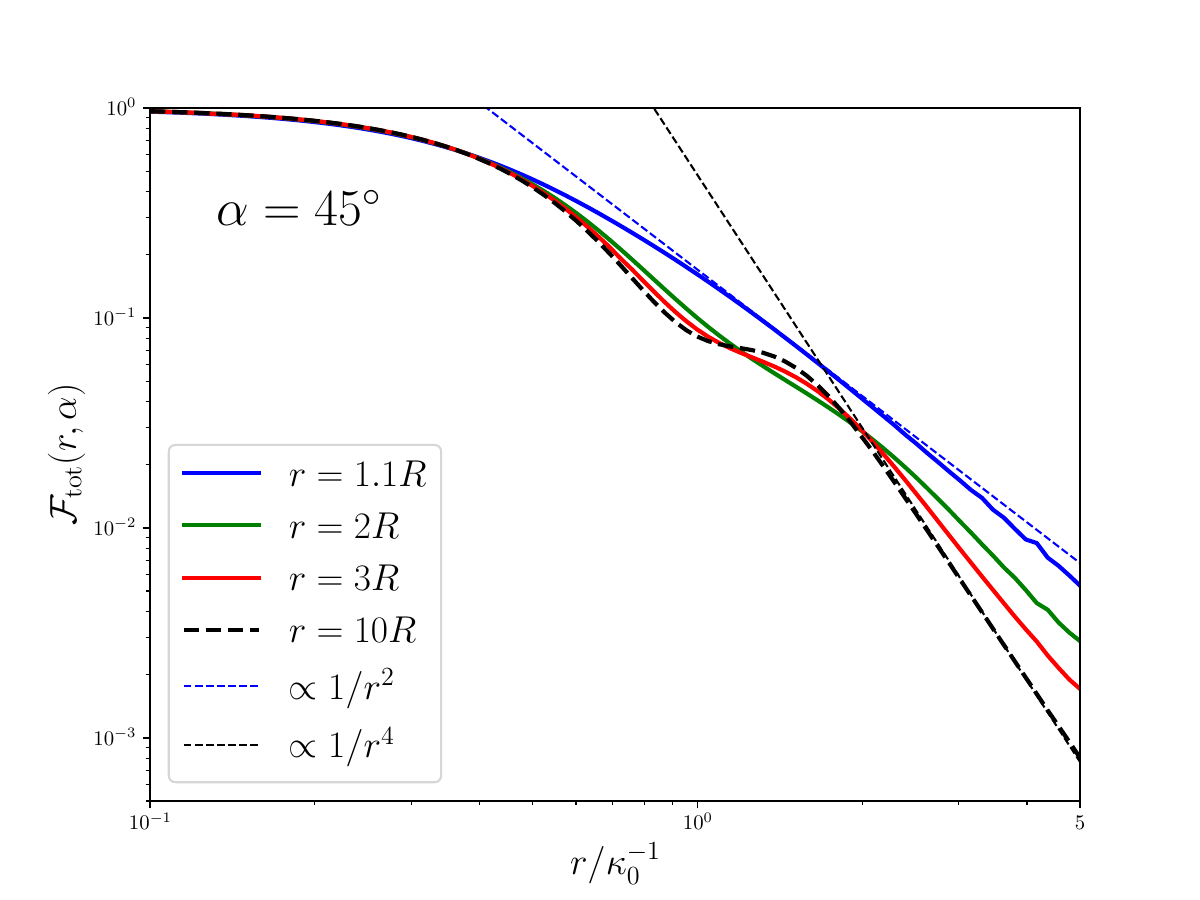}
\caption{\label{fig:Ftot}The scaling behavior of the total form factor ${\cal F}_{\rm tot}$ from a finite-size object with a radius $R$ at $\alpha=0$ (left) or $\alpha = 45^{\circ}$ (right). The thick black dashed lines ($r=10R$) correspond to the point-object limit.}
\end{figure}

\begin{figure}[t]
    \centering
    \includegraphics[scale=0.7]{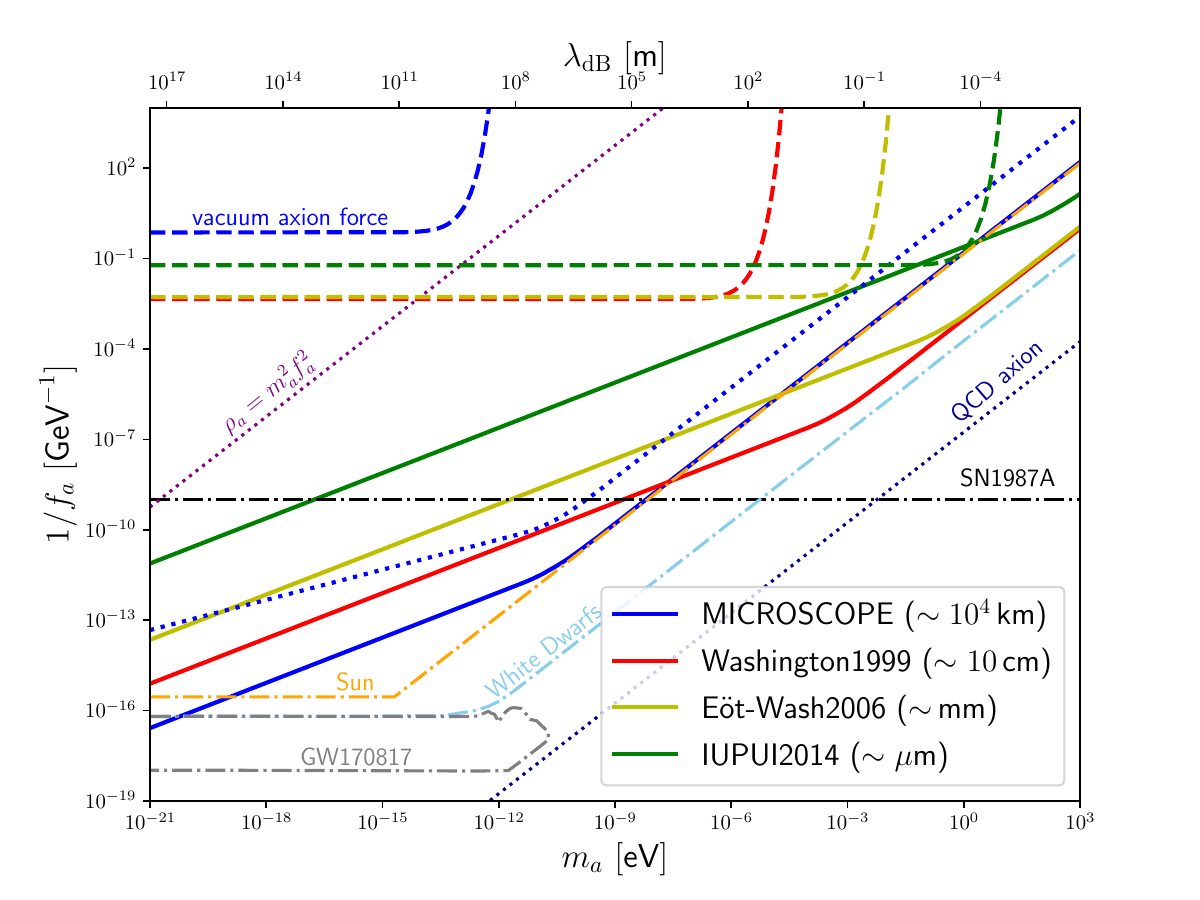}
\caption{\label{fig:DMfa}Fifth-force bounds on the axion couplings from the axion forces. The solid lines correspond to the search for the axion force induced from the axion DM background $\vbkg$ by fixing $\rho_a = \rho_{\rm DM}=0.4~{\rm GeV/cm^3}$ and $\mu=\mu_{\rm QCD}=15~{\rm MeV}$, where different colors denote the  experiments at different length scales. As a comparison, the blue dotted line is obtained by fixing $\mu=\sqrt{m_a f_a}$ and using the MICROSCOPE sensitivity.
The dashed lines with different colors correspond to the search for the vacuum two-axion force $V_{2a}$ by assuming $\rho_a = 0$.  The dark-blue dotted line represents the relation of the QCD axion: $m_a f_a \approx 5.7 \times 10^{-3}~{\rm GeV^2}$~\cite{GrillidiCortona:2015jxo}. The purple dotted line represents the effective theory bound in Eq.~(\ref{eq:EFT bound}), above which the correction to $\vbkg$ from axion self-interactions might be non-negligible. 
The black dash-dotted line denotes the generic astrophysical bound from SN1987A~\cite{Raffelt:1987yt,Turner:1987by,Burrows:1988ah,Raffelt:1990yz}. The orange, light blue, and gray dash-dotted lines denote other astrophysical constraints from the Sun, the white dwarf, and the gravitational wave signals~\cite{Hook:2017psm,Zhang:2021mks,Balkin:2022qer}.}
\end{figure}

In Fig.~\ref{fig:Ftot}, we show the scaling behavior of ${\cal F}_{\rm tot}$ for different  sizes of the object: $r=1.1 R$, $r=2 R$, $r=3 R$ and $r=10 R$. The result of ${\cal F}_{\rm tot}$ with $r=10 R$ already agrees well with that of ${\cal F}_{\rm PS}$, so $r=10 R$ can be taken as the point-source limit, where the finite-size effect is irrelevant. For larger objects, we find that the finite-size effect does not change the scaling behavior of the axion force as long as $\alpha \neq 45^{\circ}$. For example, the source with $r=1.1R$ only decreases ${\cal F}_{\rm tot}$ by about a factor of 2 compared to the point-source limit for $\alpha=0$, as shown in the left panel of Fig.~\ref{fig:Ftot}. However, the finite-size effect becomes important when $\alpha$ is close to $45^{\circ}$. This is because at long distances $r\gg R$, the partial cancellation still exists and leads to ${\cal F}_{\rm tot}\propto 1/r^4$, similar to the point-source limit, but as the target gets closer to the source, the cancellation no longer occurs for most of the points in the source, for which the effective angle is away from $45^{\circ}$. As a result, the finite size enhances ${\cal F}_{\rm tot}$ compared to the point-source limit and makes it again scale as $1/r^2$. This is shown in the right panel of Fig.~\ref{fig:Ftot}.

\subsubsection{Fifth-force bounds}

\begin{figure}[t]
    \centering
    \includegraphics[scale=0.55]
    {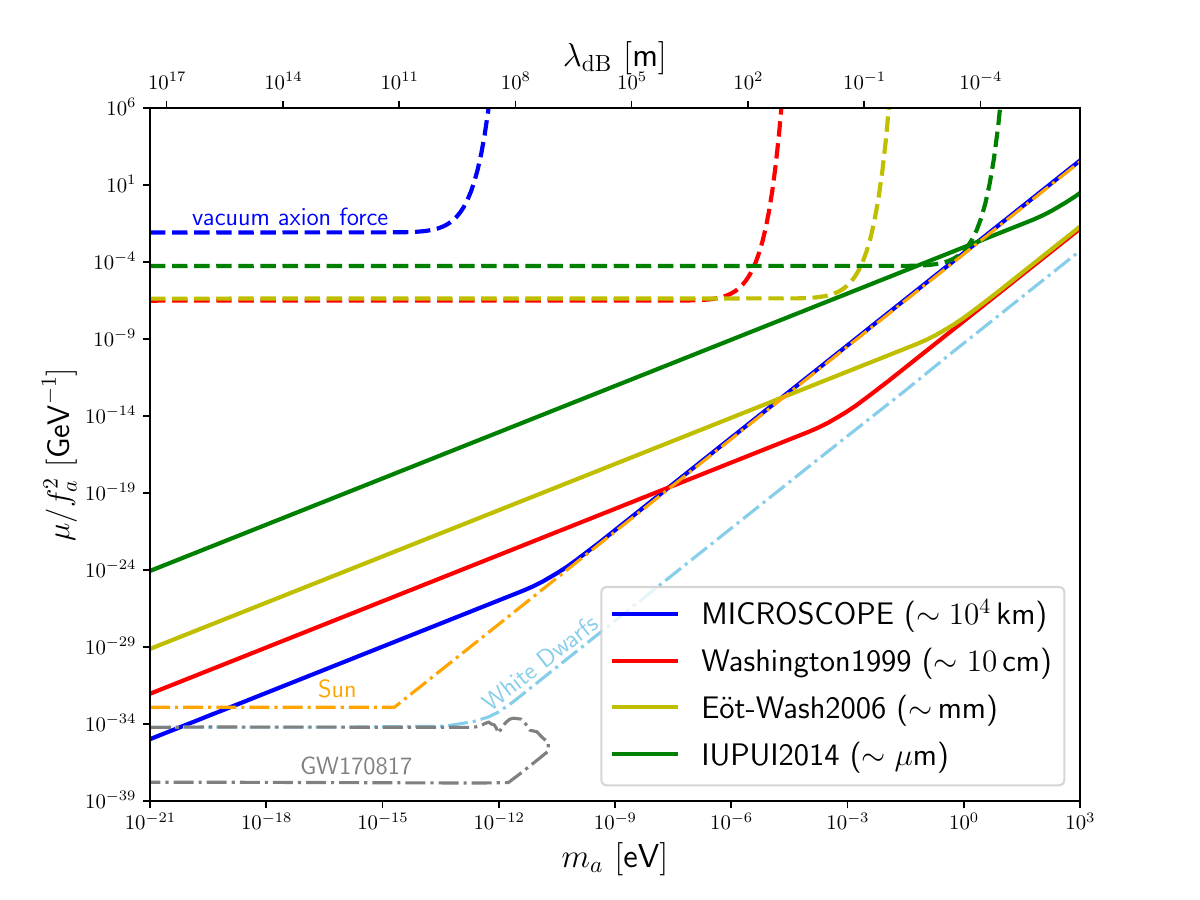}
\caption{\label{fig:DMfa-mu}Same conventions as Fig.~\ref{fig:DMfa}, but with $\mu$ as a free parameter.}
\end{figure}

\begin{figure}[t]
    \centering
    \includegraphics[scale=0.55]
    {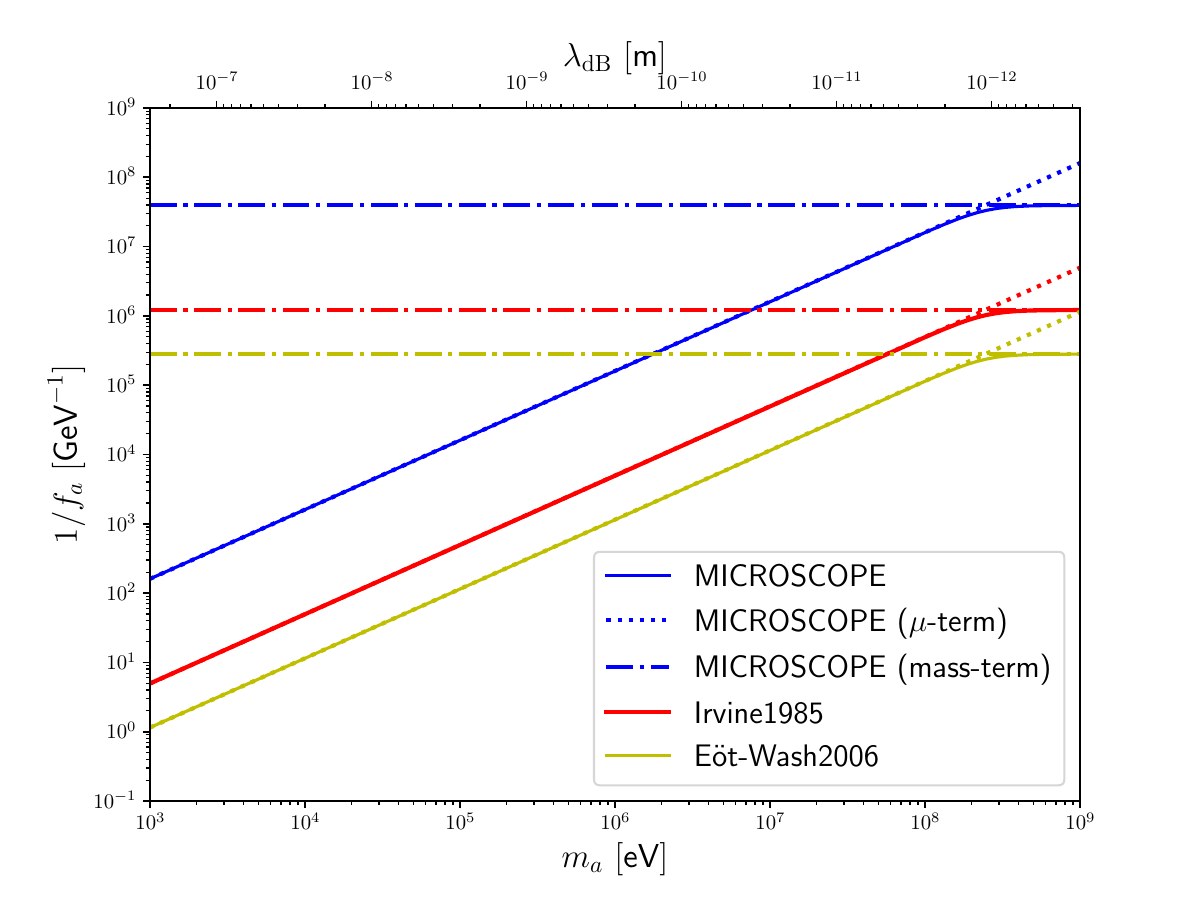}
\caption{\label{fig:DMfa-heavymass}Same conventions as Fig.~\ref{fig:DMfa} but in the heavy axion mass region, where the astrophysical bound is relaxed. The dotted (dash-dotted) lines correspond to the contribution to $\vbkg$ from $\mu$-term (mass term), while the solid lines include both contributions.}
\end{figure}

Then we calculate the experimental bounds on the axion couplings from the fifth-force searches for the axion forces. We map Eqs.~(\ref{eq:deltaVmu})-(\ref{eq:deltaVmass}) to the experimental sensitivities in Tab.~\ref{table:deltav/v}, taking into account the decoherence effect from ${\cal F}_{\rm tot}$ in Eq.~(\ref{eq:Ftotdef}). For concreteness, we take $\alpha=0$ to maximize the background effect and fix $r=1.1R$ (corresponding to the setup of MICROSCOPE~\cite{MICROSCOPE:2022doy}).

The results are shown in Fig.~\ref{fig:DMfa} and Fig.~\ref{fig:DMfa-mu}. We plot the bounds separately from the background-induced axion force $\vbkg$ (solid lines) and the vacuum two-axion force $V_{2a}$ (dashed lines). For $\vbkg$, we fix $\rho_a = \rho_{\rm DM} = 0.4~{\rm GeV/cm^3}$. 
Note that for $m_a \ll {\rm GeV}$, the contribution to $\vbkg$ from the mass term (\ref{eq:deltaVmass}) is negligible compared to that from the $\mu$-term (\ref{eq:deltaVmu}). The bounds from $V_{2a}$ remain constant in the small axion mass limit, but the sensitivities decrease exponentially when the axion mass is greater than the inverse of the experimental length scale, as indicated by Eq.~(\ref{eq:V2a-large-r}). On the other hand, for a fixed energy density, the sensitivity of the bounds from $\vbkg$ is enhanced for smaller axion masses due to the increasing occupation number of background axions. However, from a theoretical point of view, the sensitivity cannot increase infinitely by keeping decreasing the axion mass, otherwise the axion classical field value is too large to invalidate the axion effective coupling in Eq.~(\ref{eq:muterm}). As a conservative bound, the axion field value cannot exceed the cutoff scale $f_a$ --- this corresponds to the purple dotted line in Fig.~\ref{fig:DMfa}, above which the contributions from higher-order axion couplings might be relevant. When the experimental length scale is greater than the de Brogile wavelength of the background axions $\db\sim 10^3/m_a$, the sensitivities are additionally reduced by the decoherence effect from ${\cal F}_{\rm tot}$, which corresponds to the turning points of the solid lines in Figs.~\ref{fig:DMfa} and \ref{fig:DMfa-mu}. A similar decoherence behavior was also obtained in \cite{VanTilburg:2024xib}.
In most of the parameter space, the contribution from $\vbkg$ by assuming an axion DM background dominates over that from $V_{2a}$. 

It can be seen from Fig.~\ref{fig:DMfa} that for the axion DM lighter than about $10^{-9}{\rm eV}$, and when the effective coupling $\mu$ is not much more suppressed compared to $\mu_{\rm QCD}$, the bound from $\vbkg$ is stronger than the generic astrophysical bound from SN1987A: $f_a \gtrsim 10^{9}~{\rm GeV}$~\cite{Raffelt:1987yt,Turner:1987by,Burrows:1988ah,Raffelt:1990yz} (black dash-dotted line). For light QCD axions, the quadratic coupling to nucleons can induce an effective negative mass-squared term in dense stellar environments, potentially flipping the sign of the axion potential. In such cases, the axion field acquires a nonzero expectation value inside the star, which can affect stellar stability and modify the gravitational-wave signals from binary systems~\cite{Hook:2017psm}. These effects lead to new astrophysical bounds that are significantly stronger than the traditional SN1987A limit in the low-mass region~\cite{Hook:2017psm,Zhang:2021mks,Balkin:2022qer,Gomez-Banon:2024oux,Springmann:2024ret,Kumamoto:2024wjd,Bartnick:2025lbg,Bartnick:2025lbv}. The corresponding constraints are shown in Figs.~\ref{fig:DMfa} and \ref{fig:DMfa-mu} as orange (Sun~\cite{Hook:2017psm}), light blue (White Dwarfs~\cite{Balkin:2022qer}), and gray (GW170817~\cite{Zhang:2021mks}) dash-dotted lines.

The Lyman-$\alpha$ forest places a generic lower bound on the mass of ultralight DM: $m_a \gtrsim 10^{-21}~{\rm eV}$~\cite{Irsic:2017yje,Armengaud:2017nkf,Kobayashi:2017jcf,Rogers:2020ltq}. For axion DM heavier than $10^{-21}~{\rm eV}$, our bounds do not surpass the existing astrophysical constraints. For lighter mass, $m_a \lesssim 10^{-21}~{\rm eV}$, axions are allowed only if they constitute a subdominant fraction of the total DM, $\rho_a=\xi\,\rho_{\rm DM}$ with $\xi\lesssim 0.2$, where the Lyman-$\alpha$ constraints lose sensitivity~\cite{Kobayashi:2017jcf}.\footnote{High-redshift observables still impose complementary constraints. For instance, Ref.~\cite{Winch:2024mrt} excluded the region $m_a < 10^{-26}~{\rm eV}$ for axion fraction $\xi > 0.22$, while Ref.~\cite{Lazare:2024uvj} found $m_a < 10^{-23}~{\rm eV}$ disfavored for $\xi > 0.16$ and $m_a < 10^{-26}~{\rm eV}$ disfavored for $\xi > 0.01$.}
In this regime, the solid curves in Figs.~\ref{fig:DMfa} and \ref{fig:DMfa-mu} can be extrapolated toward lighter masses with nearly the same slope. From Eq.~(\ref{eq:deltaVoverVgrav}), for a fixed experimental sensitivity, the decay constant scales $f_a \propto \rho_a^{1/4}\propto \xi^{1/4} \rho_{\rm DM}^{1/4}$; hence the bounds on $f_a$ depend only weakly on the axion fraction $\xi$. Consequently, our projected limits could in principle exceed the astrophysical bounds. However, because axions cannot constitute the dominant component of DM in this mass range, we do not display these extrapolated regions in Figs.~\ref{fig:DMfa} and \ref{fig:DMfa-mu}.

In Fig.~\ref{fig:DMfa-heavymass}, we show the bounds for heavier axions with masses in the range ${\rm keV}<m_a<{\rm GeV}$. In this regime, the corresponding astrophysical constraints are largely relaxed.
It should be noted that such heavy axions are generally unstable if they couple to photons. To allow axions to serve as DM and to estimate the maximal possible background effect, we therefore assume in Fig.~\ref{fig:DMfa-heavymass} that the axion-photon coupling is absent.
The vacuum axion force has a very short range and is thus totally negligible in practical fifth-force experiments. 
On the other hand, $\vbkg$ is only quadratically suppressed due to the decoherence effect: ${\cal F}_{\rm tot} \propto 1/(m_a^2 r^2)$, where $r$ is the experimental length scale. In the heavy axion mass region, the contribution to $\vbkg$ from the mass term (\ref{eq:deltaVmass}) is non-negligible and is shown with dash-dotted lines (with $c_N=1$) in Fig.~\ref{fig:DMfa-heavymass}. The contribution from the $\mu$-term (\ref{eq:deltaVmu}) is shown with dotted lines, and the solid lines represent the combined effects.

\subsubsection{Comparison with CP-violating axion coupling}
In the absence of CP-violating axion coupling, the tree-level axion force (\ref{eq:Va}) is always spin dependent. However, a Yukawa-like one-axion force appears if there exists a small CP-violating axion coupling ${\cal L}\supset \gCPV \,a \bar{\psi}\psi$ with $\gCPV \ll 1$: $V_{\cancel{\rm CP}}(r) = -\gCPV^2\,e^{-m_a r}/(4\pi r)$. For $\psi$ to be the nucleon and $a$ to be the QCD axion, it was found~\cite{Okawa:2021fto}: 
\begin{align}
\gCPV \sim 1.5 \times 10^{-12} \left(\frac{\rm GeV}{f_a}\right) \left(\frac{\theta_{\rm ind}}{10^{-10}} \right),\label{eq:gCPV}   
\end{align}
where $\theta_{\rm ind}$ denotes the additional contribution to the QCD theta term induced by the CP-violating axion coupling, which is heavily constrained due to the non-measurement of the neutron's electric dipole moment, $\theta_{\rm ind} \lesssim 10^{-10}$. 

It is then interesting to compare the CP-violating axion force $V_{\cancel{\rm CP}}$ with the background-induced axion force $V_{\rm bkg}$ we computed above, because both effects are spin-independent. Since $\VCPV$ becomes exponentially suppressed at $r \gtrsim 1/m_a$, while $\vbkg\sim 1/r$ for $1/m_a \lesssim r\lesssim 10^{3}/m_a$ and $\vbkg\sim \db^2/r^3$ for $r\gg 10^3/m_a$, it is clear that $\vbkg$ will dominate over $\VCPV$ at $r \gg 1/m_a$. Therefore, we only compare them at short distances $r \lesssim 1/m_a$, where both potentials scale as $1/r$. For $r\lesssim 1/m_a$, we have
\begin{align}
\frac{\vbkg}{\VCPV}= \frac{\mu^2 \rho_a}{f_a^4 m_a^2 \gCPV^2}\;.
\end{align}
For the QCD axion, plugging into $\mu \approx 15~{\rm MeV}, m_a f_a \approx 5.7 \times 10^{-3}~{\rm GeV^2}, \rho_a \approx 0.4~{\rm GeV}/{\rm cm}^3$, and using Eq.~(\ref{eq:gCPV}), we obtain
\begin{align}
 \frac{\vbkg}{\VCPV} \sim  \left(\frac{3\times 10^{-19}}{\theta_{\rm ind}}\right)^2.  
\end{align}

Therefore, we conclude that for the QCD axion, $\vbkg$ dominates over $\VCPV$ at $r\gg 1/m_a$ regardless of the value of $\theta_{\rm ind}$, while at $r\lesssim 1/m_a$, the background effect is dominant only with an extremely small induced theta angle, $\theta_{\rm ind}\lesssim 10^{-19}$.

\subsection{Atomic spectroscopy}
\label{subsec:atom}
Over the last two decades, significant advancements in quantum technology have rendered precision atomic physics a promising approach to probing new physics at low-energy scales (see \cite{Safronova:2017xyt} for a review). Atomic systems usually have better sensitivities compared to traditional experiments at macroscopic scales to search for the fifth force and are particularly useful for probing the interactions mediated by particles that couple feebly to ordinary matter. For example, a long-range spin-spin interaction can lead to a hyperfine splitting in the atomic ground state, which has been precisely measured~\cite{Klaft:1994zz,Liu:1999iz,ullmann2017high}. Moreover, any interaction that fundamentally violates parity can contribute to atomic parity violation,  an observable that has been well measured in Cesium~\cite{Wood:1997zq,Bouchiat:1997mj,Wieman:2019vik}. The utilization of atomic systems for detecting the long-range force mediated by neutrinos was recently discussed in \cite{Dzuba:2017cas,Ghosh:2019dmi,Munro-Laylim:2022fsv,Ghosh:2024ctv}.

In this work, we are interested in the spin-independent axion force induced by an axion background, which does not contribute to hyperfine splitting or atomic parity violation. However, the spin-independent axion force is coherent in the sense that its magnitude is proportional to the number of nucleons in atoms coupled to the axion. As a result, it contributes differently in atoms with different numbers of nucleons and can be probed by the atomic isotope shift~\cite{Berengut:2017zuo}. 

As a simple example, we use the $1S-2S$ isotope shift measured in the Hydrogen-Deuterium system~\cite{Parthey:2010aya}:
\begin{align}
\Delta \nu_{\rm exp} \equiv \nu^{\rm D}_{1S-2S}-\nu_{1S-2S}^{\rm H}-\Delta \nu_{\rm HFS}= 670 994 334 606 (15)~{\rm Hz}\;,\label{eq:fexp}
\end{align}
where $\nu_{1S-2S}^{\rm D}$ and $\nu_{1S-2S}^{\rm H}$ denote the transition frequency between $1S$ and $2S$ states in Deuterium and Hydrogen, respectively, and $\Delta \nu_{\rm HPS}$ is the hyperfine
structure correction. Note that the spin-independent axion force does not contribute to $\Delta \nu_{\rm HFS}$. On the other hand, the theoretical prediction from the Standard Model interaction reads~\cite{Parthey:2010aya}:
\begin{align}
\Delta \nu_{\rm th} = 670 999 566.90(66)(60)~{\rm kHz}\;.\label{eq:fth}    
\end{align}
Combining Eqs.~(\ref{eq:fexp}) and (\ref{eq:fth}) we have
\begin{align}
\left|\delta \nu_{\rm HD}\right|\equiv \left|\Delta \nu_{\rm exp}-\Delta \nu_{\rm th}\right| \approx 5.2~{\rm MHz} \approx 2.2\times 10^{-8}~{\rm eV}\;. \label{eq:deltaf}   
\end{align}
As a conservative limit, the contribution from axion forces should not exceed Eq.~(\ref{eq:deltaf}).

In the following, we calculate the axion contribution to the isotope shift in the Hydrogen-Deuterium system. We assume that axions comprise DM and that the axion force is dominated by the quadratic coupling in Eq.~(\ref{eq:muterm}). We include contributions from both the background-induced axion force $\vbkg$ and the vacuum two-axion force $V_{2a}$. Note that the tree-level axion force in Eq.~(\ref{eq:Vamassless}) does not contribute  because the tensor structure vanishes for $S$-state matrix elements:
\begin{align}
\bigg\langle nS\bigg| \frac{3\left(\vecsigma_e\cdot\hat{\vecr}\right)\left(\vecsigma_N\cdot\hat{\vecr}\right)-\left(\vecsigma_e\cdot\vecsigma_N\right)}{r^3} \bigg|nS\bigg\rangle \equiv 0\;,    
\end{align}
where $n$ is any positive integer, $\vecsigma_e$ and $\vecsigma_N$ are the spin operators of the electron and the nucleon in the atom. 

For the sub-eV axion DM, its de Broglie wavelength $\db \sim 0.02~{\rm cm}~({\rm eV}/m_a)$ is much larger than the atomic length scale. As a result, it is an extremely good approximation to take the coherent limit. Using Eq.~(\ref{eq:Vcoh}), the background-induced axion force between the electron and the nucleon reads 
\begin{align}
V_{\rm bkg}^{eN}(r) = -\frac{\mu_e \mu_N \rho_a}{4\pi r f_a^4 m_a^2}\;,    
\end{align}
where $\mu_e$ and $\mu_N$ correspond to the quadratic coupling in Eq.~(\ref{eq:muterm}) with $\psi$ replaced by the electron or the nucleon, respectively. (A natural UV realization of the quadratic coupling between an ultralight axion and the electron can be found in, e.g., \cite{Delaunay:2025pho}.)
It is straightforward to calculate the induced matrix elements:
\begin{align}
\Big\langle nS \Big|V_{\rm bkg}^{eN}\Big|nS \Big\rangle  =   -\frac{\mu_e \mu_N \rho_a}{4\pi f_a^4 m_a^2 a_0 n^2}\;, \label{eq:VbkgnS}
\end{align}
where $a_0 \approx 1/(3.7~{\rm keV})$ is the Bohr radius. 

On the other hand, the vacuum two-axion force $V_{2a}^{eN}$ is given by Eq.~(\ref{eq:V2a}) with $\mu^2\to\mu_e \mu_N$. However, since $V_{2a}^{eN} \propto 1/r^3$ at short distances, the $S$-state matrix elements induced by $V_{2a}^{eN}$ are divergent, indicating the failure of the effective coupling (\ref{eq:muterm}) at high-energy scales. To estimate the effect from the vacuum two-axion force, we put a cutoff at $r=1/f_a$ in the radial integral, leading to
\begin{align}
\Big\langle 1S \Big|V_{2a}^{eN}\Big|1S \Big\rangle  &=  -\frac{\mu_e \mu_N m_a}{8\pi^3 f_a^4 a_0^3} \int_{1/f_a}^{\infty} {\rm d}r\,K_1\left(2m_a r\right) e^{-2r/a_0}\nonumber\\
&\approx -\frac{\mu_e \mu_N}{16\pi^3 f_a^4 a_0^3}\left[\log\left(\frac{f_a a_0}{2}\right)-\gamma_{\rm E}\right]+{\cal O}\left(\frac{1}{f_a a_0}\right),\label{eq:V2a1S}\\
\Big\langle 2S \Big|V_{2a}^{eN}\Big|2S \Big\rangle  &=  -\frac{\mu_e \mu_N m_a}{256\pi^3 f_a^4 a_0^3} \int_{1/f_a}^{\infty} {\rm d}r\,K_1\left(2m_a r\right)\left(2-\frac{r}{a_0}\right)^2 e^{-r/a_0}\nonumber\\
&\approx -\frac{\mu_e \mu_N}{128\pi^3 f_a^4 a_0^3}\left[\log\left(f_a a_0\right)-\gamma_{\rm E}-\frac{3}{2}\right]+{\cal O}\left(\frac{1}{f_a a_0}\right),\label{eq:V2a2S}
\end{align}
where $\gamma_{\rm E}\approx 0.577$ is Euler's constant. Note that in the second line of Eqs.~(\ref{eq:V2a1S}) and (\ref{eq:V2a2S}), we have assumed $ f_a\gg 1/a_0 \approx 3.7~{\rm keV}$; in this case, the matrix elements only depend logarithmically on the cutoff scale.

\begin{figure}[t]
    \centering
    \includegraphics[scale=0.45]
    {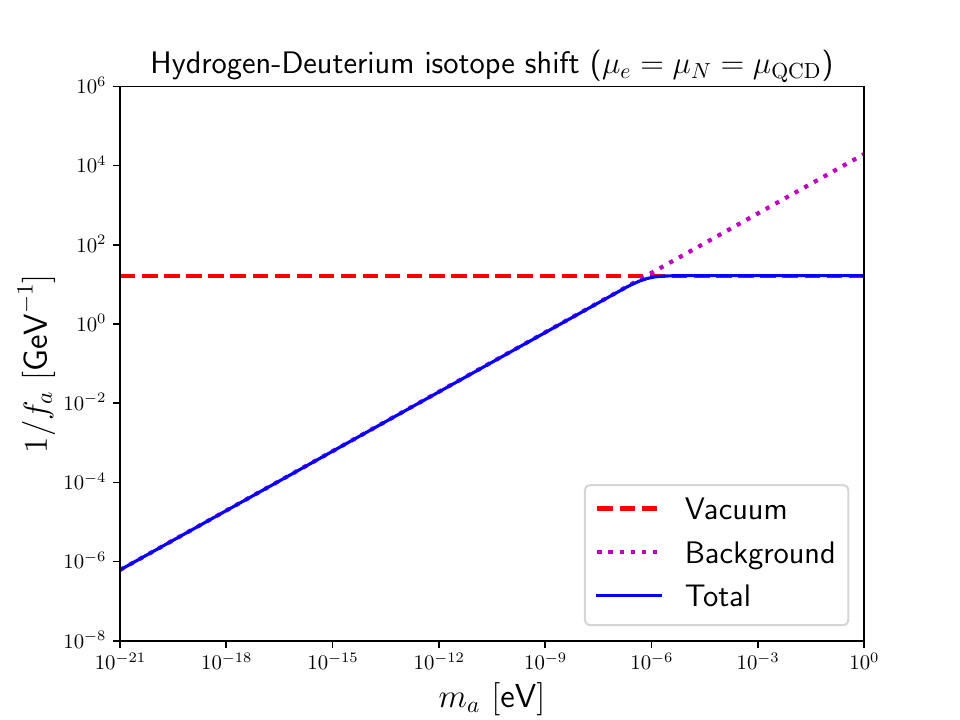}\quad
    \includegraphics[scale=0.45]
    {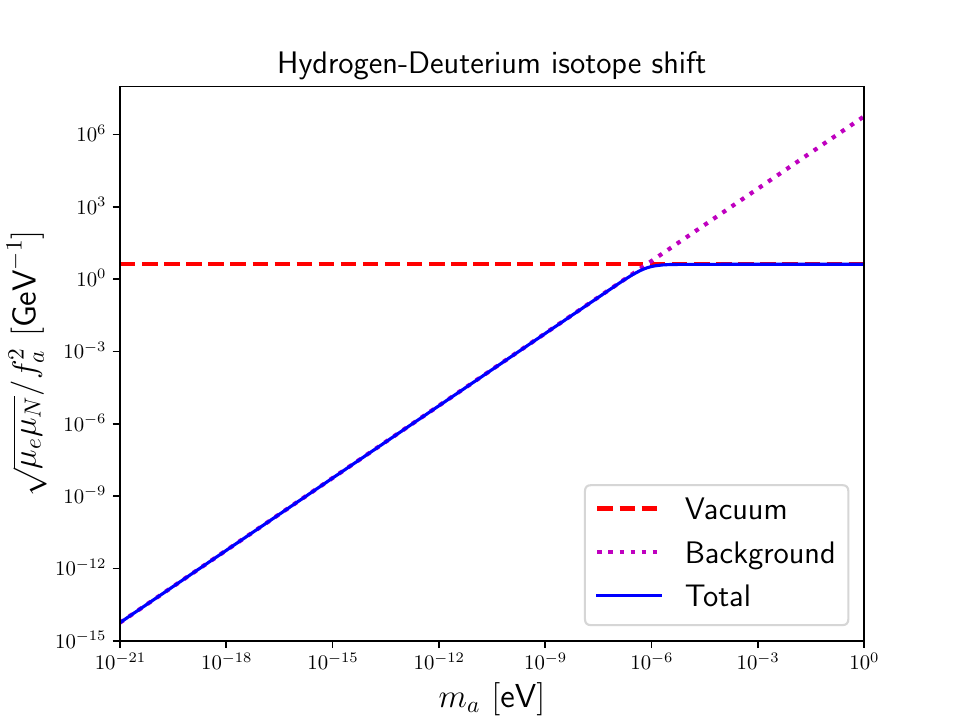}
\caption{\label{fig:atom}Constraints on the axion coupling from the isotope shift measured in the Hydrogen-Deuterium system. The dashed lines and the dotted lines correspond to the contribution from the vacuum two-axion force $V_{2a}$ and the background-induced axion force $\vbkg$ (with $\rho_a = \rho_{\rm DM} = 0.4~{\rm GeV/cm^3}$ fixed), respectively, while the solid lines include both contributions. In the left panel we fix $\mu_e = \mu_N = \mu_{\rm QCD} = 15~{\rm MeV}$ and put the bound on $1/f_a$; in the right panel we take $\mu_e$ and $\mu_N$ as free parameters and put the model-independent bound on $\sqrt{\mu_e \mu_N}/f_a^2$.}
\end{figure}

Combining Eqs.~(\ref{eq:VbkgnS})-(\ref{eq:V2a2S}), we obtain the isotope shift in the Hydrogen-Deuterium system caused by the axion force:
\begin{align}
\delta \nu_{\rm axion} \approx  \frac{3 \mu_e \mu_N \rho_a}{16\pi f_a^4 m_a^2 a_0}  + \frac{7 \mu_e \mu_N}{128\pi^3 f_a^4 a_0^3}\left[\log\left(f_a a_0\right)+\frac{3}{14}-\gamma_{\rm E}-\frac{8}{7}\log2\right],\label{eq:deltafaxion} 
\end{align}
where we have neglected the difference between protons and neutrons and assumed both of their couplings to the axion are described by $\mu_N$. The first (second) term in Eq.~(\ref{eq:deltafaxion}) corresponds to the contribution from $\vbkg$ ($V_{2a}$). To get a  bound on the axion coupling, we assume that there are no new-physics contributions to the isotope shift other than the axion force and ask $|\delta \nu_{\rm axion}| \leq |\delta \nu_{\rm HD}|$. The corresponding bound is shown in Fig.~\ref{fig:atom}. 

In addition to the isotope shift, the axion force can also contribute to the $1S-2S$ transition frequency of non-hadronic atoms such as muonium (if axions couple to charged leptons), which has been measured with high precision~\cite{Meyer:1999cx}. The ongoing Mu-MASS experiment could further improve the sensitivity of the measurement of the muonium $1S-2S$ transition frequency by three orders of magnitude~\cite{Crivelli:2018vfe}. Apart from that, the entanglement witness, which was originally designed to probe quantum gravity~\cite{Bose:2017nin,Marshman:2019sne,Bose:2022uxe}, is another promising quantum technology for probing axion-mediated forces~\cite{Rufo:2025rps}.

\subsection{Solar axion flux}
\label{subsec:solar}
\begin{figure}[t]
    \centering
    \includegraphics[scale=0.45]
    {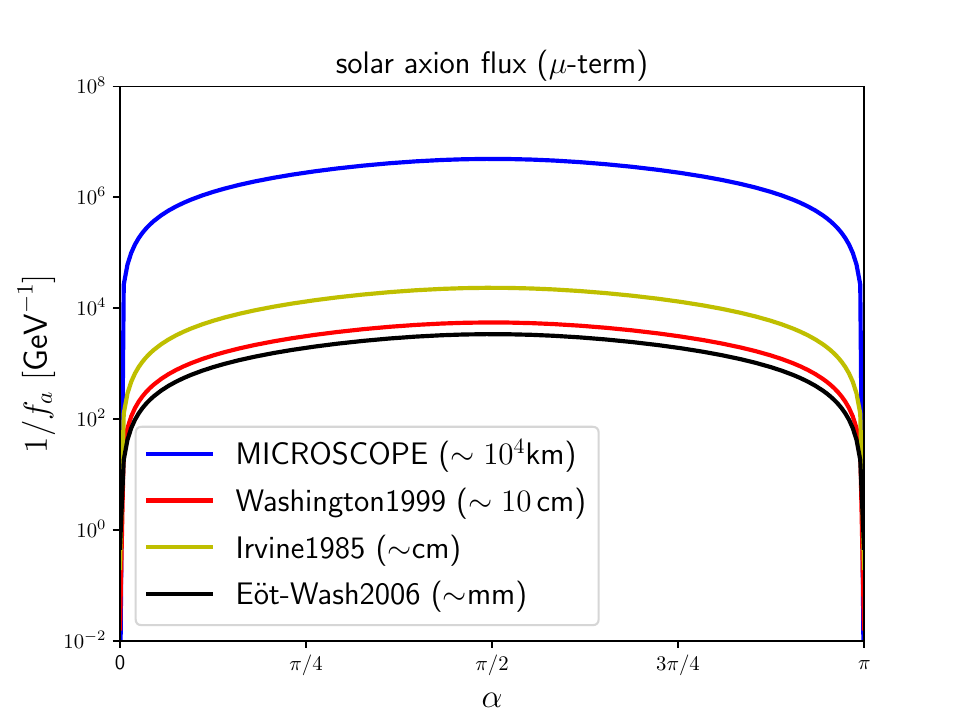}
    \includegraphics[scale=0.45]
    {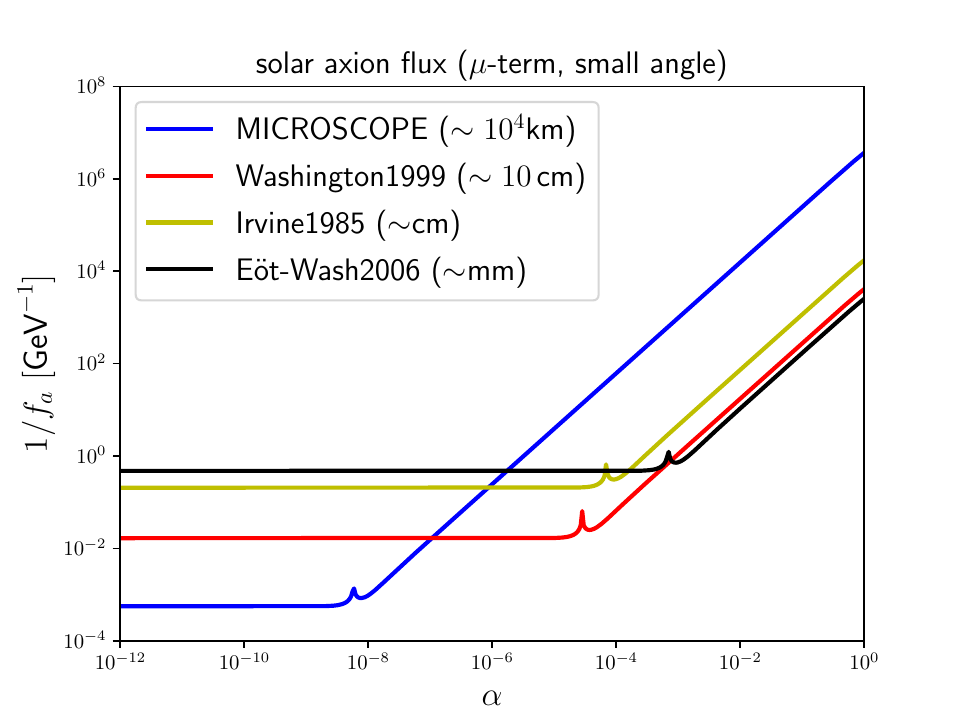}\\
    \includegraphics[scale=0.45]
    {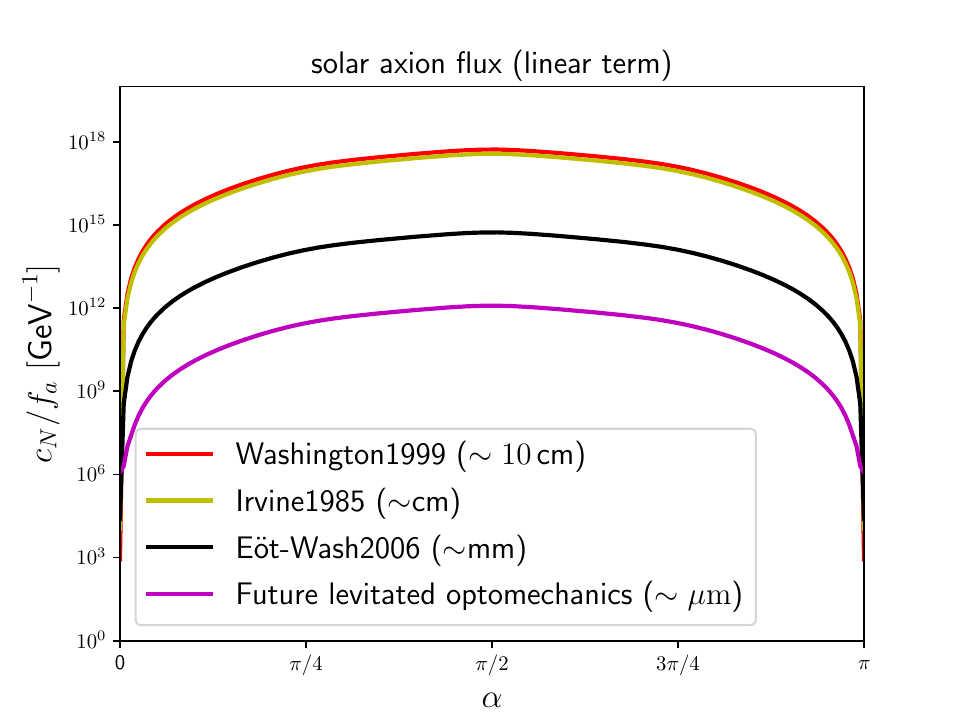}
    \includegraphics[scale=0.45]
    {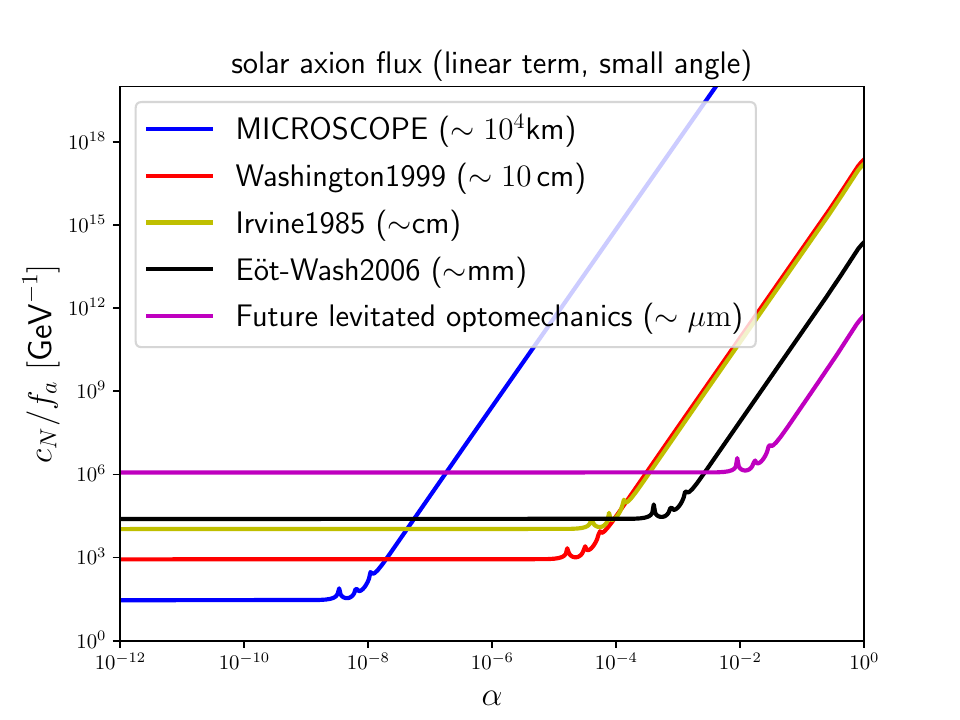}
\caption{\label{fig:solarflux}Constraints on the axion coupling from the axion force induced by the solar axion flux. The bounds are applicable for relativistic solar axions ($m_a \ll {\rm keV}$).
The diagrams in the first line and second line correspond to the axion force in Eq.~(\ref{eq:Vfluxmu}) with $\mu=\mu_{\rm QCD}$ and that in Eq.~(\ref{eq:Vfluxk0}) with $c_1=c_2 = c_N$ and $m_1=m_2=m_N$, respectively. The sensitivitis and length scales of specific experiments are taken from Tab.~\ref{table:deltav/v}. In the left panel, we show the constraints for $\alpha \in [0,\pi)$, while the right panel zooms in the constraints at small values of the angle (coherent limit). The cusps correspond to the places where the axion forces change the sign.}
\end{figure}

Apart from DM relic abundance, the Sun is another typical source of axion background. Axions can be copiously produced in the Sun through their coupling to photons ${\cal L} \supset -g_{a\gamma\gamma}aF\tilde{F}/4$, via the Primakoff process~\cite{Primakoff:1951}: $\gamma + \psi \to a +\psi$. Here $\gamma$ is a photon, $F$ and $\tilde{F}$ are the electromagnetic tensor and its dual, respectively, and $\psi$ is a charged particle. 

The first semi-analytical spectrum for solar axion flux was derived in \cite{vanBibber:1988ge}, where the typical energy scale is $\kappa_0 \sim {\cal O}$(keV). For the practical calculation in this work, we use the solar axion spectrum fitted by the CAST collaboration~\cite{CAST:2007jps}:
\begin{align}
\frac{{\rm d}\Phi}{{\rm d}\kappa} = 6.02 \times 10^{10}~{\rm cm}^{-2} {\rm s}^{-1} {\rm keV}^{-1} \left(\frac{g_{a\gamma\gamma}}{10^{-10}~{\rm GeV}^{-1}}\right)^{2} \left(\frac{\kappa}{{\rm keV}}\right)^{2.481}\exp\left(-\frac{\kappa}{1.205~{\rm keV}}\right).\label{eq:solar-spectrum}
\end{align}
The spectrum in Eq.~(\ref{eq:solar-spectrum}) is valid for relativistic axions (i.e., $m_a\ll {\rm keV}$), see \cite{Wu:2024fsf} for the correction of the spectrum from the axion mass. In the following, we take $g_{a\gamma\gamma}=10^{-10}~{\rm GeV}^{-1}$ to saturate the CAST bound on the axion-photon coupling and investigate the constraints on the axion-nucleon coupling from fifth-force experiments.

The general expression of the axion force in the background of a relativistic axion flux has been derived in Eqs.~(\ref{eq:Vfluxmu})-(\ref{eq:Vfluxk0}). Substituting the spectrum (\ref{eq:solar-spectrum}) into Eqs.~(\ref{eq:Vfluxmu})-(\ref{eq:Vfluxk0}) and performing the integral, we obtain the axion force induced by the solar axion background. The calculation details are provided in App.~\ref{app:solar}, and the constraints are plotted in Fig.~\ref{fig:solarflux}. 

Sensitivities are enhanced in the small-$\alpha$ limit, where all axions in the flux can coherently contribute to $\vbkg$. From the experimental point of view, such small angles are difficult to realize in current experiments, given that both the target and the source have a finite size. Therefore, the enhancement in Fig.~\ref{fig:solarflux} should be understood only as the ideal situation, rather than one that is practically available.
When $\alpha^2 \gtrsim \kappa_0 r \sim (r/0.1\,{\rm nm})$, the decoherence effect starts to appear and sensitivities are decreased, corresponding to the turning points of the curves in the right panel of Fig.~\ref{fig:solarflux}. For $\alpha^2 \gg \kappa_0 r$, sensitivities are suppressed by some inverse power of $(r/0.1\,{\rm nm})$  [see Eqs.~(\ref{eq:D1larger})-(\ref{eq:D2larger}) for detailed expressions] compared to the coherent limit, where $r$ is the length scale of the experiment. 

From Fig.~\ref{fig:solarflux}, we conclude that the fifth-force bound from the solar axion flux is much weaker than the astrophysics bound. Even in the most optimistic case (coherent limit), the best constraint is $f_a\gtrsim 10^3~{\rm GeV}$.

\section{Comparison with the literature}
\label{sec:compare}
The detection of light DM using background effects has attracted a great deal of attention in recent years. In this regard, it is useful to compare our results with those in the literature.

The background-induced force mediated by quadratically coupled scalars was first calculated in \cite{Ferrer:1998rw,Ferrer:2000hm}, which assumed a thermal distribution of background scalars. Recently, the formalism  was generalized to an arbitrary non-thermal background in \cite{VanTilburg:2024xib,Barbosa:2024zfz} and was applied to the detection of quadratically coupled light mediators by assuming they form a non-relativistic DM background. The background-induced axion force $\vbkg$ calculated in this work includes both linear and quadratic couplings, as well as the relativistic correction from axion backgrounds. In particular, we find that the result of $\vbkg$ induced by a pure quadratic coupling in the non-relativistic limit agrees with the scalar case calculated in \cite{VanTilburg:2024xib,Barbosa:2024zfz}. 

The decoherence effect on the background-induced force due to phase-space and finite-size spread was first noticed in \cite{Ghosh:2022nzo} for fermionic mediators. Later, \cite{VanTilburg:2024xib} studied this effect in great detail for bosonic mediators
with an asymmetric decoherence factor ${\cal D}_{\rm asy}=\cos\left(\absk r-\veck\cdot\vecr\right)$ that comes from the use of retarded propagator. In this work, we derived a symmetric decoherence factor ${\cal D}=[\cos\left(\absk r-\veck\cdot\vecr\right)+\cos\left(\absk r+\veck\cdot\vecr\right)]/2$ by explicitly computing the scattering amplitudes mediated by two axions using the Feynman propagator. Our result agrees with that of \cite{VanTilburg:2024xib} in both the isotropic limit and the coherent limit; in the decoherent region with anisotropic backgrounds, our phase-space form factor shares a similar quadratic suppression ${\cal F}_{\rm PS}\propto\db^2/r^2$ as that in \cite{VanTilburg:2024xib} but differs by an ${\cal O}(1)$ numerical factor. Fundamentally, the discrepancy appears because of the use of different propagators. As a result, our potential has a parity symmetry: $\vbkg(\vecr)=\vbkg(-\vecr)$, and corresponds to a non-dissipative two-body system that does not  exchange momentum with the background --- 
particle 1 absorbs momentum $\veck$ from the background, mediating the interaction to particle 2 via an off-shell mediator, and then particle 2 returns the \emph{same} momentum $\veck'=\veck$ to the background. Note that this momentum conservation condition $\veck'=\veck$ is automatically satisfied in the quantum description, as enforced by Eq.~(\ref{eq:ensemble}).
On the other hand, the potential obtained in \cite{VanTilburg:2024xib} is not invariant under parity, which corresponds to the case where the two-body system keeps absorbing/emitting momentum from/to the background.
In addition to the phase-space suppression, we also explicitly calculated the decoherence effect from the finite-size spread of the object. We find that the finite-size effect does not change the scaling behavior of axion forces in the general case. This qualitative behavior agrees with the result in \cite{VanTilburg:2024xib} obtained from the asymmetric decoherence factor.

All the formalisms adopted in \cite{Ferrer:1998rw,Ferrer:2000hm,VanTilburg:2024xib,Barbosa:2024zfz} as well as in this work used quantum field theories to compute the relevant scattering amplitudes and then mapped them onto the non-relativistic potential. However, the background effect can also be captured by performing a pure quantum-mechanical-like calculation and solving the scalar equation of motion with fixed boundary conditions~\cite{Hees:2018fpg,Banerjee:2022sqg,Bauer:2024hfv,Gan:2025nlu}. Our result of $\vbkg$ induced by a pure quadratic coupling agrees with their results in the coherent region.

Finally, we comment on the possible influence of the matter effect on the axion force.
When deriving the experimental bounds in Sec.~\ref{subsec:axion-DM}, we have fixed the energy density of the axion background to be the  DM density in the galaxy, $\rho_a = \rho_{\rm DM}\approx 0.4~{\rm GeV/cm^3}$, that is, we assumed the axion field value on the surface of the Earth the same as the average value in the galaxy. However, this might not be the case because the matter effect of Earth may change the axion field value around the Earth as well as its spatial gradient through the axion coupling to ordinary matter, as recently pointed out in \cite{Banerjee:2025dlo,delCastillo:2025rbr} (see also \cite{Bauer:2024yow,Bauer:2024hfv}). The effective axion density around the Earth is then shifted: $\rho_a \to \rho_a^{\rm eff}$, so is its gradient. Note that the physical observables induced by the axion force depend on both the axion field value and its spatial gradient. As a result, the bounds derived in Sec.~\ref{subsec:axion-DM} in the strongly coupled (i.e., small $f_a$) region may be modified by the Earth matter effect in a nontrivial way. We leave the detailed analysis for future studies.

\section{Conclusions}
\label{sec:conclusions}

In this work, we studied the spin-independent force mediated by two axions in the presence of an axion background, which is the leading long-range coherent effect caused by axions. 

We started from the effective axion interactions including both linear and quadratic couplings to the SM fermions and derived a general expression for the background-induced axion force $\vbkg$, which is applicable to arbitrary axion backgrounds. We find that the magnitude of this force is controlled by the degree of shift symmetry breaking. There are three relevant effects to break the shift symmetry. First, the axion mass behaves as an order parameter that breaks the shift symmetry and leads to a nonzero $\vbkg$, but this effect is suppressed for light axions. 
Going beyond the mass term, the shift symmetry is explicitly broken by the axion quadratic coupling to SM fermions, whose magnitude can be much larger than the axion mass and results in a significant enhancement to $\vbkg$. In the absence of quadratic coupling, the shift symmetry is also effectively broken by the axion background, which is essentially the relativistic correction to the NR amplitude; in an energetic axion background, this effect dominates over the axion mass contribution and enhances $\vbkg$.

In the presence of shift symmetry breaking,  $\vbkg$ has a generic $1/r$ scaling behavior as long as $r \ll \db$, the de Broglie wavelength of the background axions. Moreover, $\vbkg$ is proportional to the number density of background axions, indicating that particles within the volume of $\db^3$ coherently contribute to $\vbkg$. This is a significant effect if axions are cold DM that has a huge occupation number. At distances $r\gtrsim \db$, the background axions generically have different phases and thus contribute destructively to $\vbkg$, making it more suppressed than $1/r$. We calculated this effect with a symmetric decoherence factor by including the finite spread from both the phase space and the configuration space. We find that for most realistic axion backgrounds, the suppression is some power of $\db/r$, as opposed to an exponential suppression $e^{-2m_a r}$ to the vacuum two-axion force $V_{2a}$. This reflects the intrinsic difference between the quantum force and the background-induced force: $V_{2a}$ is a pure quantum effect, where both propagators are off-shell; this effect is exponentially suppressed when the distance exceeds the inverse mass of the progagator. On the other hand, $\vbkg$ is essentially a classical effect since one of the propagators is on-shell; the scattering between external particles and background particles extends the range of this effect, allowing it to go much further than the inverse mass of the propagator. As a result, $\vbkg$ is much more significant on macroscopic length scales compared to its vacuum counterpart $V_{2a}$.

The background-induced axion force has rich phenomena given that axion backgrounds exist in a large class of well-motivated scenarios. We calculated its effect in fifth-force detection experiments by separately assuming the existence of axion DM background and solar axion flux. We find that the effect induced by the DM background of ultralight axions can impose strong constraints on the axion coupling, comparable to existing astrophysical bounds.
The axion force can also cause an energy shift for the atomic states and can be probed using atomic spectroscopy, which is a promising method with sensitivities that will be significantly improved in the near future. 
Another interesting application that is not discussed in this work is to investigate the effect of this long-range force at galactic scales in the early universe --- this effect is expected to be significant because the axion DM number density is much higher in the early universe than at present. 

In conclusion, our results demonstrate that the background effect could be very important for detecting axions and axion DM, and would also be helpful in searching for other exotic light particles beyond the SM.

\section*{Acknowledgement}
We would like to thank Abhishek Banerjee, Sergio Barbosa, Kai Bartnick, Brando Bellazzini, Itay Bloch, Jeff Dror, Francesc Ferrer, Sylvain Fichet, Xucheng Gan, Anson Hook, Da Liu, Di Liu, Xuheng Luo, Maxim Perelstein, Ken Van Tilburg, and Kevin Zhou for helpful discussions. We thank Itay Bloch and Maxim Perelstein for insightful comments on the draft. We thank Sergio Barbosa, Brando Bellazzini, Sylvain Fichet, Anson Hook, Ken Van Tilburg, and Kevin Zhou for discussions on the form of the decoherence factor.
We thank Xucheng Gan and Xuheng Luo for discussions on the finite-size decoherence effect. We thank Gilad Perez and Konstantin Springmann for discussions on the Earth matter effect.
This work is supported in part by the NSF grant PHY-2309456.

\begin{appendix}
\section{Axion forces at the tree level}
\label{app:vacuum}
Although we focus on the two-axion forces throughout this work, we would like to briefly review the one-axion effect in this appendix, which, as we show below, is always spin-dependent due to the pseudoscalar nature of the coupling.

\begin{figure}[t]
\centering
\includegraphics[scale=1.0]{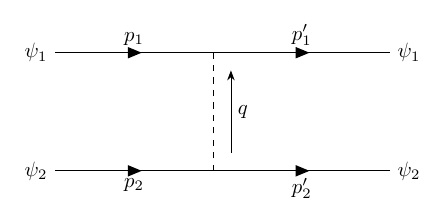}
\caption{\label{fig:tree}The Feynman diagram of one-axion exchange between two fermions.}
\end{figure}

The linear coupling between the axion and fermions in Eq.~(\ref{eq:Leff}) leads to tree-level $t$-channel scattering $\psi_1 (p_1)+\psi_2(p_2) \to \psi_1(p_1') + \psi_2(p_2')$ by exchanging one axion (see Fig.~\ref{fig:tree}). The amplitude reads
\begin{align}
    i {\cal M}_{\rm tree} = \frac{c_{1} c_{2}}{4 f_a^2} \bar{u}(p_1')\slashed{q}\gamma_5 u(p_1)\bar{u}(p_2')\slashed{q}\gamma_5 u(p_2)\frac{i}{q^2-m_a^2}\;,
\end{align}
where $c_i \equiv c_{\psi_i}$ (for $i=1,2$) and $u(p)$ denotes the wavefunction of external fermions with momentum $p$. Using the equation of motion of external fermions, it is reduced to
\begin{align}
    {\cal M}_{\rm tree} = -\frac{c_{1}c_{2}}{4 f_a^2}\frac{(2m_1)(2m_2)}{q^2 - m_a^2} \left[\bar{u}(p_1')\gamma_5 u(p_1)\right]\left[\bar{u}(p_2')\gamma_5 u(p_2) \right]. 
\end{align}
Under the NR approximation $q^\mu \approx (0,\vecq)$, we have
\begin{align}
   \bar{u}(p_1')\gamma_5 u(p_1) = \vecsigma_1 \cdot \vecq + {\cal O}(\vecq^2)\;,\qquad
   \bar{u}(p_2')\gamma_5 u(p_2) = -\vecsigma_2 \cdot \vecq + {\cal O}(\vecq^2)\;,\label{eq:pseudo-contraction}
\end{align}
where $\vecsigma_i \equiv \xi_i^\dagger \vecsigma \xi_i $ (for $i=1,2$) has been defined, $\vecsigma = (\sigma_1,\sigma_2,\sigma_3)$ is the vector of Pauli matrices, and $\xi_i$ denotes the two-component spinor of $\psi_i$ that satisfies the Dirac equation. This leads to the NR amplitude:
\begin{align}
  \frac{{\cal M}_\text{tree,NR}}{(2m_1)(2m_2)} = \frac{c_{1}c_{2}}{4 f_a^2}\frac{1}{\vecq^2+m_a^2}\left(\vecsigma_1 \cdot \vecq \right)\left(\vecsigma_2 \cdot \vecq \right).
\end{align}
Substituting the amplitude into Eq.~(\ref{eq:fourier}), we obtain (for $r\neq 0$):
\begin{align}
\label{eq:one-axion}
V_{a}(\vecr)
&=\frac{c_{1}c_{2}}{4 f_a^2}\left(\vecsigma_1\cdot \nabla\right)\left(\vecsigma_2\cdot \nabla\right) \frac{e^{-m_a r}}{4\pi r}\nonumber\\
 & = \frac{c_{1}c_{2}}{4 f_a^2}\frac{e^{-m_a r}}{4\pi r^3}\left[\left(\vecsigma_1\cdot\hat{\vecr}\right)\left(\vecsigma_2\cdot\hat{\vecr}\right)\left(3+3m_a r + m_a^2 r^2\right)-\left(\vecsigma_1\cdot\vecsigma_2\right)\left(1+m_a r\right)\right],
\end{align}
where $\hat{\vecr}\equiv \vecr/r$ is the unit vector. In the axion massless limit, it is reduced to
\begin{align}
V_{a}(\vecr) =   \frac{c_{1}c_{2}}{4 f_a^2}\frac{1}{4\pi r^3}  \left[3\left(\vecsigma_1\cdot\hat{\vecr}\right)\left(\vecsigma_2\cdot\hat{\vecr}\right)-\left(\vecsigma_1\cdot\vecsigma_2\right)\right],\quad m_a \to 0\;.\label{eq:Vamassless}
\end{align}
The result of the one-axion force in Eq.~(\ref{eq:one-axion}) is well known in the literature~\cite{Moody:1984ba,Daido:2017hsl}.\footnote{We notice that our result in Eq.~(\ref{eq:one-axion}) agrees with that in Ref.~\cite{Daido:2017hsl} but differs from that in Ref.~\cite{Moody:1984ba} by an overall minus sign.} In history, Eq.~(\ref{eq:one-axion}) was first used to describe the interaction between two nucleons mediated by the exchange of a neutral pseudoscalar meson~\cite{Brueckner:1953zzb,sakurai1967advanced}.

Therefore, the one-axion force depends on the spins of the external fermions, which is the result of the pseudoscalar-type contraction in Eq.~(\ref{eq:pseudo-contraction}). Furthermore, the one-axion force (\ref{eq:one-axion}) vanishes when performing the spin average over the unpolarized macroscopic objects. This is because
\begin{align}
\left(\vecsigma_1 \cdot \nabla\right)\left(\vecsigma_2 \cdot \nabla\right) &= (\xi_1^{*})_\alpha\, (\xi_1)_\beta \,(\sigma^i)_{\alpha \beta} \,  (\xi_2^{*})_x\,(\xi_2)_y\,(\sigma^j)_{xy}\,\partial_i \partial_j =
\langle \alpha, x | \sigma^i \otimes \sigma^j | \beta,y\rangle \partial_i \partial_j\;,
\end{align}
where $\alpha,\beta$ and $x,y$ are indices in the two-dimensional spinor space. After taking the spin average, it becomes
\begin{align}
\langle \alpha, x | \sigma^i \otimes \sigma^j | \beta,y\rangle \partial_i \partial_j  \overset{\text{spin average}}{=} {\rm Tr}\left(\sigma^i \otimes \sigma^j\right) \partial_i \partial_j = 0\;,
\end{align}
where in the last step we have used the fact that the trace of the tensor product of any two Pauli matrices vanishes.

As a result, the axion \emph{cannot} mediate long-range forces among unpolarized macroscopic objects at the tree level, which is precisely the reason why the usual fifth-force and equivalence-principle test experiments cannot apply constraints on the axion linear coupling via Fig.~\ref{fig:tree}. In order to get a spin-independent axion force, one needs to include at least two axions as the mediator.

\section{Pseudoscalar basis versus derivative basis}
\label{app:mass-vs-derivative}
In this appendix, we briefly review axion effective interactions with fermions. In particular, we pay attention to the structure of couplings in two different axion bases. 

The axion appears as the angular mode of a complex scalar in the UV models. After the spontaneous $U(1)$ symmetry breaking, its effective coupling to some fermion $\psi$ can be written as:
\begin{align}
    \label{eq:Lag}
    {\cal L}  = -m_\psi \bar{\psi} e^{i c_\psi \gamma_5 a/f_a} \psi  = - m_\psi \bar{{\psi}}_{\rm L} \psi_{\rm R} e^{i c_\psi a/f_a} + {\rm h.c.}\;,
\end{align}
where $\psi_{\rm L,R}\equiv \frac{1}{2}(1\mp\gamma_5)\psi$, $m_\psi$ is the mass of $\psi$, and $c_\psi$ is a dimensionless parameter depending on $\psi$.
The above coupling has a shift symmetry:
\begin{align}
a \to a + c_\psi f_a \theta\;,\quad
\psi_{\rm L} \to  e^{i c_\psi \theta/2} \psi_{\rm L}\;,\quad
\psi_{\rm R} \to  e^{-i c_\psi \theta/2} \psi_{\rm R}\;,
\end{align}
with $\theta$ an arbitrary constant. The axion-fermion interactions are obtained by series expanding the exponent in Eq.~(\ref{eq:Lag}):
\begin{align}
\label{eq:Lpseudo}
{\cal L}_{\rm int}^{\rm p} = - i c_\psi m_\psi \frac{a}{f_a}  \bar{\psi} \gamma_5 \psi  + \frac{1}{2} c_\psi^2 m_\psi \frac{a^2}{f_a^2} \bar{\psi}\psi + {\cal O}\left(\frac{a^3}{f_a^3}\right) \quad (\text{pseudoscalar basis})\;.
\end{align}
This is the structure of couplings in the \emph{pseudoscalar basis}, where the name comes from the leading pseudoscalar coupling $a\bar{\psi}\gamma_5\psi$ in Eq.~(\ref{eq:Lpseudo}). 

Alternatively, one can change to the \emph{derivative basis} by performing a chiral rotation in Eq.~(\ref{eq:Lag}) to absorb the axion dependence:
\begin{align}
    \psi_{\rm L} \to e^{i c_\psi a/(2 f_a)}\psi_{\rm L}\;,\quad
    \psi_{\rm R} \to e^{-i c_\psi a/(2 f_a)}\psi_{\rm R}\;,
\end{align}
which, after rotation, makes (\ref{eq:Lag}) a pure mass term $m_\psi \bar{\psi}\psi$. The axion-fermion interaction appears from the kinetic term of $\psi$ in the new basis:
\begin{align}
\label{eq:Lderi}
{\cal L}_{\rm int}^{\rm d} = \frac{\partial_\mu a}{2f_a} c_\psi  \bar{\psi}\gamma^\mu \gamma_5 \psi \quad (\text{derivative basis})\;,
\end{align}
where the axion shift invariance $a \to a + {\rm constant}$ is obvious.

The amplitude predicted by Eqs.~(\ref{eq:Lpseudo}) and (\ref{eq:Lderi}) should agree for any given process,\footnote{There are an infinite number of interacting terms in Eq.~(\ref{eq:Lpseudo}). To compare with the observable predicted by Eq.~(\ref{eq:Lderi}), one should expand the interaction in (\ref{eq:Lpseudo}) up to ${\cal O}((a/f_a)^n)$ for a process involving $n$ axions.} meaning that the physical observable should not be dependent on the basis that is selected. For the two-axion force that is the subject of this work, one would expect that the interaction in (\ref{eq:Lpseudo}), which has both linear and quadratic couplings, would provide the same result as (\ref{eq:Lderi}), which only has linear coupling. In the following section, we shall demonstrate that this is indeed the case.

\section{Compton scattering of axions}
\label{app:compton}
\begin{figure}[t]
    \centering
    \subfigure[$t$-channel]{
    \includegraphics[scale=0.7]
    {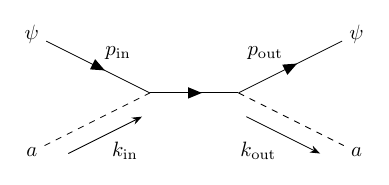}
    \label{subfig:t-channel}
    }
    \hspace{0.5cm}
    \subfigure[$u$-channel]{
    \includegraphics[scale=0.7]
    {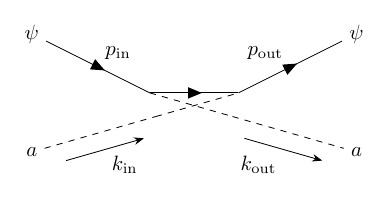}
    \label{subfig:u-channel}
    }
    \hspace{0.5cm}
    \subfigure[contact]{
    \includegraphics[scale=0.75]
    {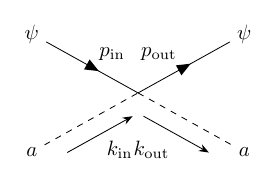}
    \label{subfig:contact}
    }
    \caption{Feynman diagrams for the Compton scattering $\psi(p_{\rm in})+a(k_{\rm in})\to \psi(p_{\rm out})+a(k_{\rm out})$  between axions and fermions. Note that under the pseudoscalar basis, all three diagrams are relevant, while under the derivative basis there are only $t$-channel and $u$-channel diagrams.}
    \label{fig:compton}
\end{figure}

The amplitude relevant to the two-axion force can be calculated by the multiplication of two tree-level Compton scatterings [see Eq.~(\ref{eq:amplitude-tot})]
\begin{align}
\psi \left(p_{\rm in}\right) + a \left(k_{\rm in}\right) \to  \psi \left(p_{\rm out}\right) + a \left(k_{\rm out}\right)\;, 
\end{align}
as shown in Fig.~\ref{fig:compton}. In this section, we calculate the amplitude explicitly in both the pseudoscalar basis and the derivative basis and show that the results agree with each other.

In the pseudoscalar basis (\ref{eq:Lpseudo}), there are both linear and quadratic couplings, so all three diagrams ($t$-channel, $u$-channel, and contact) in Fig.~\ref{fig:compton} contribute. Their amplitudes read:
\begin{align}
{\cal M}_t^{\rm p} &=  \frac{c_{\psi}^{2}}{f_a^{2}}m_{\psi}^{2}\bar{u}\left(p_{\text{out}}\right)\frac{\slashed{k}_{\rm in}}{k_{\rm in}^{2}+2k_{\rm in}\cdot p_{\rm in}}u\left(p_{\text{in}}\right), \label{eq:Mtp} \\
{\cal M}_u^{\rm p} &=  \frac{c_{\psi}^{2}}{f_a^{2}}m_{\psi}^{2}\bar{u}\left(p_{\text{out}}\right)\frac{-\slashed{k}_{\rm out}}{k_{\rm out}^{2}-2k_{\rm out}\cdot p_{\rm in}}u\left(p_{\text{in}}\right),\label{eq:Mup}\\
{\cal M}_c^{\rm p} &=   -\frac{c_{\psi}^{2}}{f_a^{2}}m_{\psi}\bar{u}\left(p_{\text{out}}\right)u\left(p_{\text{in}}\right).\label{eq:Mcp}
\end{align}
So the total amplitude in the pseudoscalar basis is given by
\begin{align}
    \mathcal{M}_C^{\rm p}&=\mathcal{M}_t^{\rm p}+\mathcal{M}_u^{\rm p}+\mathcal{M}_c^{\rm p}\nonumber\\
    &=\frac{c_{\psi}^{2}}{f_a^{2}}m_{\psi}^{2}\bar{u}\left(p_{\text{out}}\right)\left(\frac{\slashed{k}_{\rm in}}{k^{2}_{\rm in}+2k_{\rm in}\cdot p_{\text{in}}}-\frac{\slashed{k}_{\rm out}}{k^{2}_{\rm out}-2k_{\rm out}\cdot p_{\text{in}}}-\frac{1}{m_{\psi}}\right)u\left(p_{\text{in}}\right),
\label{eq:amplitude-compton-ps}
\end{align}

On the other hand, there is only linear coupling in the derivative basis (\ref{eq:Lderi}), so only the $t$-channel and $u$-channel diagrams in Fig.~\ref{fig:compton} are relevant. Their amplitudes are found to be:
\begin{align}
    \mathcal{M}_{t}^{\rm d}\left(p_{\rm in},p_{\rm out};k_{\rm in},k_{\rm out}\right)&=\frac{c_{\psi}^{2}}{f^{2}_a}m_{\psi}^{2}\bar{u}\left(p_{\text{out}}\right)\left(\frac{\slashed{k}_{\rm in}}{k^{2}_{\rm in}+2k_{\rm in}\cdot p_{\text{in}}}-\frac{1}{2m_{\psi}}+\frac{\slashed{k}_{\rm out}}{4m_{\psi}^{2}}\right)u\left(p_{\text{in}}\right),\label{eq:Mtd}\\
    \mathcal{M}_{u}^{\rm d}\left(p_{\rm in},p_{\rm out};k_{\rm in},k_{\rm out}\right)&=\frac{c_{\psi}^{2}}{f^{2}_a}m_{\psi}^{2}\bar{u}\left(p_{\text{out}}\right)\left(\frac{-\slashed{k}_{\rm out}}{k^{2}_{\rm out}-2k_{\rm out}\cdot p_{\text{in}}}-\frac{1}{2m_{\psi}}+\frac{-\slashed{k}_{\rm in}}{4m_{\psi}^{2}}\right)u\left(p_{\text{in}}\right).\label{eq:Mud}
\end{align}
It is obvious that the $t$-channel and $u$-channel amplitudes satisfy the crossing symmetry
\begin{align}
{\cal M}_t^{\rm d}\left(p_{\rm in},p_{\rm out};k_{\rm in},k_{\rm out}\right)={\cal M}_u^{\rm d}\left(p_{\rm in},p_{\rm out};-k_{\rm out},-k_{\rm in}\right).\label{eq:crossing}    
\end{align}
Adding them together, one obtains the total amplitude in the derivative basis:
\begin{align}
    \mathcal{M}_C^{\rm d}&=\mathcal{M}_t^{\rm d}+\mathcal{M}_u^{\rm d}\nonumber\\
    &=\frac{c_{\psi}^{2}}{f^{2}_a}m_{\psi}^{2}\bar{u}\left(p_{\text{out}}\right)\left(\frac{\slashed{k}_{\rm in}}{k_{\rm in}^{2}+2k_{\rm in}\cdot p_{\text{in}}}-\frac{\slashed{k}_{\rm out}}{k^{2}_{\rm out}-2k_{\rm out}\cdot p_{\text{in}}}-\frac{1}{m_{\psi}}\right)u\left(p_{\text{in}}\right),
\label{eq:amplitude-compton-deri}
\end{align}
which agrees with the result in the mass basis (\ref{eq:amplitude-compton-ps}). Note that in Eq.~(\ref{eq:amplitude-compton-deri}) we have used the momentum conservation $k_{\rm out}-k_{\rm in}=p_{\rm in}-p_{\rm out}$ and the equation of motion of $\psi$.

As a result, we have confirmed that, as anticipated, the amplitude of the axion Compton scattering computed in the derivative basis and the pseudoscalar basis agree up to ${\cal O}(1/f_a^2)$. Furthermore, since the axion propagators in Eq.~(\ref{eq:amplitude-tot}) clearly do not depend on the axion basis, we conclude that the two-axion force and its background correction are also independent of the basis.

\section{Derivation of $V_{\rm bkg}$ from coherent scattering}
\label{app:coherent-scattering}

In this appendix, we provide an equivalent approach to deriving the background-induced axion force without using the formalism of modified propagator introduced in Sec.~\ref{sec:formalism}. The new approach essentially relies on the coherent scattering between background axions and external fermions. A similar derivation with only quadratic couplings (corresponding to the bubble diagram) can be found in~\cite{VanTilburg:2024xib}.

More specifically, our goal is to derive Eq.~(\ref{eq:amplitude-bkg-simplified}) without using the modified propagator in Eq.~(\ref{eq:mod-propagator}). We perform the calculations in the derivative basis, where only the two box diagrams in Fig.~\ref{fig:Feyn} are relevant. When in the medium of axions, the background effect contributes in such a way that the background axions are first scattered with $\psi_1$ (or $\psi_2$), then the interaction is mediated to $\psi_2$ (or $\psi_1$) via an off-shell axion, and finally scattered back to the background, as shown in Fig.~\ref{fig:background boxes}.

The background effect is determined by the ensemble average of the axion fields. We expand the axion field in the standard way:
\begin{equation}    a\left(t,\vecx\right)=\int\frac{\text{d}^3\veck}{(2\pi)^3}\frac{1}{\sqrt{2E_{\veck}}}\left(a_{\veck}^{}e^{-ik\cdot x}+a_{\veck}^\dagger e^{ik\cdot x}\right),
\end{equation}
where $E_{\veck}\equiv(\veck^2+m_a^2)^{1/2}$, and $a_{\veck}^{}$ and $a_{\veck}^\dagger$ are the annihilation and creation operators, respectively. Obviously $\langle a(t,\vecx)\rangle = 0$, where $\langle \cdots \rangle$ denotes the ensemble average. However, the mean square (or equivalently, the variance) of the axion field is nonzero 
\begin{align}
\left\langle a^2\left(t,\vecx\right)\right\rangle = \int \frac{\text{d}^3\veck}{(2\pi)^3}\frac{1}{2E_{\veck}} f(\veck)\;,\label{eq:variance}
\end{align}
where $f(\veck)$ is normalized such that $\int\text{d}^3\veck f(\veck)/(2\pi)^3$ is equal to the number density of background axions. To get Eq.~(\ref{eq:variance}), one needs to use the ensemble average of the number operator:
 \begin{align}
     \left\langle a_{\veck}^\dagger a_{\veck^{\prime}}^{}\right\rangle=(2\pi)^3f(\veck)\delta^{(3)}\left(\veck-\veck^{\prime}\right).\label{eq:ensemble}
 \end{align}

As a result, the one-axion effect cannot be affected by the axion background (since $\langle a(t,\vecx) \rangle = 0$),  while the two-axion effect can. For example, the $t$-channel Compton scattering diagram in Fig.~\ref{subfig:t-channel}, when put into the background, becomes
\begin{align}
\left\langle  {\cal M}_t \right \rangle &=  -\left\langle a^2 \right\rangle\times \frac{c_\psi^2}{4 f_a^2}\bar{u}\left(p_{\rm out}\right) \slashed{k}_{\rm out}\gamma_5 \frac{1}{\slashed{p}_{\rm in}+\slashed{k}_{\rm in}-m_\psi}\slashed{k}_{\rm in}  \gamma_5 u\left(p_{\rm in}\right)\nonumber\\
& = \int \frac{\text{d}^3\veck}{(2\pi)^3}\frac{1}{2E_{\veck}} f(\veck)\,{\cal M}_t\;,
\end{align}
where in the second line we have used Eq.~(\ref{eq:variance}) and ${\cal M}_t$ is given by Eq.~(\ref{eq:Mtd}) with the identification $k_{\rm in}=k$ and $k_{\rm out} = k+q$.

Then we proceed to calculate the ensemble average of the box diagrams in Fig.~\ref{fig:background boxes}, each of which can be written as the multiplication of two Compton scatterings. The amplitudes of the first line in Fig.~\ref{fig:background boxes} read (from left to right): 
\begin{align}
\langle\mathcal{M}_{tt1}\rangle&=\int\frac{\text{d}^{3}\veck}{\left(2\pi\right)^{3}}\frac{f(\veck)}{2E_{\veck}}\frac{-1}{(k+q)^2-m_a^2} \mathcal{M}_{t}\left(1\right)\mathcal{M}_{t}\left(2\right),\\
\langle\mathcal{M}_{tt2}\rangle&=\int\frac{\text{d}^{3}\veck}{\left(2\pi\right)^{3}}\frac{f(\veck+\vecq)}{2E_{\veck+\vecq}}\frac{-1}{k^2-m_a^2} \mathcal{M}_{t}\left(1\right)\mathcal{M}_{t}\left(2\right),\\
\langle\mathcal{M}_{uu1}\rangle&=\int\frac{\text{d}^{3}\veck}{\left(2\pi\right)^{3}}\frac{f(\veck)}{2E_{\veck}}\frac{-1}{(k+q)^2-m_a^2} \mathcal{M}_{u}\left(1\right)\mathcal{M}_{u}\left(2\right),\\
\langle\mathcal{M}_{uu2}\rangle&=\int\frac{\text{d}^{3}\veck}{\left(2\pi\right)^{3}}\frac{f(\veck+\vecq)}{2E_{\veck+\vecq}}\frac{-1}{k^2-m_a^2} \mathcal{M}_{u}\left(1\right)\mathcal{M}_{u}\left(2\right),
\end{align}
where Eq.~(\ref{eq:crossing}) is used to transform between ${\cal M}_t$ and ${\cal M}_u$, and we have also used the following notations for brevity:
\begin{align}
{\cal M}_{j}\left(1\right)\equiv {\cal M}_{j} \left(p_1,p_1',-k,-(k+q)\right),\qquad {\cal M}_{j}\left(2\right)\equiv {\cal M}_{j}\left(p_2,p_2',k,k+q\right).
\end{align}

Therefore, the ensemble average of the amplitude of the ``Box 1'' diagram in Fig.~\ref{fig:Feyn} is given by
\begin{align}
\langle\mathcal{M}_\text{box,1}\rangle &=\langle\mathcal{M}_{tt1}\rangle + \langle\mathcal{M}_{tt2}\rangle + \langle\mathcal{M}_{uu1}\rangle + \langle\mathcal{M}_{uu2}\rangle \nonumber\\
& =-\int\frac{\text{d}^{3}\veck}{\left(2\pi\right)^{3}}\left[\frac{f(\veck)}{2E_{\veck}}\frac{1}{\left(k+q\right)^2-m_a^2}+\frac{f(\veck+\vecq)}{2E_{\veck+\vecq}}\frac{1}{k^2-m_a^2} \right]\left[ \mathcal{M}_{t}\left(1\right)\mathcal{M}_{t}\left(2\right)+(t\leftrightarrow u)\right].
\end{align}

Similarly, the amplitudes in the second line of Fig.~\ref{fig:background boxes} are given by (from left to right):
\begin{align}
\langle\mathcal{M}_{tu1}\rangle&=\int\frac{\text{d}^{3}\veck}{\left(2\pi\right)^{3}}\frac{f(\veck)}{2E_{\veck}}\frac{-1}{(k+q)^2-m_a^2}\mathcal{M}_{t}\left(1\right)\mathcal{M}_{u}\left(2\right),\\
\langle\mathcal{M}_{tu2}\rangle&=\int\frac{\text{d}^{3}\veck}{\left(2\pi\right)^{3}}\frac{f(\veck+\vecq)}{2E_{\veck+\vecq}}\frac{-1}{k^2-m_a^2}\mathcal{M}_{t}\left(1\right)\mathcal{M}_{u}\left(2\right),\\
\langle\mathcal{M}_{ut1}\rangle&=\int\frac{\text{d}^{3}\veck}{\left(2\pi\right)^{3}}\frac{f(\veck)}{2E_{\veck}}\frac{-1}{(k+q)^2-m_a^2}\mathcal{M}_{u}\left(1\right)\mathcal{M}_{t}\left(2\right),\\
\langle\mathcal{M}_{ut2}\rangle&=\int\frac{\text{d}^{3}\veck}{\left(2\pi\right)^{3}}\frac{f(\veck+\vecq)}{2E_{\veck+\vecq}}\frac{-1}{k^2-m_a^2}\mathcal{M}_{u}\left(1\right)\mathcal{M}_{t}\left(2\right),
\end{align}
leading to
\begin{align}
\langle\mathcal{M}_\text{box,2}\rangle &=\langle\mathcal{M}_{tu1}\rangle + \langle\mathcal{M}_{tu2}\rangle + \langle\mathcal{M}_{ut1}\rangle + \langle\mathcal{M}_{ut2}\rangle \nonumber\\
& =-\int\frac{\text{d}^{3}\veck}{\left(2\pi\right)^{3}}\left[\frac{f(\veck)}{2E_{\veck}}\frac{1}{\left(k+q\right)^2-m_a^2}+\frac{f(\veck+\vecq)}{2E_{\veck+\vecq}}\frac{1}{k^2-m_a^2} \right]\left[ \mathcal{M}_{t}\left(1\right)\mathcal{M}_{u}\left(2\right)+(t\leftrightarrow u)\right].
\end{align}

The total amplitude is then given by
\begin{align}
{\cal M}_{\rm bkg}   &=  \langle\mathcal{M}_\text{box,1}\rangle + \langle\mathcal{M}_\text{box,2}\rangle \nonumber\\
&=-\int\frac{\text{d}^{3}\veck}{\left(2\pi\right)^{3}}\left[\frac{f(\veck)}{2E_{\veck}}\frac{1}{\left(k+q\right)^2-m_a^2}+\frac{f(\veck+\vecq)}{2E_{\veck+\vecq}}\frac{1}{k^2-m_a^2} \right]{\cal M}_C \left(1\right)  {\cal M}_C \left(2\right),\label{eq:Mbkg-app}
\end{align}
where ${\cal M}_C = {\cal M}_t + {\cal M}_u$ from Eq.~(\ref{eq:compton}) is used. Finally, by performing the shift $k\to k-q$
for the second term in the bracket of Eq.~(\ref{eq:Mbkg-app}), we recover the result in Eq.~(\ref{eq:amplitude-bkg-simplified}) as promised. Note that $k^0_{}=E^{}_{\veck}$ is automatically satisfied because background axions are on-shell.

\section{Relevant integrals}
\label{app:integral}
In this appendix, we provide details to calculate the integrals relevant to the background-induced axion force. Using Eq.~(\ref{eq:schematical}), the amplitude in Eq.~(\ref{eq:amplitude-bkg-simplified}) can be recast into a compact form: 
\begin{align}
\mathcal{M}_{\text{bkg}}(q)=-\frac{c_{1}^{2}c_{2}^{2}}{f_{a}^{4}}m_{1}^{2}m_{2}^{2}\int\frac{\text{d}^3\veck}{(2\pi)^3}\frac{f(\veck)}{2E_{\veck}} \left[\frac{{\cal J}(k)}{\left(k+q\right)^{2}-m_{a}^{2}} +\frac{{\cal J}(k-q)}{\left(k-q\right)^{2}-m_{a}^{2}}\right]\Bigg|_{k^0 = E_{\veck}},    
\end{align}
where
\begin{align}
{\cal J}(k)&=   \bar{u}\left(p_{1}^{\prime}\right)\gamma_{\mu}u\left(p_{1}\right)\bar{u}\left(p_{2}^{\prime}\right)\gamma_{\nu}u\left(p_{2}\right)\left[J_{\text{box,1}}^{\mu\nu}(k)+J_{\text{box,2}}^{\mu\nu}(k)\right]\nonumber\\
&+\bar{u}\left(p_{1}^{\prime}\right)\gamma_{\mu}u\left(p_{1}\right)\bar{u}\left(p_{2}^{\prime}\right)u\left(p_{2}\right)J_{\text{tri,1}}^{\mu} (k)+\bar{u}\left(p_{1}^{\prime}\right)u\left(p_{1}\right)\bar{u}\left(p_{2}^{\prime}\right)\gamma_{\mu}u\left(p_{2}\right)J_{\text{tri,2}}^{\mu}(k)\nonumber\\
&+\bar{u}\left(p_{1}^{\prime}\right)u\left(p_{1}\right)\bar{u}\left(p_{2}^{\prime}\right)u\left(p_{2}\right)J_{\rm bub}(k)\;, \label{eq:J-factor}
\end{align}
and
\begin{align}
J_{\text{box,1}}^{\mu\nu}(k)&=\frac{-k^{\mu}}{k^{2}-2p_{1}\cdot k}\,\frac{k^{\nu}}{k^{2}+2p_{2}\cdot k}+\left(-k\leftrightarrow k+q\right),\\
J_{\text{box,2}}^{\mu\nu}(k) & =\frac{-k^{\mu}}{k^{2}-2p_{1}\cdot k}\,\frac{-\left(k+q\right)^{\nu}}{\left(k+q\right)^{2}-2p_{2}\cdot\left(k+q\right)}+\left(-k\leftrightarrow k+q\right),\\
J_{\text{tri,1}}^{\mu}(k)&=-\frac{1}{m_{2}}\frac{-k^{\mu}}{k^{2}-2p_{1}\cdot k}+\left(-k\leftrightarrow k+q\right),\\
J_{\text{tri,2}}^{\mu}(k)&=-\frac{1}{m_{1}}\frac{k^{\mu}}{k^{2}+2p_{2}\cdot k}+\left(-k\leftrightarrow k+q\right),\\
J_{\text{bub}}(k) &=\frac{1}{m_{1}m_{2}}\;.\label{eq:Jbub}
\end{align}
Here those $J$-terms correspond to the contributions from different diagrams in Fig.~\ref{fig:Feyn} in the pseudoscalar basis, and each of them corresponds to one term in Eq.~(\ref{eq:Mbkg-split}), respectively. Note that terms in Eq.~(\ref{eq:J-factor}) that are proportional to $\slashed{q}$ will vanish after using the equation of motion $\bar{u}(p_i')\slashed{q}u(p_i)=0$.

The background-induced axion force is given by the Fourier transform of the amplitude in the NR limit $q^\mu \approx (0,\vecq)$ [see Eq.~(\ref{eq:fourier})]:
\begin{align}
V_{\rm bkg}(r) = - \frac{c_1^2 c_2^2}{4 f_a^4}m_1 m_2 \int\frac{\text{d}^3\veck}{(2\pi)^3}\frac{f(\veck)}{2E_{\veck}}\int\frac{\text{d}^3\vecq}{(2\pi)^3}e^{i\vecq\cdot\vecr} \left[\frac{{\cal J}_{\rm NR}(k)}{\vecq^2+2\veck\cdot\vecq} +\frac{{\cal J}_{\rm NR}(k-q)}{\vecq^2-2\veck\cdot\vecq}\right].\label{eq:fourier-app}
\end{align}
Expanding the wavefunctions up to the leading order of velocity of external fermions, the ${\cal J}$-factor in Eq.~(\ref{eq:J-factor}) becomes
\begin{align}
{\cal J}_{\rm NR} = 4m_1 m_2 \left[\sum_{i=1}^2 \left(J_\text{box,$i$}^{00}+J_\text{tri,$i$}^{0}\right)+J_{\rm bub}\right]+{\cal O}\left(v\right),    
\end{align}
where $J_{\rm bub}=1/(m_1 m_2)$ and in the NR limit, by keeping the leading order of momentum transfer, we have
\begin{align}
J_\text{box,1}^{00}(k) &=\frac{2E_{\veck}^2\left(4m_1 m_2 E_{\veck}^2-m_a^4\right)}{\left(4m_1^2 E_{\veck}^2-m_a^4\right)\left(4m_2^2 E_{\veck}^2-m_a^4\right)}+{\cal O}\left(\vecq^2\right), \\
J_\text{box,2}^{00}(k) &=\frac{2E_{\veck}^2\left(4m_1 m_2 E_{\veck}^2+m_a^4\right)}{\left(4m_1^2 E_{\veck}^2-m_a^4\right)\left(4m_2^2 E_{\veck}^2-m_a^4\right)}+{\cal O}\left(\vecq^2\right),  \\    
J_\text{tri,1}^{0}(k) &= -\frac{4m_1 E_{\veck}^2}{m_2\left(4m_1^2 E_{\veck}^2-m_a^4\right)}+{\cal O}\left(\vecq^2\right),\\    
J_\text{tri,2}^{0}(k) &= -\frac{4m_2 E_{\veck}^2}{m_1\left(4m_2^2 E_{\veck}^2-m_a^4\right)}+{\cal O}\left(\vecq^2\right),
\end{align}
leading to
\begin{align}
    \sum_{i=1}^2 \left(J_\text{box,$i$}^{00}+J_\text{tri,$i$}^{0}\right) &= \frac{-16 E_{\veck}^4 m_1 m_2+ 4 E_{\veck}^2 m_a^4 \left(m_1/m_2 + m_2/m_1\right)}{\left(4 m_1^2 E_{\veck}^2-m_a^4\right)\left(4m_2^2 E_{\veck}^2 - m_a^4\right)} + {\cal O}\left(\vecq^2\right)\nonumber\\
    &= -\frac{1}{m_1 m_2} + \frac{m_a^8}{16 m_1^3 m_2^3 E_{\veck}^4} + {\cal O}\left(m_a^{12}\right),\label{eq:Jsum}
\end{align}
where in the second line we have assumed $m_a\ll m_i$ and expanded the result as a series of $m_a^2/m_i^2$.
Note that the leading term $-1/(m_1 m_2)$ in Eq.~(\ref{eq:Jsum}) is exactly cancelled when added to $J_{\rm bub}$ in Eq.~(\ref{eq:Jbub}), making the remaining terms all suppressed by the axion mass.

Hence, in the NR limit, we arrive at
\begin{align}
{\cal J}_{\rm NR}(k) = {\cal J}_{\rm NR}(k-q) =  \frac{4m_a^8}{\left(4 m_1^2 E_{\veck}^2-m_a^4\right)\left(4m_2^2 E_{\veck}^2 - m_a^4\right)}+ {\cal O}\left(\vecq^2\right).
\label{eq:JNR}
\end{align}
Note that Eq.~(\ref{eq:JNR}) is exact up to ${\cal O}(\vecq^0)$, i.e., we did not assume the hierarchy between $m_a$ and $m_i$. Substituting Eq.~(\ref{eq:JNR}) back to Eq.~(\ref{eq:fourier-app}) and neglecting the ${\cal O}(\vecq^2)$ terms, we obtain
\begin{align}
    V_{\rm bkg}(r) &= - \frac{c_1^2 c_2^2}{f_a^4}m_1 m_2 \int\frac{\text{d}^3\veck}{(2\pi)^3}\frac{f(\veck)}{2E_{\veck}}\frac{m_a^8}{\left(4 m_1^2 E_{\veck}^2-m_a^4\right)\left(4m_2^2 E_{\veck}^2 - m_a^4\right)}\nonumber\\
    &\qquad\qquad\qquad\times\int\frac{\text{d}^3\vecq}{(2\pi)^3}e^{i\vecq\cdot\vecr} \left[\frac{1}{\vecq^2+2\veck\cdot\vecq} +\frac{1}{\vecq^2-2\veck\cdot\vecq}\right].\label{eq:Vbkg-app}
\end{align}

The remaining thing is to work out the Fourier transform of $\vecq$. To this end, we restore the $i\epsilon$ description (Feynman propagator) to the axion propagators  and perform the shift $\vecq \to \vecq-\veck$ (or $\vecq \to \vecq+\veck$) for the first (or second) term in the bracket:
\begin{align}
    &\int\frac{\text{d}^3\vecq}{(2\pi)^3}e^{i\vecq\cdot\vecr} \left[\frac{1}{\vecq^2+2\veck\cdot\vecq-i\epsilon} +\frac{1}{\vecq^2-2\veck\cdot\vecq-i\epsilon}\right]\nonumber\\
    &=e^{-i\veck\cdot \vecr} \int\frac{\text{d}^3\vecq}{(2\pi)^3}e^{i\vecq\cdot\vecr} \frac{1}{\vecq^2 - \veck^2
    -i\epsilon} + e^{i\veck\cdot \vecr} \int\frac{\text{d}^3\vecq}{(2\pi)^3}e^{i\vecq\cdot\vecr} \frac{1}{\vecq^2 - \veck^2 -i\epsilon}\nonumber\\
    &=-\frac{i}{4\pi^2 r}\left[\left(e^{-i\veck\cdot \vecr}  + e^{i\veck\cdot \vecr}\right) \int_{-\infty}^\infty {\rm d}\absq \frac{\absq e^{i\absq r}}{\absq^2-\veck^2-i\epsilon} \right]\nonumber\\
    &=\frac{1}{4\pi r}\left[\cos\left(\absk r-\veck\cdot\vecr\right)+\cos\left(\absk r+\veck\cdot\vecr\right)\right],\label{eq:Feynman-prop}
\end{align}
where in the final step the real part has been implicitly taken. 

Substituting Eq.~(\ref{eq:Feynman-prop}) into Eq.~(\ref{eq:Vbkg-app}), one obtains the final result of the background-induced axion force with the shift-invariant interaction:
\begin{align}
V_{\rm bkg}(r) = -\frac{c_1^2 c_2^2}{4\pi r}\frac{m_1 m_2}{f_a^4} \int\frac{\text{d}^3\veck}{(2\pi)^3}\frac{f(\veck)}{2E_{\veck}}&\frac{m_a^8}{\left(4 m_1^2 E_{\veck}^2-m_a^4\right)\left(4m_2^2 E_{\veck}^2 - m_a^4\right)}\nonumber\\
&\times\left[\cos\left(\absk r-\veck\cdot\vecr\right)+\cos\left(\absk r+\veck\cdot\vecr\right)\right].
\end{align}

\subsection{Relativistic corrections}
\label{appsub:relativistic-correction}
In the above calculations, we have taken the NR limit; that is, we neglected all the ${\cal O}(\vecq^2)$ terms in Eq.~(\ref{eq:JNR}). Due to the shift symmetry, cancellation occurs among the ${\cal O}(\vecq^0)$ terms, causing the remaining term to be suppressed by the axion mass.

In the following, we explicitly work out the next-leading-order correction of the ${\cal J}$-factor in Eq.~(\ref{eq:JNR}) under the NR approximation. To simplify the calculation, we introduce
\begin{align}
P_1\equiv p_1+p_1^\prime\;,\qquad P_2\equiv p_2+p_2^\prime\;.
\end{align}
Recall that the full solution of the Dirac equation in the Pauli-Dirac basis is given by
\begin{equation}
u\left(p_i\right)=\sqrt{E_i+m_i}
\left(\begin{array}{c}
\xi_i\\
\frac{\vecsigma\cdot\vecp_i}{E_i + m_i}\xi_i
\end{array}\right),\quad i=1,2\;,
\end{equation}
where $\xi_i$ is the two-component spinor of $\psi_i$.
Then the contraction of the wavefunctions gives
\begin{equation}
\begin{aligned}
\bar{u}\left(p_i^{\prime}\right)u(p_i)=\alpha_i\alpha_i^\prime-\frac{\vecP_i^{2}-\vecq^{2}\pm2\vecq\cdot\left(\vecP_i\times\boldsymbol{\sigma}_{i}\right)}{4\alpha_i\alpha_i^\prime}\;,
\end{aligned}
\end{equation}
with $\alpha_i\equiv \sqrt{E_i+m_{i}}$, $\alpha_i^{\prime}\equiv \sqrt{E_i^{\prime}+m_{i}}$ being defined. For signs, `$+$' corresponds to $i=1$ while `$-$' corresponds to $i=2$. Similarly,
\begin{align}
\bar{u}\left(p_i^{\prime}\right)\gamma_{0}^{}u(p_i) &= \alpha_i\alpha_i^\prime+\frac{\vecP_i^{2}-\vecq^{2}\pm2\vecq\cdot\left(\vecP_i\times\boldsymbol{\sigma}_{i}\right)}{4\alpha_i\alpha_i^\prime}\;,\\
\bar{u}\left(p_i^{\prime}\right)\gamma_{j}^{}u(p_i) &=-\frac{\alpha_i}{2\alpha_i^{\prime}}\left[\vecP_i\pm\vecq-\left(\vecP_i\pm\vecq\right)\times\boldsymbol{\sigma}_{i}\right]-\frac{\alpha_i^{\prime}}{2\alpha_i}\left[\vecP_i\mp\vecq+\left(\vecP_i\mp\vecq\right)\times\boldsymbol{\sigma}_{i}\right].
\end{align}

Then we substitute them back to Eq.~(\ref{eq:J-factor}) and expand the result as the inverse power of the fermion mass. After some straightforward calculations, one obtains 
\begin{align}
{\cal J}(k) & =   \frac{\left[\left(m_a^2-\vecq\cdot\veck\right)^2-E_{\veck}^2\vecq^2\right]^2}{4 m_1^2 m_2^2 E_{\veck}^4}  + {\cal O}\left(\frac{m_a^6}{m_i^6},\frac{\vecq^6}{m_i^6}\right),\label{eq:high order J(k)}\\
{\cal J}(k-q)&=\frac{\left[\left(m_a^2+\vecq\cdot\veck\right)^2-E_{\veck}^2\vecq^2\right]^2}{4 m_1^2 m_2^2 E_{\veck}^4}  + {\cal O}\left(\frac{m_a^6}{m_i^6},\frac{\vecq^6}{m_i^6}\right).\label{eq:high order J(k-q)}
\end{align}
As a check, by taking $\vecq^2 \to 0$, Eqs.~(\ref{eq:high order J(k)})-(\ref{eq:high order J(k-q)}) are reduced to Eq.~(\ref{eq:JNR}) at the leading order. 
We are interested in the effect that is not suppressed by the axion mass, so we set
$m_a=0$ for simplicity. Then Eqs.~(\ref{eq:high order J(k)})-(\ref{eq:high order J(k-q)}) are simplified to
\begin{align}
{\cal J}(k) ={\cal J}(k-q)=\frac{\left[\left(\vecq\cdot\hat{\veck}\right)^2-\vecq^2\right]^2}{4 m_1^2 m_2^2}  + {\cal O}\left(\frac{\vecq^6}{m_i^6}\right),
\end{align}
where the hat denotes the unit vector $\hat{\veck}\equiv \veck/|\veck|$.
Keeping the leading term and performing the Fourier transform, we get
\begin{align}
    V_{\rm bkg}(r) &=- \frac{c_1^2 c_2^2}{64\pi m_1m_2f_a^4} \int\frac{\text{d}^3\veck}{(2\pi)^3}\frac{f(\veck)}{E_{\veck}}\left[\nabla^2-\left(\hat{\veck}\cdot\nabla\right)^2\right]^2\frac{\cos\left(\absk r\right)\cos\left(\veck\cdot\vecr\right)}{r}\;.
\end{align}
Noticing that
\begin{equation}
    \left(\delta_{ij}-\hat{k}_i\hat{k}_j\right)k_i=\left(\delta_{ij}-\hat{k}_i\hat{k}_j\right)k_j=0\;,
\end{equation}
one can obtain
\begin{equation}
    \begin{aligned}
        V_{\rm bkg}(r) =-& \frac{c_1^2 c_2^2}{64\pi m_1m_2f_a^4} \int\frac{\text{d}^3\veck}{(2\pi)^3}\frac{f(\veck)}{E_{\veck}}\cos\left(\veck\cdot\vecr\right)\\
        &\times\left\{\cos\left(\absk r\right)\left[\frac{9-90c^2+105c^4}{r^5}+\frac{|\veck|^2\left(4-21c^2+45c^4\right)}{r^3}+\frac{|\veck|^4\left(1-3c^2+3c^4\right)}{r}\right]\right.\\
        &\qquad\left.+\sin\left(\absk r\right)\left[\frac{|\veck|\left(9-90c^2+105c^4\right)}{r^4}-\frac{|\veck|^3\left(2+2c^2-18c^4\right)}{r^2}\right]\right\},
        \label{eq:relativistic}
    \end{aligned}
\end{equation}
where $c\equiv\cos(\hat{\veck}\cdot\hat{\vecr})$ has been defined. At long distances, only the $1/r$ component is dominant, so we have
\begin{align}
V_{\rm bkg}(r) \approx - \frac{c_1^2 c_2^2}{64\pi r m_1m_2f_a^4} \int\frac{\text{d}^3\veck}{(2\pi)^3}\frac{f(\veck)}{E_{\veck}}|\veck|^4\cos\left(\veck\cdot\vecr\right) \cos\left(\absk r\right)\left(1-3c^2+3c^4\right).
\end{align}

\section{Decoherence effects on the axion force}
\label{app:decoherence}
In real fifth-force experiments, both the momentum space and the configuration space have some finite spread, which will suppress the axion force at long distances. In this appendix, we provide calculation details for this decoherence effect.

\subsection{Point-like object}
We first consider the decoherence effect between two point-like objects.
Without loss of generality, we choose $\vecv_a$ to be aligned with the $z$-axis and $\vecr$ in the $x$-$z$ plane. More specifically, we choose the following coordinates:
\begin{align}
\vecv_a = v_a\left(0,0,1\right),\qquad
\vecr = r\left(s_\alpha,0,c_\alpha\right),\qquad
\veck=\kappa\left(s_\theta c_\varphi,s_\theta s_\varphi,c_\theta\right),\label{eq:coordinate1}
\end{align}
where $c_x\equiv \cos x$, $s_x \equiv \sin x$. Then we have
\begin{align}
\veck \cdot \vecr &= \kappa r\left(s_\alpha s_\theta c_\varphi+c_\alpha c_\theta\right),\qquad
\int {\rm d}^3 \veck = \int_0^\infty {\rm d}\kappa \kappa^2 \int_0^{\pi}{\rm d}\theta\,s_\theta \int_0^{2\pi}{\rm d}\varphi\;,\nonumber\\
\left|\veck - m_a \vecv_a\right|^2 &= \kappa^2 + m_a^2 v_a^2 -2 \kappa m_a v_a c_\theta\;,
\end{align}
Then the phase-space form factor in Eq.~(\ref{eq:FPS}) can be explicitly written as
\begin{align}
{\cal F}_{\rm PS}\left(r,\alpha\right) &= \left(2\pi\right)^{-3/2}\kappa_0^{-3}\int_0^\infty {\rm d}\kappa \kappa^2 \exp\left(-\frac{\kappa^2 +m_a^2 v_a^2}{2\kappa_0^2}\right)\cos\left(2\kappa r\right)\nonumber\\
&\times\int_0^\pi {\rm d}\theta\,s_\theta \exp\left(\frac{\kappa m_a v_a}{\kappa_0^2}c_\theta\right)\int_0^{2\pi}{\rm d}\varphi \cos\left[\kappa r \left(s_\alpha s_\theta c_\varphi + c_\alpha c_\theta\right)
\right].  \label{eq:FPSapp} 
\end{align}
Integrating out $\varphi$ and using $m_a v_a =\sqrt{2}\kappa_0$, one obtains
\begin{align}
{\cal F}_{\rm PS}\left(r,\alpha\right) = \frac{e^{-1}}{\sqrt{2 \pi}\,\kappa_0^3}\int_0^\infty {\rm d}\kappa \kappa^2\cos\left(\kappa r\right)e^{-\frac{\kappa^2}{2\kappa_0^2}}\int_{-1}^1 {\rm d}z\, e^{\frac{\sqrt{2}\kappa z}{\kappa_0}}\cos\left(\kappa r c_\alpha z\right)J_0\left(\kappa r s_\alpha \sqrt{1-z^2}\right),  
\end{align}
where $z\equiv c_\theta$ and $J_0$ is the Bessel function of the first kind. For the special case of $\alpha=0$, i.e., when the DM wind is parallel to $\vecr$, we can work out the integral analytically:
\begin{align}
{\cal F}_{\rm PS}\left(r,0\right) &= \frac{2\,e^{-1}}{\sqrt{2\pi}\,\kappa_0^2\left(2+\kappa_0^2 r^2\right)}\int_0^\infty {\rm d}\kappa \kappa \cos\left(\kappa r\right)e^{-\frac{\kappa^2}{2\kappa_0^2}}\nonumber\\
&\qquad\qquad\qquad\times\left[\kappa_0 r \cosh\left(\frac{\sqrt{2} \kappa}{\kappa_0}\right)\sin\left(\kappa r\right)+\sqrt{2}\sinh\left(\frac{\sqrt{2}\kappa}{\kappa_0}\right)\cos\left(\kappa r\right)\right]\nonumber\\
&=\frac{1}{2+\kappa_0^2 r^2}\left\{1+e^{-2\kappa_0^2 r^2}\left[\left(1+\kappa_0^2 r^2\right)\cos\left(2\sqrt{2}\kappa_0 r\right)
-\frac{\kappa_0 r}{\sqrt{2}}\sin\left(2\sqrt{2}\kappa_0 r\right)\right]
\right\}.
\end{align}
It is easy to check that ${\cal F}_{\rm PS} \to 1$ for $\kappa_0 r\ll 1$ and ${\cal F}_{\rm PS} \to 1/(\kappa_0^2 r^2)$ for $\kappa_0 r \gg 1$.

\subsection{Finite-size object}

It is straightforward to generalize Eq.~(\ref{eq:FPSapp}) to arbitrary finite-size objects by integrating over the configuration space. In the following, we take the geometric setup in Fig.~\ref{fig:finite-size} as an example, where the source is a ball with a radius $R$ and the target is approximated to be point-like. 

We choose the center of the ball $O$ as the reference point, and the vector from $O$ to the target is denoted by $\vecr$. The angle between $\vecr$ and the preferred direction of the axion DM wind is $\alpha$.
For an arbitrary point located at $\vecr'$ in the ball, its contribution to $\vbkg$ is known, and the form factor is described by Eq.~(\ref{eq:FPSapp}) with the replacement $r \to |\vecr -\vecr'|$ and $\alpha \to \alpha_{\text{eff}}$, where $\alpha_{\rm eff}$ denotes the angle between $\vecr-\vecr'$ and $\vecv_a$. The total form factor, including both phase-space and finite-size decoherence effects, is obtained by integrating ${\cal F}_{\rm PS}$ over the ball [see Eq.~(\ref{eq:Ftotdef})]:
\begin{align}
{\cal F}_{\rm tot}\left(r,\alpha\right) = \frac{3r}{4\pi R^3}\int_{\rm ball} {\rm d}^3\vecr'\,\frac{{\cal F}_{\rm PS}\left(\left|\vecr-\vecr'\right|,\alpha_{\rm eff}\right)}{\left|\vecr -\vecr'\right|}\;. \label{eq:Ftotdefapp}   
\end{align}
We choose the coordinate of $\vecr'$ such that
\begin{align}
\vecr'=r'\left(s_\beta c_\gamma,s_\beta s_\gamma,c_\beta\right),\qquad
\int_{\rm ball} {\rm d}^3 \vecr' = \int_0^R{\rm d}r' r'^2 \int_{0}^{\pi}{\rm d}\beta\,s_\beta\int_0^{2\pi}{\rm d}\gamma\;.\label{eq:coordinate2}
\end{align}
Combining Eqs.~(\ref{eq:coordinate2}) and (\ref{eq:coordinate1}) and using the geometric relation, one obtains
\begin{align}
{\cal F}_{\rm tot} \left(r,\alpha\right) = \frac{3r}{4\pi R^3}\int_0^R{\rm d}r' r'^2 \int_{0}^{\pi}{\rm d}\beta\,s_\beta\int_0^{2\pi}{\rm d}\gamma \frac{{\cal F}_{\rm PS}\left(\left|\vecr-\vecr'\right|,\alpha_{\rm eff}\right)}{\left|\vecr -\vecr'\right|}\;, \label{eq:Ftotcompute}   
\end{align}
where 
\begin{align}
\left|\vecr-\vecr^\prime\right|=\sqrt{r^2+r^{\prime2}-2rr^\prime(s_\alpha s_\beta c_\gamma+c_\alpha c_\beta)}\;,\qquad
\alpha_{\rm eff} = \arccos\left(\frac{rc_\alpha-r^\prime c_\beta}{\left|\vecr-\vecr^\prime\right|}\right).\label{eq:alphaeff}
\end{align}

\section{Axion forces from solar axion flux}
\label{app:solar}
In this appendix, we calculate the axion force induced by solar axion flux. The general formula of $\vbkg$ for a directional axion beam is given in Eqs.~(\ref{eq:Vfluxmu})-(\ref{eq:Vfluxk0}), while the energy spectrum of the solar axion flux is described by Eq.~(\ref{eq:solar-spectrum}).

The integrals in Eqs.~(\ref{eq:Vfluxmu})-(\ref{eq:Vfluxk0}) can be analytically worked out using the following identity:
\begin{align}
    \int_0^\infty\text{d}\kappa\, \kappa^b e^{-c\kappa}\cos\left(d \kappa\right)=\left(c^2+d^2\right)^{-\frac{b+1}{2}}\cos\left[(b+1)\arctan\left(\frac{d}{c}\right)\right]\Gamma\left(b+1\right),
\end{align}
where $b, c, d$ are arbitrary positive constants. 

Comparing the axion force with gravity, we obtain
\begin{align}
\frac{V^\slashed{\rm shift}_\text{solar}}{V_{\rm grav}} &\approx 10^{-15}\left(\frac{\mu}{\mu_{\rm QCD}}\right)^2\left(\frac{2.5 \times 10^{3}~{\rm GeV}}{f_a}\right)^4 {\cal  D}_1(\alpha)\;,\\
\frac{V^\text{rel}_\text{solar}}{V_{\rm grav}} &\approx 10^{-15}\left(\frac{48~{\rm MeV}}{f_a/c_N}\right)^4 \left(1-3\cos^2\alpha+3\cos^4\alpha\right) {\cal  D}_2(\alpha)\;,
\end{align}
where the decoherence factors are given by:
\begin{align}
{\cal D}_1(\alpha) &\approx \left(1+1.5 \,r_{\si{\angstrom}}^2\sin^4\frac{\alpha}{2}\right)^{-1.2}\cos\left[2.5\arctan\left(1.2\,r_{\si{\angstrom}}\sin^2\frac{\alpha}{2}\right)\right]\nonumber\\
&+\left(1+1.5 \,r_{\si{\angstrom}}^2\cos^4\frac{\alpha}{2}\right)^{-1.2}\cos\left[2.5\arctan\left(1.2\,r_{\si{\angstrom}}\cos^2\frac{\alpha}{2}\right)\right],\label{eq:D1}\\
{\cal D}_2(\alpha) &\approx \left(1+1.5 \,r_{\si{\angstrom}}^2\sin^4\frac{\alpha}{2}\right)^{-3.2}\cos\left[6.5\arctan\left(1.2\,r_{\si{\angstrom}}\sin^2\frac{\alpha}{2}\right)\right]\nonumber\\
&+\left(1+1.5 \,r_{\si{\angstrom}}^2\cos^4\frac{\alpha}{2}\right)^{-3.2}\cos\left[6.5\arctan\left(1.2\,r_{\si{\angstrom}}\cos^2\frac{\alpha}{2}\right)\right], \label{eq:D2}
\end{align}
with 
\begin{align}
r_{\si{\angstrom}}\equiv \frac{r}{0.1\,{\rm nm}}\;.    
\end{align}

Note that the typical energy scale of solar axion flux is $\kappa_0 \sim {\rm keV}$, with the de Broglie wavelength $\db \sim 1/\kappa_0\sim 0.2\,{\rm nm}$. So, for $r\gg 0.1\,{\rm nm}$, the decoherence effect will suppress the sensitivity. This can be seen from Eqs.~(\ref{eq:D1})-(\ref{eq:D2}). For $r_{\si{\angstrom}} \alpha^2 \gg 1$, ${\cal D}_1$ and ${\cal D}_2$ are suppressed by the inverse power of $r_{\si{\angstrom}}$:
\begin{align}
{\cal D}_1\left(\alpha \gg 1/\sqrt{r_{\si{\angstrom}}}\right) &\approx -0.43 \times  r_{\si{\angstrom}}^{-2.4} \left[\left(\sin\frac{\alpha}{2}\right)^{-4.8}+\left(\cos\frac{\alpha}{2}\right)^{-4.8}
\right],\label{eq:D1larger}\\
{\cal D}_2\left(\alpha \gg 1/\sqrt{r_{\si{\angstrom}}}\right) &\approx -0.19 \times  r_{\si{\angstrom}}^{-6.4} \left[\left(\sin\frac{\alpha}{2}\right)^{-12.8}+\left(\cos\frac{\alpha}{2}\right)^{-12.8}
\right]. \label{eq:D2larger} 
\end{align}
On the other hand, in the small angle limit, $\alpha \ll 1/\sqrt{r_{\si{\angstrom}}}$, we have ${\cal D}_1, {\cal D}_2 \sim {\cal O}(1)$.

\end{appendix}


\bibliographystyle{JHEP}
\bibliography{ref}

\end{document}